\documentclass[a4paper,12pt,dvipsnames,usenames]{article}
\pdfoutput=1

\usepackage{etex} \usepackage{jheppub} 
\usepackage{graphicx} \graphicspath{{fig/}{./Plots/}}

\usepackage{tikz} 

\usepackage{amsmath} \usepackage{amssymb} \usepackage{array} \usepackage{bm} \usepackage{color,xcolor,colortbl} \usepackage[utf8]{inputenc} \usepackage{paralist} \usepackage[section]{placeins} \usepackage{slashed}
\usepackage[final,textsize=footnotesize,color=ForestGreen!05]{todonotes}
\usepackage{units} \usepackage{varioref} \usepackage{xspace} \usepackage{longtable}

\usepackage{footnote} \makesavenoteenv{tabular} \makesavenoteenv{table}

\newcommand{\tikzcolor}[2]{\tikz[baseline]{\node[fill=#1!20,anchor=base] (a){\ensuremath{#2}}; }} \newcommand{\inElena}[1]{{\tikzcolor{blue}{#1}}} \newcommand{\ket}[1]{\left|#1\right\rangle} \newcommand{\bra}[1]{\left\langle#1\right|} \newcommand{\parfrac}[2]{\left(\frac{#1}{#2}\right)} \newcommand{\BR}{\mathop{\text{BR}}} \newcommand{\Jpsi}{{\ensuremath{J/\psi}}}  \newcommand{\ship}{{SHiP}\xspace} \newcommand{\GeV}{\:\unit{GeV}}    \newcommand{\CM}{\mathcal{M}}  %

\usepackage[colorlinks=true,citecolor=blue]{hyperref}

\title{Phenomenology of GeV-scale Heavy Neutral Leptons}
\author[1]{Kyrylo~Bondarenko,}
\author[1]{Alexey~Boyarsky,}
\author[2,3]{Dmitry~Gorbunov,}
\author[4]{Oleg~Ruchayskiy}

\affiliation[1]{Intituut-Lorentz, Leiden University, Niels Bohrweg 2, 2333 CA Leiden, The Netherlands} 
\affiliation[2]{Institute for Nuclear Research of the Russian Academy of Sciences, Moscow 117312, Russia} 
\affiliation[3]{Moscow Institute of Physics and Technology, Dolgoprudny 141700, Russia}
\affiliation[4]{Discovery Center, Niels Bohr Institute, Copenhagen University, Blegdamsvej 17, DK-2100 Copenhagen, Denmark} 

\emailAdd{bondarenko@lorentz.leidenuniv.nl}
\emailAdd{boyarsky@lorentz.leidenuniv.nl}
\emailAdd{gorby@inr.ac.ru}
\emailAdd{oleg.ruchayskiy@nbi.ku.dk}

\abstract{We review and revise phenomenology of the GeV-scale heavy neutral
leptons (HNLs). We extend the previous analyses by including more channels of HNLs production and
decay and provide with more refined treatment, including QCD corrections for the HNLs
of masses $\mathcal{O}(1)$\,GeV. We summarize  the relevance of individual production and
decay channels for different masses, resolving a few discrepancies in the literature.
Our final results are directly suitable for sensitivity studies of
particle physics experiments (ranging from proton beam-dump to the LHC) aiming
at searches for heavy neutral leptons.}
\preprint{INR-TH-2018-014}
\arxivnumber{1805.08567}

\begin{document}

\maketitle
\flushbottom

\section{Introduction: heavy neutral leptons}
\label{sec:hnl-phenomenology}

We review and revise phenomenology of the heavy neutral leptons (HNLs) with masses in the GeV range.
The interest to these particles has recently increased, since it was recognized that they are capable of resolving 3 major observational BSM phenomena: neutrino oscillation, baryon asymmetry of the universe and dark matter~\cite{Asaka:2005an,Asaka:2005pn} (for review see e.g.~\cite{Boyarsky:2009ix,Drewes:2013gca}, \cite[Chapter 4]{Alekhin:2015byh} and references therein).

Several particle physics experiments, that put the searches for heavy neutral leptons among their scientific goals, have been proposed in the recent years: DUNE~\cite{Adams:2013qkq}, NA62~\cite{Mermod:2017ceo,CortinaGil:2017mqf,Drewes:2018gkc} SHiP~\cite{Alekhin:2015byh,Anelli:2015pba}, CODEX-b~\cite{Gligorov:2017nwh}, MATHUSLA~\cite{Chou:2016lxi,Curtin:2017izq,Evans:2017lvd,Helo:2018qej}, FASER~\cite{Feng:2017uoz,Feng:2017vli,Kling:2018wct}.
The searches for HNLs (also often called ``Majorana neutrinos'' or ``sterile
neutrinos'') have been performed and are ongoing at the experiments LHCb, CMS,
ATLAS, T2K, Belle (see
e.g.~\cite{Liventsev:2013zz,Aaij:2014aba,Khachatryan:2015gha,Aad:2015xaa,Sirunyan:2018mtv,Izmaylov:2017lkv}) with many more proposals for novel ways to search for them~\cite{Asaka:2012bb,Asaka:2013jfa,Blondel:2014bra,Asaka:2014kia,Canetti:2014dka,Das:2014jxa,Gago:2015vma,Antusch:2015mia,Das:2015toa,Asaka:2015qma,Asaka:2015oia,Asaka:2016rwd,Cvetic:2016fbv,Caputo:2016ojx,Rasmussen:2016njh,Das:2017zjc,Antusch:2017hhu,Caputo:2017pit,Das:2017rsu,Chun:2017spz,Yue:2017mmi,Antusch:2017pkq,Das:2018hph,Cvetic:2018elt}.
This interest motivates the current revision.
The information relevant for sensitivity studies of the GeV-scale HNLs is scattered around the research literature~\cite{Johnson:1997cj,Gribanov:2001vv,Gorbunov:2007ak,Ramazanov:2008ph,Atre:2009rg,Helo:2010cw,Abada:2013aba,Helo:2013esa,Elena,Cvetic:2016fbv,Cvetic:2018elt,Rasmussen:2016njh} and is sometimes controversial.
We collect all relevant phenomenological results and present them with the unified notation, discussion of the relevance of the individual channels and references to the latest values of phenomenological parameters (meson form factors) that should be used in practical application.
The relevance of individual channels depending on the masses of HNLs is present in the resulting Table~\ref{tab:decaychannels}.
We also discuss existing discrepancies in the literature, pointing out the way of obtaining the correct results and analyze new channels of production and new modes of decay neglected in the previous literature.

\subsection{General introduction to heavy neutral leptons}

Heavy neutral leptons or sterile neutrinos $N$ are singlets with respect to
the SM gauge group and couple to the gauge-invariant combination
$(\bar L^c_\alpha \cdot \tilde H)$ (where $L_{\alpha}$, $\alpha=1,\dots,3$,
are SM lepton doublet, $\tilde{H}_i=\varepsilon_{ij} H^*_j$ is conjugated SM
Higgs doublet) as follows
\begin{equation}
  \label{ship_introduction:eq:1}
  \mathcal{L}_\text{Neutrino portal} = F_{\alpha}(\bar L_\alpha\cdot
  \tilde H)N + h.c.\;, 
\end{equation}
with $F_{\alpha}$ denoting dimensionless Yukawa couplings.  The name ``sterile
neutrino'' stems from the fact that the
interaction~\eqref{ship_introduction:eq:1} fixes SM gauge charges of $N$ to be
zero.  After electroweak symmetry breaking the SM Higgs field gains nonzero
vacuum expectation value $v$ and interaction \eqref{ship_introduction:eq:1}
provides heavy neutral leptons and SM (or \emph{active}) neutrinos --- with
the mixing mass term ($v=246$\,GeV)
\[
  M^D_{\alpha}\equiv F_{\alpha} v/\sqrt{2}\,.
\]
The truly neutral nature of $N$ allows one to introduce for it a Majorana mass term,
consistent with the SM gauge invariance, 
resulting in the HNL Lagrangian at GeV scale
\begin{equation}
  \label{eq:4}
  \mathcal{L}_{\text{HNL}} = 
  i \bar{N} \slashed\partial N + 
  \left( M^D_{\alpha}\bar{\nu}_{\alpha} N - \frac{M_{N}}{2} \bar{N}^c N + h.c. \right).
\end{equation}
The mass eigenstates of the active-plus-sterile sector are the mixtures of
$\nu$ and $N$, but with small mixing angles and large splitting between mass
scales of sterile and active neutrinos.  The heavy mass eigenstates are
``almost sterile neutrinos'' while light mass eigenstates are ``almost active
neutrinos''.  In what follows we keep the same terminology for the mass states
as for the gauge states.  As a result of mixing, HNL couples to the SM fileds
in the same way as active neutrinos,
\begin{equation}
  \label{eq:10}
  \mathcal{L}_{int} = \frac{g}{2\sqrt 2} W^+_\mu \overline{N^c} \sum_\alpha U_{\alpha}^* \gamma^\mu (1-\gamma_5) \ell^-_\alpha + \frac{g}{4 \cos\theta_W}Z_\mu \overline{N^c} \sum_\alpha U_{\alpha}^* \gamma^\mu (1-\gamma_5) \nu_\alpha + \text{h.c.}\;,
\end{equation}
except the coupling is strongly suppressed by the small \emph{mixing angles}
\begin{equation}
  \label{eq:3}
  U_{\alpha} = M^D_{\alpha} M^{-1}_{N}
\end{equation}
In~\eqref{eq:10} $\ell_\alpha$ are charged leptons of the three SM
generations.

The number of model parameters increase with the number of HNLs (see e.g.
reviews~\cite{Boyarsky:2009ix,Drewes:2013gca}).  In particular in the model
with 2 sterile neutrinos there are 11 free parameters and in the case of 3
sterile neutrinos there are 18 parameters~\cite{Boyarsky:2009ix}.  Not all of
them play important role in phenomenology.  The collider phenomenology is
sensitive only to masses of the HNL(s) and absolute values of mixing angles,
$|U_{\alpha}|$. When sterile neutrinos are not degenerate in mass, in all the
processes they are produced and decay independently, without oscillations
between themselves, in contrast to the behavior of active neutrinos~\cite{Anamiati:2017rxw,Antusch:2017ebe,Cvetic:2018elt}. So, from
the phenomenological point of view it is enough to describe only 1 sterile
neutrino, which needs only 4 parameters: sterile neutrino mass $M_N$ and
sterile neutrino mixings with all three active neutrinos $U_{\alpha}$,
Eq.~\eqref{eq:4}.

The papers is organized as follows: in Section~\ref{sec:hnl_production} we
review the different HNL production channels; in Section~\ref{sec:decays} we
discuss the most relevant HNL decay channels. The summary and final discussion
is given in the section~\ref{sec:summary}.
Appendices provide necessary technical clarifications.


\section{HNL production in proton fixed target  experiments} 
\label{sec:hnl_production}

In fixed target experiments (such as NA62, SHiP or DUNE) the initial
interaction is proton-nuclei collision. In such collisions HNLs can be
produced in a number of ways:
\begin{compactenum}[\it a)]
\item Production from hadron's decays; 
\item Production from Deep Inelastic
  Scattering (DIS) p-nucleon interaction;
\item Production from the coherent proton-nucleus scattering.
\end{compactenum}
Below we provide overview of each of the channels summarizing previous results and emphasizing novel points.

\subsection{Production from hadrons}
\label{sec:production-from-hadrons}

The main channels of HNL production from hadrons are via decays of sufficiently long-lived hadrons, i.e.\ the lightest hadrons of each flavour\footnote{\label{fn:1}Such hadrons decay \textit{only} through weak interactions with relatively small decay width (as compared to electromagnetic or strong interaction).
  As the probability of HNL production from the hadron's decay is inversely proportional to the hadron's decay width, the HNL production from the lightest hadrons is significantly more efficient.}.
In the framework of the Fermi theory, the decays are inferred by the weak charged currents.
One can also investigate the hidden flavored mesons $\Jpsi(c\bar{c},3097)$, $\Upsilon (b\bar{b},9460)$ as sources of HNLs.
These mesons are short-lived, but 1.5--2 times heavier than the corresponding open flavored mesons, giving a chance to produce heavier HNLs.

\begin{figure}[!t]
 \centering
  \includegraphics[width=0.9\textwidth]{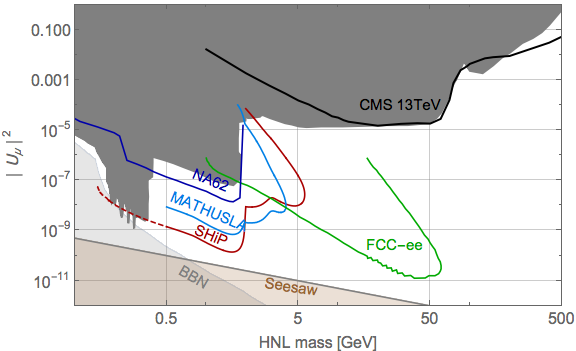}
  \caption{Existing limits and future prospects for searches for HNLs.
    Only mixing with muon flavour is shown.
    For the list of previous experiments (gray area) see~\cite{Alekhin:2015byh}.
    Black solid line is the recent bounds from the CMS 13~TeV run~\cite{Sirunyan:2018mtv}.
    The sensitivity estimates from prospective experiments are based
    on~\cite{Blondel:2014bra} (FCC-ee), \cite{Drewes:2018gkc} (NA62),
    \cite{SHiP:2018xqw} (SHiP) and \cite{Curtin:2018mvb} (MATHUSLA@LHC).
    The sensitivity of SHiP below kaon mass (dashed line) is based on the number of
    HNLs produced in the decay of $D$-mesons only and does not take into
    account contributions from kaon decays, see~\cite{SHiP} for details.
    The primordial nucleosynthesis bounds on HNL lifetime are from~\cite{Dolgov:2000jw}.
    The Seesaw line indicates the parameters obeying the seesaw relation $|U_\mu|^2\sim m_\nu/M_N$, where for active neutrino mass we substitute $m_\nu = \sqrt{\Delta m_{\text{atm}}^2} \approx \unit[0.05]{eV}$~\cite{Alekhin:2015byh}.}
  \label{fig:HNLbounds}
\end{figure}

As the region of HNL masses below that of the kaon is strongly bounded by the previous experiments (see~\cite{Alekhin:2015byh} for details, reproduced in Fig.~\ref{fig:HNLbounds}), in what follows we concentrate on production channels for HNL masses $M_N>0.5\GeV$.

\begin{figure}[!htb]
  \centering
  \includegraphics[width=0.45\textwidth]{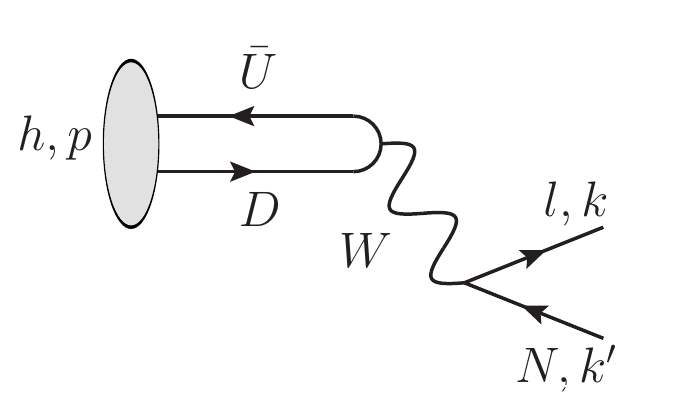}~  \includegraphics[width=0.45\textwidth]{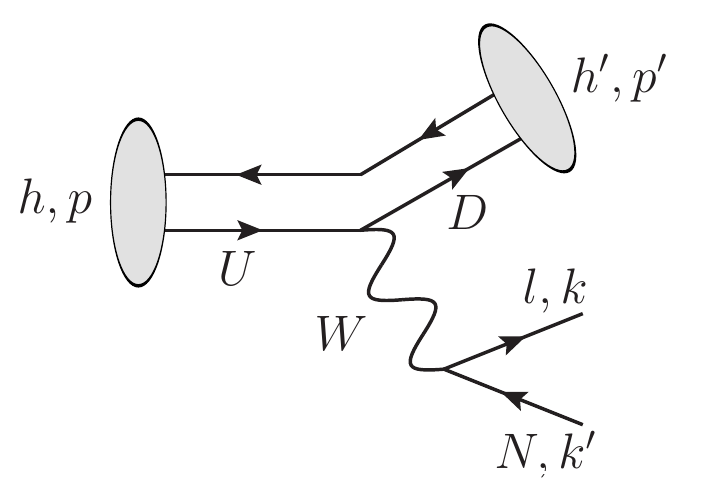}
  \caption{\emph{Left:} The diagram of leptonic decay of the meson $h$
    with 4-momentum $p$. \emph{Right:} The diagram of semileptonic
    decay of the meson $h$ with 4-momentum $p$ into meson $h'$ with
    4-momentum $p'$. In both diagrams the transferred to the lepton
    pair 4-momentum is $q = k + k'$.}
  \label{fig:h2l_decay}
\end{figure}

HNLs are produced in meson decays via either 2-body purely leptonic decays (left panel of Fig.~\ref{fig:h2l_decay}) or semileptonic decays (right panel of Fig.~\ref{fig:h2l_decay})~\cite{Shrock:1980ct,Shrock:1980vy}.
The branching fractions of leptonic decays have been found e.g.\ in~\cite{Johnson:1997cj,Gorbunov:2007ak}.
For the semileptonic decays only the processes with a single pseudo-scalar or vector meson in the final state have been considered so far~\cite{Gorbunov:2007ak} (see also~\cite{Abada:2013aba} and \cite{Cvetic:2016fbv})
\begin{align}
  \label{eq:20}
  h \to h'_P \ell N\\
  h \to h'_V \ell N
\end{align}
(where $h'_P$ is a \emph{pseudo-scalar} and $h'_V$ is a \emph{vector} meson)
and their branching ratio has been computed. We reproduce these
computations in the Appendix~\ref{sec:production} paying special
attention to the treatment of form factors.

Finally, to calculate the number of produced HNLs one should ultimately know  the
\textit{production fraction}, $f(\bar q q \to h)$ -- the
  probability to get a given hadron from the corresponding heavy quark.
The latter can either be determined experimentally or computed from Pythia simulations (as e.g. in~\cite{Elena}).

\subsubsection{Production from light unflavored and strange mesons}

\begin{figure}[!t]
  \centering
  \includegraphics[width=0.7\textwidth]{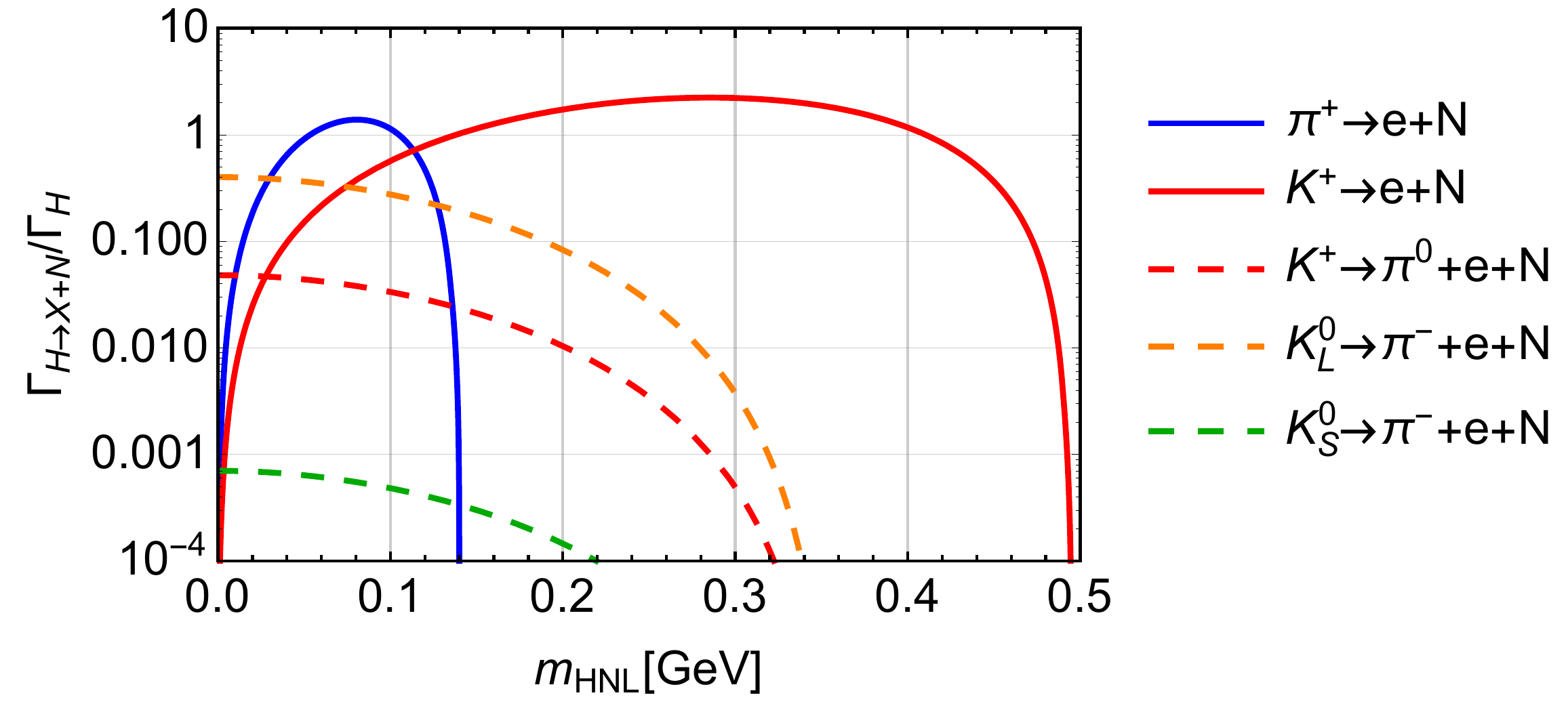}
  \caption[Decays of pions and kaons to HNLs]{Light mesons decay width to HNLs related to the measured value of the total decay width for pions and kaons correspondingly. In this Figure we take $U_e=1$, $U_\mu=U_{\tau}=0$. The ratio for two-body decay channels exceeds 1 due to the helicity enhancement when a massive HNL is present in the final state instead of neutrino.}
  \label{fig:piKbranching}
\end{figure}

Among the light unflavored and strange mesons the relevant mesons for the HNL
production are:\footnote{The particle lists here and below is given in the format 'Meson name(quark contents, mass in MeV)').} $\pi^+(u\bar{d},139.6)$, $K^+(u\bar{s},494)$, $K^0_S(d\bar{s},498)$ and $K^0_L(d\bar{s},498)$. 

The only possible production channel from the $\pi^+$ is the two body decay \mbox{$\pi^+\to \ell^+_{\alpha}N$} with $\ell = e,\mu$. The production from $K^+$ is possible through the two-body decay of the same type.
There are also 3-body decays $K^+\to\pi^0 \ell^+_{\alpha}N$ and $K^0_{L/S}\to\pi^- \ell^+_{\alpha}N$.

The resulting branching ratios for corresponding mesons are shown in Fig.~\ref{fig:piKbranching}.
For small HNL masses the largest branching ratio is that of $K^0_{L}\to\pi^- \ell^+_{\alpha}N$ due to the helicity suppression in the two-body decays and small $K^0_{L}$ decay width.

\subsubsection{Production from charmed mesons}

The following charmed mesons are most relevant for the HNL production: 
$D^0(c\bar{u},1865)$, $D^+(c\bar{d},1870)$,
$D_s(c\bar{s},1968)$. 

$D^0$ is a neutral meson and therefore its decay through the charged current interaction necessarily involves a meson in a final state.
The largest branching is to $K$ meson, owing to the CKM suppression $|V_{cd}|/|V_{cs}|\approx 0.22$.
Then the mass of the resulting HNL is limited as $M_N < M_D - M_K \approx 1.4\GeV$.
For the charmed \textit{baryons} the same argument is applicable: they
should decay into baryons and the most probable is strange baryon,
hence 
$M_N < M_{\Lambda_c} - M_{\Lambda} \approx 1.2\GeV$.
Therefore these channels are open only for HNL mass \textit{below} $\sim 1.4\GeV$.

\begin{figure}[!t]
  \centering
  \includegraphics[width=0.8\textwidth]{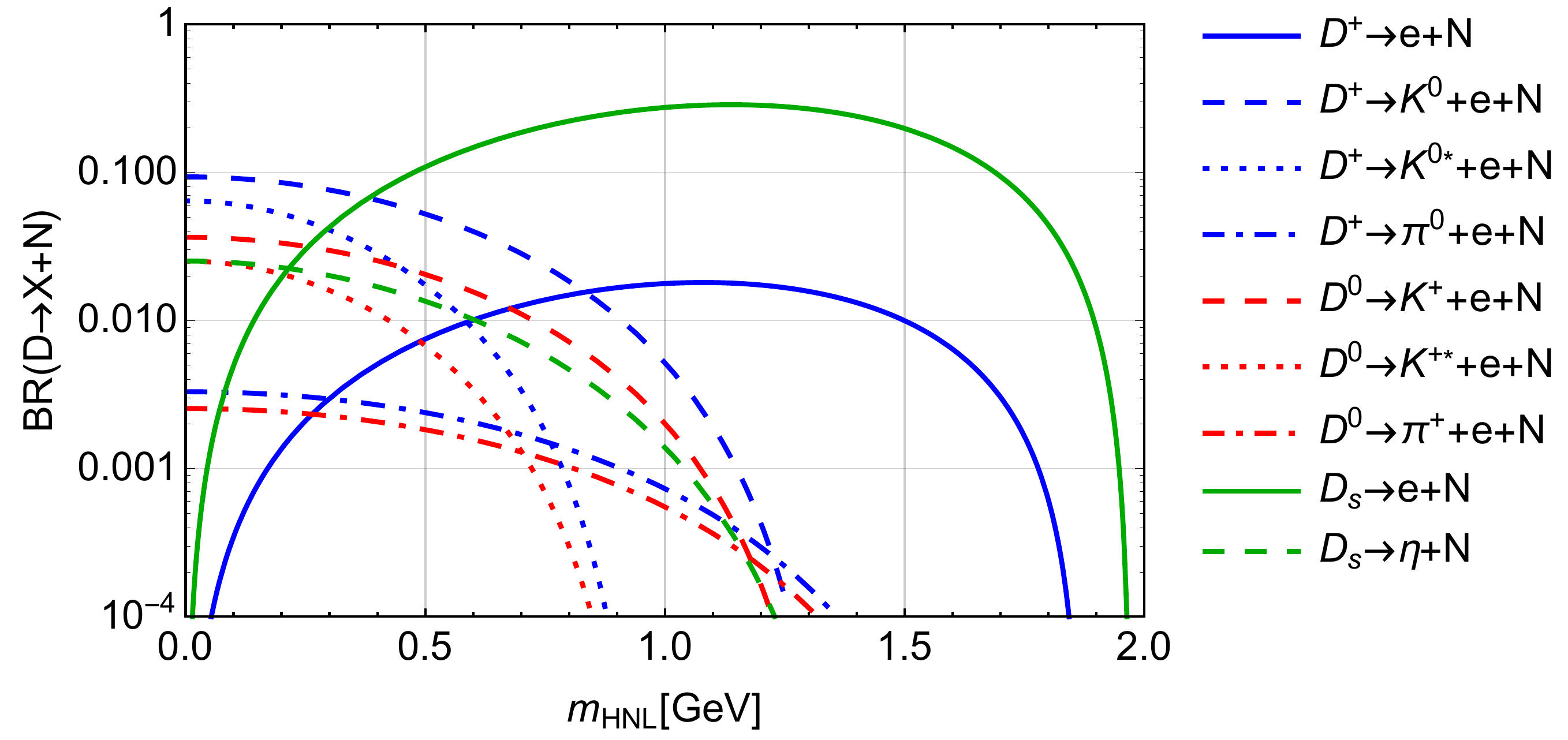}
  \caption[D-mesons to HNLs]{Dominant branching ratios of HNL production from different charmed and beauty mesons. For charged mesons two-body leptonic decays are shown, while for the neutral mesons decays are necessarily semi-leptonic. For these plots we take $U_e=1$, $U_\mu=U_{\tau}=0$.}
  \label{fig:Dbranching}
\end{figure}

Charged charmed mesons $D^\pm$ and $D_s$ would exhibit two body decays into an HNL and a charged lepton, so they can produce HNLs almost as heavy at themselves.
The branching of $D_s \to N + X$ is more than a factor 10 larger than
any similar of other $D$-mesons.
The number of $D_s$ mesons is of course suppressed as compared to $D^\pm$ and
$D^0$ mesons, however only by a factor of few\footnote{For example at SPS energy
  (400\,GeV) the production fractions of the charmed mesons are given by $
  f(D^+)=0.204$, $ f(D^0)=0.622$, $ f(D_s)=0.104$~\cite{Elena}.}.
Indeed, at energies relevant for $\bar c c$ production,
the fraction of strange quarks is already sizeable, $\chi_{\bar s s} \sim 1/7$~\cite{Olive:2016xmw}.
\emph{As a result, the two-body decays of $D_s$ mesons dominate in the HNL production from charmed mesons}, see Fig.~\ref{fig:Dbranching}.

\begin{figure}[!t]
  \centering
  \includegraphics[width=0.99\textwidth]{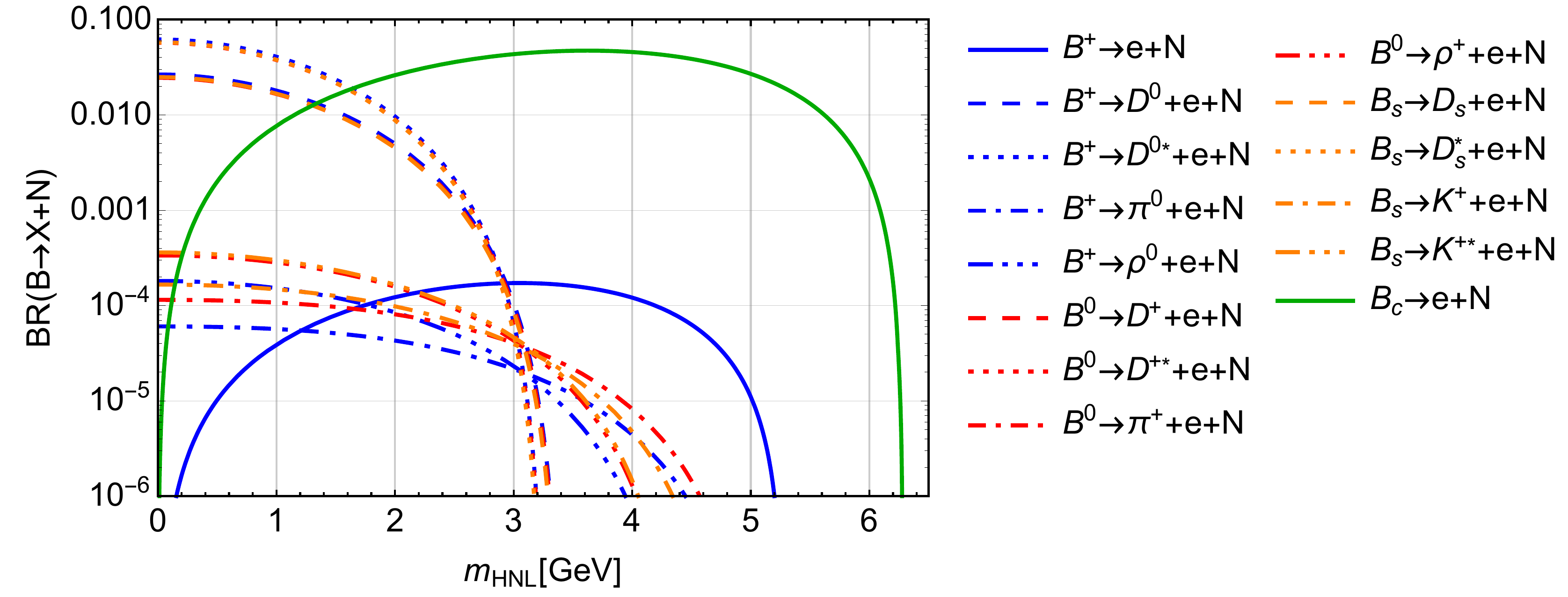}
  \caption[B-mesons to HNLs]{Dominant branching ratios of HNL production from different beauty mesons. For charged mesons two-body leptonic decays are shown, while for the neutral mesons decays are necessarily semi-leptonic. For these plots we take $U_e=1$, $U_\mu=U_{\tau}=0$.}
  \label{fig:Bbranching}
\end{figure}

\begin{footnotesize}
  \renewcommand{\arraystretch}{1.25}
\begin{table}[htb!]
\centering
    \begin{tabular}{|l|l|c|}
        \hline
        \multicolumn{2}{|c|}{Decay $B^+\to \ell^+ \nu_{\ell} X$ } & BR [\%] \\
        \hline
        \rowcolor[gray]{.9}
        \multicolumn{2}{|l|}{Inclusive branching: \;\;\;$l=e,\,\mu$}  & $11.0\pm0.3$ \\
        \hline
        Dominant one-meson channels: pseudo-scalar meson &$\strut\overline{D}^0 \ell^+ \nu_{\ell}$ & $2.27\pm0.11$\;\; \\
        \cline{2-3}
        \hphantom{Dominant one-meson channels:} vector meson&$\overline{D}^*(2007)^0 \ell^+ \nu_{\ell}$\, & $5.7\pm0.19$ \\
        \hline
        \rowcolor[gray]{.9}
        \multicolumn{2}{|l|}{Two above channels together: }&$8.0 \pm0.2$\\
        \hline
        Channels with 2 meson: &$D^- \pi^+ \ell^+ \nu_{\ell}$ & $0.42\pm0.05$ \\
       \cline{2-3}
        &$D^{*-} \pi^+ \ell^+ \nu_{\ell}$ & $0.61\pm0.06$ \\
\hline
${D}^- \pi^+ \ell^+ \nu_{\ell}$ above is
saturated by 1 meson modes &$\overline{D}^{*}_0(2420)^0\ell^+ \nu_{\ell}$ & $0.25\pm0.05$ \\
        \cline{2-3}
        &$\overline{D}_2^{*}(2460)^0 \ell^+ \nu_{\ell}$ &
        $0.15\pm0.02$ \\
        \hline
${D}^{*-} \pi^+ \ell^+ \nu_{\ell}$ is
augmented with 1 meson modes &$\overline{D}_1(2420)^0\ell^+ \nu_{\ell}$ & $0.30\pm0.02$ \\
        \cline{2-3}
        &$\overline{D}'_1(2430)^0 \ell^+ \nu_{\ell}$ & $0.27\pm0.06$ \\
       
\cline{2-3}
        &$\overline{D}^*_2(2460)^0 \ell^+ \nu_{\ell}$ & $0.1\pm0.02$\\
        \hline
\rowcolor[gray]{.9}
        \multicolumn{2}{|l|}{Hence 1-meson modes contribute
          additionally}  & $1.09\pm 0.12$\\
        \hline
        Sum of other multimeson channels, $n>1$: &$\overline{D}^{(*)} n\pi \ell^+ \nu_{\ell}$ & $0.84\pm 0.27$ \\
        \hline
\rowcolor[gray]{.9}
        \multicolumn{2}{|l|}{Inclusive branching: \;\;\;$l=\tau$}  & not known \\
        \hline
Dominant one-meson channels: pseudo-scalar meson &$\strut\overline{D}^0 \tau^+ \nu_{\tau}$ & $0.77\pm0.25$\;\; \\
        \cline{2-3}
        \hphantom{Dominant one-meson channels:} vector meson&$\overline{D}^*(2007)^0 \tau^+ \nu_{\tau}$\, & $1.88\pm0.20$ \\
        \hline        
    \end{tabular}
    \caption{Experimentally measured branching widths for the main semileptonic decay modes of the $B^+$ and $B^0$ meson~\cite{Olive:2016xmw}. Decays to pseudoscalar ($D$) and vector ($D^*$) mesons together constitute $73\%$ (for $B^+$) and 69\% (for $B^0$). Charmless channels are not shown because of their low contribution} 
    \label{tab:Bdecays}
\end{table}
\end{footnotesize}

\subsubsection{Production from beauty mesons}
  
The lightest beauty mesons are $B^-(b\bar{u},5279)$, $B^0(b\bar{d},5280)$, $B_s (b\bar{s},5367)$, $B_c (b\bar{c},6276)$.
Similarly to the $D^0$ case, neutral $B$-mesons ($B^0$ and $B_s$) decay through charged current with a meson in a final state.
The largest branching is to $D$ meson because of the values of CKM matrix elements ($|V_{cb}|/|V_{ub}|\approx 0.1$).
Thus the mass of the resulting HNL is limited: $M_N < M_B - M_D \approx 3.4\GeV$.

Charged beauty mesons $B^\pm$ and $B^\pm_c$ have two body decays into HNL and charged lepton, so they can produce HNLs almost as heavy at themselves.
Due to the CKM suppression the branching ratio of $B^+ \to N + \ell^+$
is significantly smaller than that of $B_c \to N + \ell$. However, unlike the case of $D_s$ mesons, the production fraction of $f(b \to B_c)$ has only been measured at
LHC energies, where it is reaching few${}\times 10^{-3}$~\cite{Aaij:2017kea}. At lower energies it is not known. Branching ratio of $B$-mesons into HNL for different decay channels and pure electron mixing is shown at Fig.~\ref{fig:Bbranching}.

\subsubsection{Multi-hadron final states}
\label{sec:multi-hadron-final}

$D$ and especially $B$ mesons are heavy enough to decay into HNL and multimeson final states.
While any single multi-meson channel would be clearly phase-space suppressed
as compared to 2-body or 3-body decays, considered above, one should check
that the ``inclusive'' multi-hadron decay width does not give a sizeable contribution.

To estimate relative relevance of single and multi-meson decay
channels, we first consider the branching ratios of the semileptonic
decays of $B^+$ and $B^0$ (with ordinary (massless) neutrino  $\nu_\ell$ in the final state)
\begin{equation}
  \label{eq:30}
  B \to \ell^+ \nu_\ell X \;\;  , \;\;\;l=e,\,\mu\,,
\end{equation}
where $X$ are one or many hadrons.
The results are summarized in Table~\ref{tab:Bdecays}.
Clearly, by taking into account \emph{only} single meson states we would \emph{underestimate} the total inclusive width of the process~\eqref{eq:30} by about 20\%.

\begin{figure}[!t]
  \centering \includegraphics[width=0.5\textwidth]{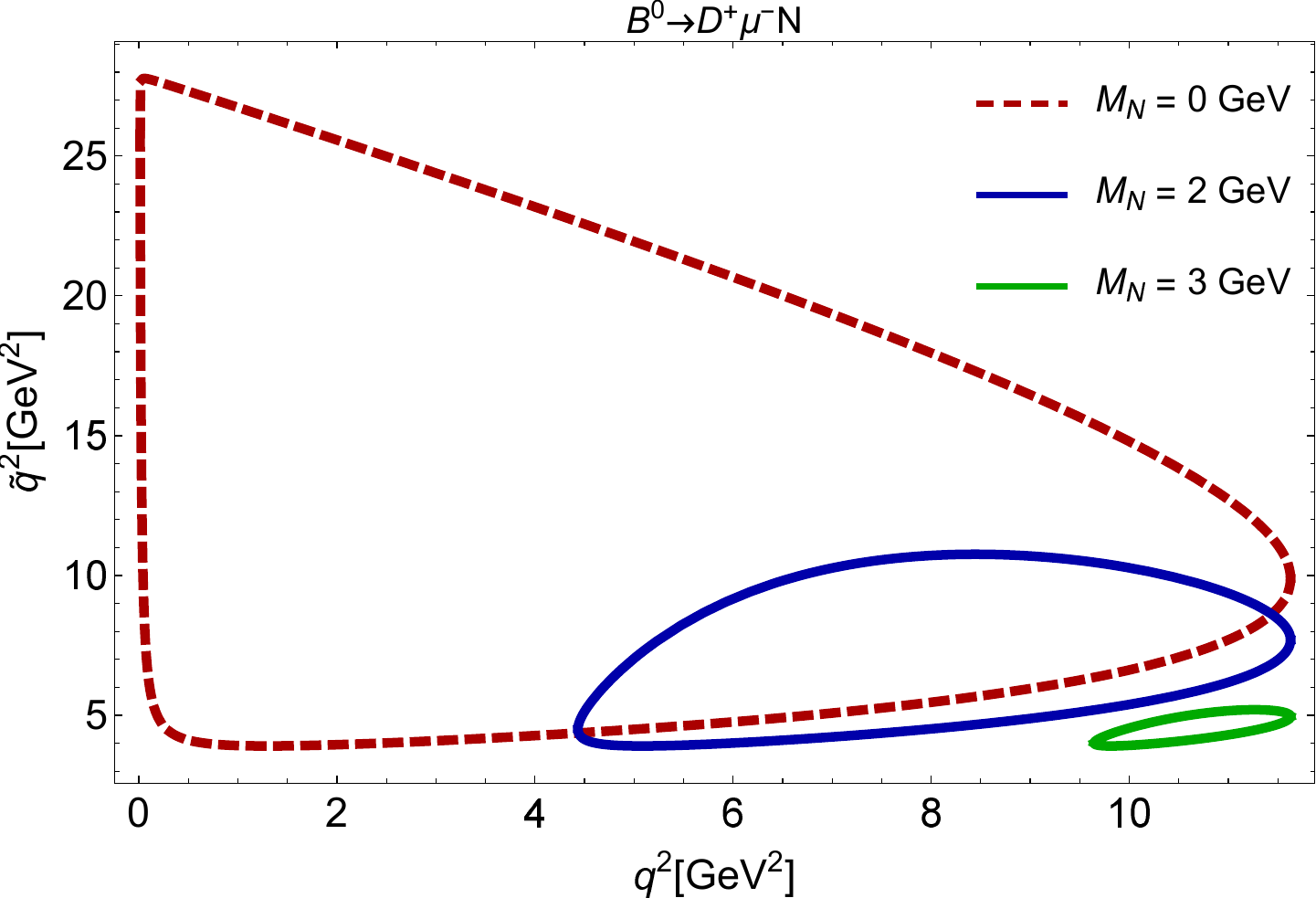}~ \includegraphics[width=0.5\textwidth]{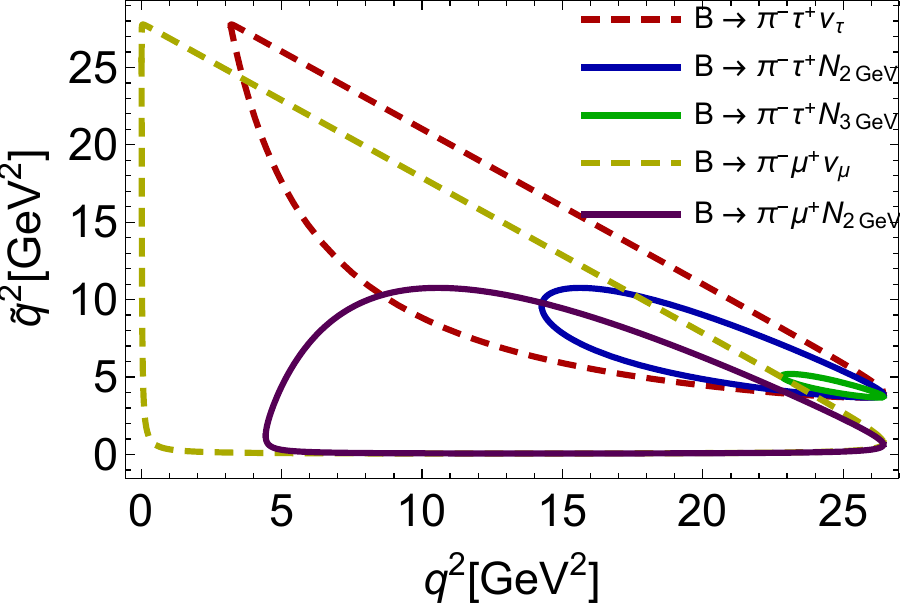}
  \caption[Dalitz plot]{Dalitz plot for the semileptonic decay $B^0
    \to D^+ \mu^- N$. Available phase-space shrinks drastically when
    HNL mass is large. $q^2$ is the invariant mass of the lepton pair,
    $\tilde q^2$ is the invariant mass of the final meson and charged lepton.}
  \label{fig:dalitz}
\end{figure}

In case of semileptonic decays in the HNL in the final state, the available phase space shrinks considereably, see Fig.~\ref{fig:dalitz}.
The effect of the mass can also be estimated by comparing the decays involving light leptons ($e/\mu$) and $\tau$-lepton in the final state.
A comparison with SM decay rates into $\tau$-lepton shows that 3-body decays into heavy sterile neutrinos are suppressed with respect to decays to light neutrinos.
\emph{Thus inclusive semi-leptonic decay of flavoured mesons to HNLs are dominated by single-meson final
states with the contributions from other state introducing small correction.}

\subsubsection{Quarkonia decays}
\label{sec:quarkonia}

Next we investigate the hidden flavored mesons $\Jpsi(c\bar{c},3097)$ and $\Upsilon (b\bar{b},9460)$ as sources of HNLs.
These mesons are short-lived, but 1.5-2 times heavier than the corresponding open flavored mesons, giving a chance to produce heavier HNLs.
We have studied these mesons in Appendix\,\ref{sec:production-from-Jpsi}, here we provide the summary of the results.

The number of HNLs produced from $\Jpsi$ decays is always subdominant to the
number of HNLs produced in $D$-meson decays (for $M_N < m_D$). Therefore, the
range of interest is $2\GeV\le M_N \le m_\Jpsi$ where this number should be compared with the number of HNLs
produced via $B$-meson decays. The resulting ratio is given by
\begin{multline}
  \label{eq:Jpsi2}
  \frac{\text{HNLs from }\Jpsi}{\text{HNLs from }B}
  =\frac{X_{c\bar c} \times f(\Jpsi)\times \BR_{\Jpsi\to  N\bar \nu}}{X_{b\bar b} \times f(B) \times \BR_{B\to N X}}=\\
 =3\times 10^{-4} \parfrac{X_{c\bar c}}{10^{-3}} \parfrac{10^{-7}}{X_{b\bar b}}
\end{multline}
where $X_{q\bar{q}}$ is the $q\bar{q}$ production rate and $f(h)$ is a production fraction for the given meson (see values for the SHiP experiment in Appendix~\ref{sec:heavy-flavour}). We have adopted $f(B) \times \text{BR}(B \to N+X) \sim 10^{-2}$ (c.f.\
Fig.~\ref{fig:Bbranching}) and used $f(\Jpsi) \sim
10^{-2}$.
The numbers in~\eqref{eq:Jpsi2} are normalized to the 400~GeV SPS proton beam.
One sees that $\Jpsi$ can play a role only below $b\bar b$ production threshold
(as $X_{b\bar b}$ tends to zero).

For experiments where sizeable number of $b\bar b$ pairs is produced one can
use the $\Upsilon$ decays to produce HNLs with $M_N \gtrsim 5\GeV$.
The number of thus produced HNLs is given by
\begin{equation}
  \label{eq:18}
  N_{\Upsilon\to N\bar \nu} \simeq 10^{-10} N_{\Upsilon} \times \parfrac{U^2}{10^{-5}}
\end{equation}
where $N_\Upsilon$ is the total number of  $\Upsilon$ mesons produced and we have normalized $U^2$ to the current experimental limit for $M_N >
5$~GeV (c.f.\ Fig.~\ref{fig:HNLbounds}).
It should be noted that HNLs with the mass of 5~GeV and $U^2 \sim 10^{-5}$
have the decay length
$c\tau \sim \unit{cm}$.

\begin{table}[!t]
  \centering
  \begin{tabular}{|l|p{0.8\textwidth}|}
    \hline
    Strange baryons& $\Lambda^0 (uds,1116)$, $\Sigma^+ (uus, 1189)$,
                     $\Sigma^- (dds, 1197)$, $\Xi^0 (uss, 1315)$, $\Xi^- (dss, 1322)$,
                     $\Omega^- (sss,1672)$\\
    \hline
    Charmed baryons &\inElena{\Lambda_c (udc,2287)}, $\Sigma^{++}_c(uuc,2453)$, $\Sigma^{0}_c(ddc,2453)$, \inElena{\Xi_c^+ (usc,2468)},
                      \inElena{\Xi_c^0 (dsc,2480)}, \inElena{\Omega^-_c(ssc,2695)}, $\Xi^{+}_{cc}(dcc,3519)$\\
    \hline
    Beauty baryons & \inElena{\Lambda_b (udb,5619)}, $\Sigma^{+}_b(uub,5811)$,
                     $\Sigma^{-}_b(ddb,5815)$, 
                     \inElena{\Xi_b^0 (usb,5792)},
                     \inElena{\Xi_b^- (dsb,5795)}, \inElena{\Omega_b^- (ssb,6071)}\\
    \hline
  \end{tabular}
  \caption{Long-lived flavoured baryons. For each quark content
    (indicated in parentheses) only
    the lightest baryon of a given quark contents (ground state,
    masses are in MeV) is
    shown, see footnote~\protect\vref{fn:1}. 
    Baryons considered in~\cite{Elena}
    have blue background. Unobserved so far baryons (such as
    $\Omega^+_{cc}(scc)$, $\Omega_{cb}(scb)$, etc.) are not listed.}
  \label{tab:baryons}
\end{table}

\subsubsection{Production from baryons}
\label{sec:prod-baryons}

Semileptonic decays of heavy flavoured baryons
(Table~\ref{tab:baryons}) produce HNLs. Baryon
number conservation implies that either proton or neutron (or other
heavier baryons) must be produced in the heavy baryon decay, which
shrink by about 1\,GeV the kinematical window for sterile
neutrino. The corresponding heavy meson decays have an obvious
advantage in this respect. Moreover, since both baryons and sterile
neutrinos are fermions, only the baryon decays into three and more
particles in the final state can yield sterile neutrinos, which
further shrinks the sterile neutrino kinematical window with respect
to the meson case, where two-body, pure leptonic decays can produce
sterile neutrinos.

Furthermore, light flavored baryons, strange baryons (see
Table~\ref{tab:baryons}) can only produce HNLs in the mass range where
the bounds are very strong already (roughly below kaon mass, see
FIG.~\ref{fig:HNLbounds}). Indeed, as weak decays change the
strangeness by 1 unit, there the double-strange $\Xi$-baryons can only
decay to $\Lambda$ or $\Sigma$ baryons (plus electron or muon and
HNL).  The maximal mass of the HNL that can be produced in this
process is \emph{smaller} than $(M_{\Xi^-} - M_{\Lambda^0}) \simeq
\unit[200]{MeV}$. Then, $\Omega^-$ baryon decays to $\Xi^0 \ell^-
  N$ with the maximal HNL mass not exceeding $M_{\Omega^-} -M_{\Xi^0}
  \simeq \unit[350]{MeV}$.  Finally, weak decays of $\Lambda$ or
$\Sigma$ baryons to $(p,n)$ can produce only HNLs lighter than $\sim
\unit[250]{MeV}$.

The production of HNL in the decays of charmed and beauty hyperons has been investigated in Ref.\,\cite{Ramazanov:2008ph}, which results have been recently checked in \cite{Mejia-Guisao:2017nzx}.
The number of such baryons is of course strongly suppressed as compared to the number of mesons with the same flavour.
At the same time the masses of HNLs produced in the decay of charmed (beauty) baryons are \emph{below} the threshold of HNL production of the corresponding charm (beauty) mesons due to the presence of a baryon in the final state.
This makes such a production channel strongly subdominant.
A dedicated studies for \ship ~\cite{Elena} and at the LHC~\cite{Mejia-Guisao:2017nzx} confirm this conclusion.
It should be noted that Refs.~\cite{Ramazanov:2008ph,Elena} use form factors from Ref.~\cite{Cheng:1995fe} which are about 20 years old.
A lot of progress has been made since then (see
e.g.~\cite{Meinel:2016dqj,Detmold:2015aaa}, where some of these form factors
were re-estimated and a factor $\sim 2$ difference with the older estimates
were established).

\subsection{HNL production from tau lepton}
\label{sec:hnl_from_tau}

At centre of mass energies well above the $\bar c c$ threshold
$\tau$-leptons are copiously produced mostly via $D_s \to \tau + X$
decays. Then 
HNLs can be produced in $\tau$ decay and these decays are important in the case of dominant mixing with $\tau$ flavour (which is the least constrained, see~\cite[Chapter 4]{Alekhin:2015byh}).
The main decay channels of $\tau$ are $\tau\to N+h_{P/V}$, $\tau\to N \ell_\alpha \bar{\nu}_\alpha$ and $\tau\to \nu_\tau \ell_\alpha N$, where $\alpha=e,\mu$.
The computations of the corresponding decays widths are similar to the processes $N\to \ell_\alpha h_{P/V}$ (c.f.\ Appendix~\ref{sec:hnl-decaying-meson}) and purely leptonic decays of HNL (see Section~\ref{sec:Nqq-cc}).
The results are 
\begin{equation}
    \Gamma(\tau\to N h_P) = \frac{G_F^2 f_h^2 m_\tau^3}{16\pi} 
	|V_{UD}|^2 |U_{\tau}|^2 
	\left[
	\left( 1 - y_N^2 \right)^2 - y_h^2(1 + y_N^2) 
	 \right] 
	\sqrt{\lambda(1, y_N^2, y_h^2)}
      \end{equation}
      \begin{equation}
        \label{eq:5}
        \Gamma(\tau\to N h_V) = \frac{G_F^2 g_h^2 m_\tau^3}{16\pi m_h^2} 
	|V_{UD}|^2 |U_{\tau}|^2 
	\left[
	\left( 1 - y_N^2 \right)^2 + y_h^2(1 + y_N^2 - 2 y_h^2) 
	 \right] 
	\sqrt{\lambda(1, y_N^2, y_h^2)}
      \end{equation}
\begin{align}
	\Gamma(\tau\to N \ell_{\alpha}\bar{\nu}_\alpha) &= \frac{G_F^2 m_\tau^5}{96\pi^3} 
	|U_{\tau}|^2
	\!\!\!\!
	\int\limits_{y_{\ell}^2}^{(1-y_N)^2} \frac{d\xi}{\xi^3}
	\left(\xi - y_{\ell}^2\right)^2
	\sqrt{\lambda(1, \xi, y_N^2)}
	\times \nonumber \\ & \times
	\left(
	\left(\xi + 2 y_{\ell}^2\right) 
	\left[1 - y_N^2\right]^2
	+ \xi \left(\xi - y_{\ell}^2 \right) \left[1 +y_N^2 -y_{\ell}^2\right] 
    - \xi y_{\ell}^4 -  2 \xi^3  \right)\nonumber \\
	 & \approx
	\frac{G_F^2 m_\tau^5}{192\pi^3} 
	|U_{\tau}|^2
	\left[
	1 - 8 y_N^2 + 8 y_N^6 - y_N^8 - 12 y_N^4\log(y_N^2)
	\right],\quad \text{for } y_l \to 0
    \\
    \Gamma(\tau\to \nu_\tau \ell_{\alpha} N) &= \frac{G_F^2 m_\tau^5}{96\pi^3} 
	|U_{\alpha}|^2
	\!\!\!\!
	\int\limits_{(y_{\ell} + y_N)^2}^{1} \frac{d\xi}{\xi^3}
	\left(1 - \xi\right)^2
	\sqrt{\lambda(\xi, y_N^2, y_{\ell}^2)}
	\times \nonumber \\ & \times
	\left(
	2 \xi^3 + \xi
	- \xi\left(1 - \xi\right) \left[1-y_N^2-y_{\ell}^2\right]
	-\left(2 + \xi\right)\left[y_N^2-y_{\ell}^2\right]^2
	\right)
	\approx \nonumber \\ & \approx
	\frac{G_F^2 m_\tau^5}{192\pi^3} 
	|U_{\alpha}|^2
	\left[
	1 - 8 y_N^2 + 8 y_N^6 - y_N^8 - 12 y_N^4\log(y_N^2)
	\right],\quad \text{for } y_l \to 0
\end{align}
where \mbox{$y_i = m_i / m_\tau$}, $V_{UD}$ is an element of CKM matrix which
corresponds to quark content of the meson $h_P$, $f_h$ and $g_h$ are 
pseudoscalar and vector meson decay constants (see Tables~\ref{tab:f_meson}
and~\ref{tab:g_meson}) and $\lambda$ is the K\"all\'en function~\cite{kallen1964elementary}:
\begin{equation}
  \label{eq:37}
  \lambda(a,b,c) = a^2 + b^2 + c^2 - 2 ab - 2 ac - 2 bc
\end{equation}
The results of this section fully agree with literature~\cite{Gorbunov:2007ak}.

\subsection{HNL production via Drell-Yan and other parton-parton scatterings}
\label{sec:production-in-pp-collision}

\begin{figure}[t]
  \centering
  \includegraphics[width=0.8\textwidth]{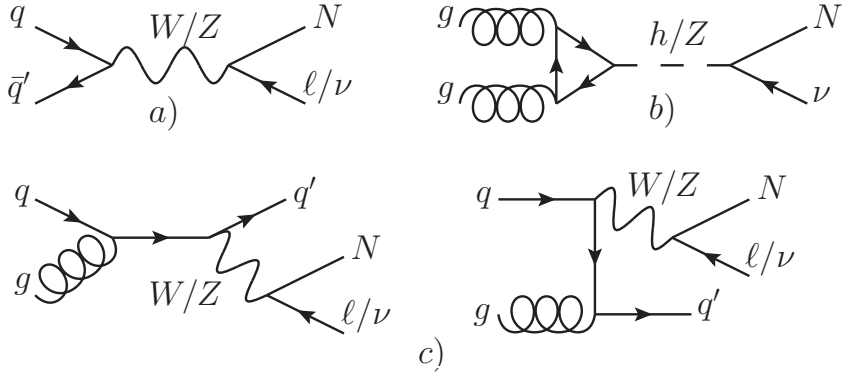} 
  \caption{HNL production channels: a) Drell-Yan-type process; b) gluon
    fusion; c) quark-gluon fusion.}
  \label{fig:HNL_direct_production}
\end{figure}

The different matrix elements for HNL production in proton-proton
collision are shown in Fig.~\ref{fig:HNL_direct_production}. Here
we are limited by the beam energy not high enough to produce real weak
bosons on the target protons. 
There
are three type of processes: Drell-Yan-type process a), gluon fusion b) and
$W\gamma/g$ fusion c). Process b) starts to play an important role
for much higher centre-of-mass energies~\cite{Degrande:2016aje,Ruiz:2017yyf}, process a) and c)
should be studied more accurately.

\begin{figure}[!t]
\centering
\includegraphics[width=0.50\textwidth]{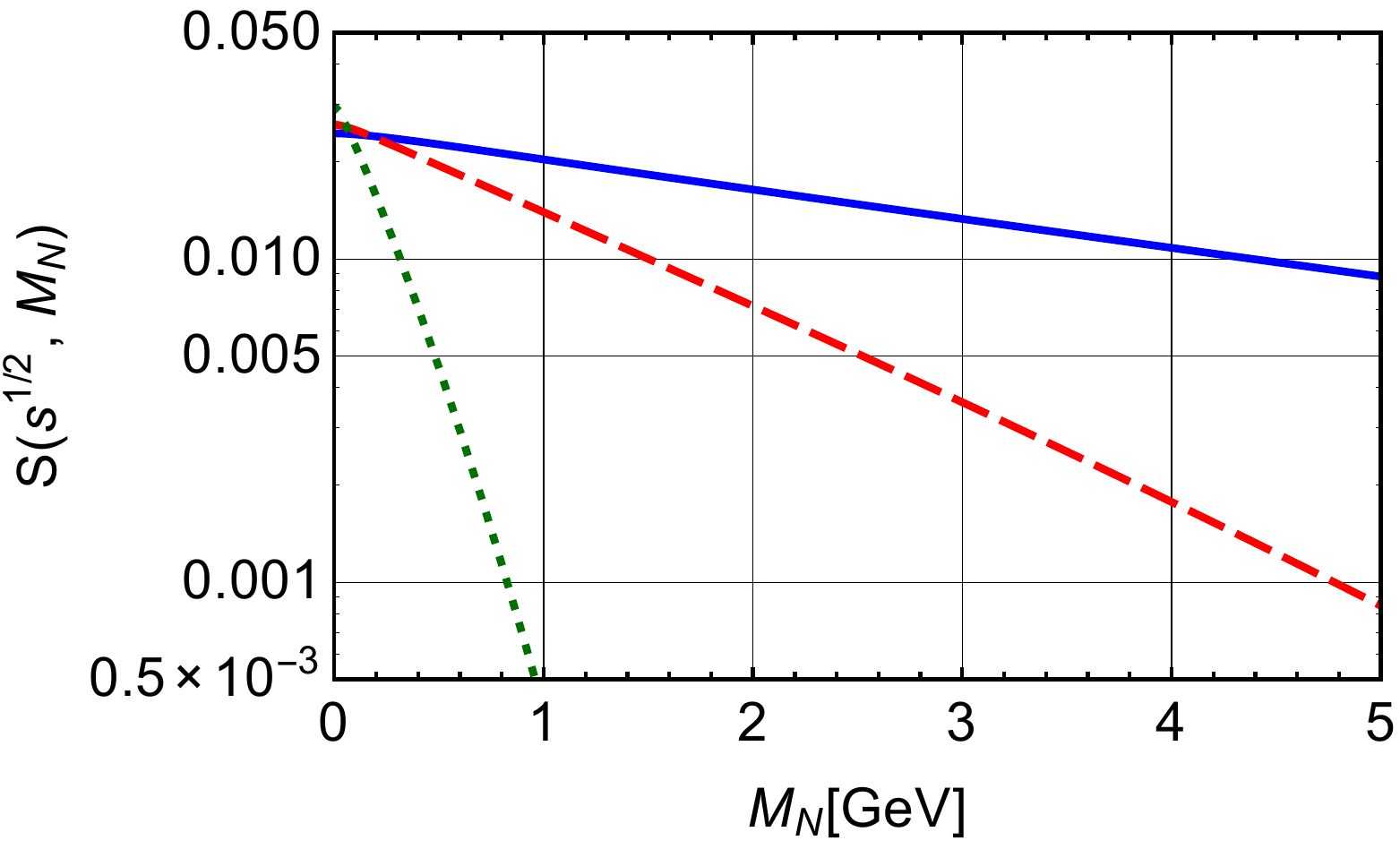}~\includegraphics[width=0.46\textwidth]{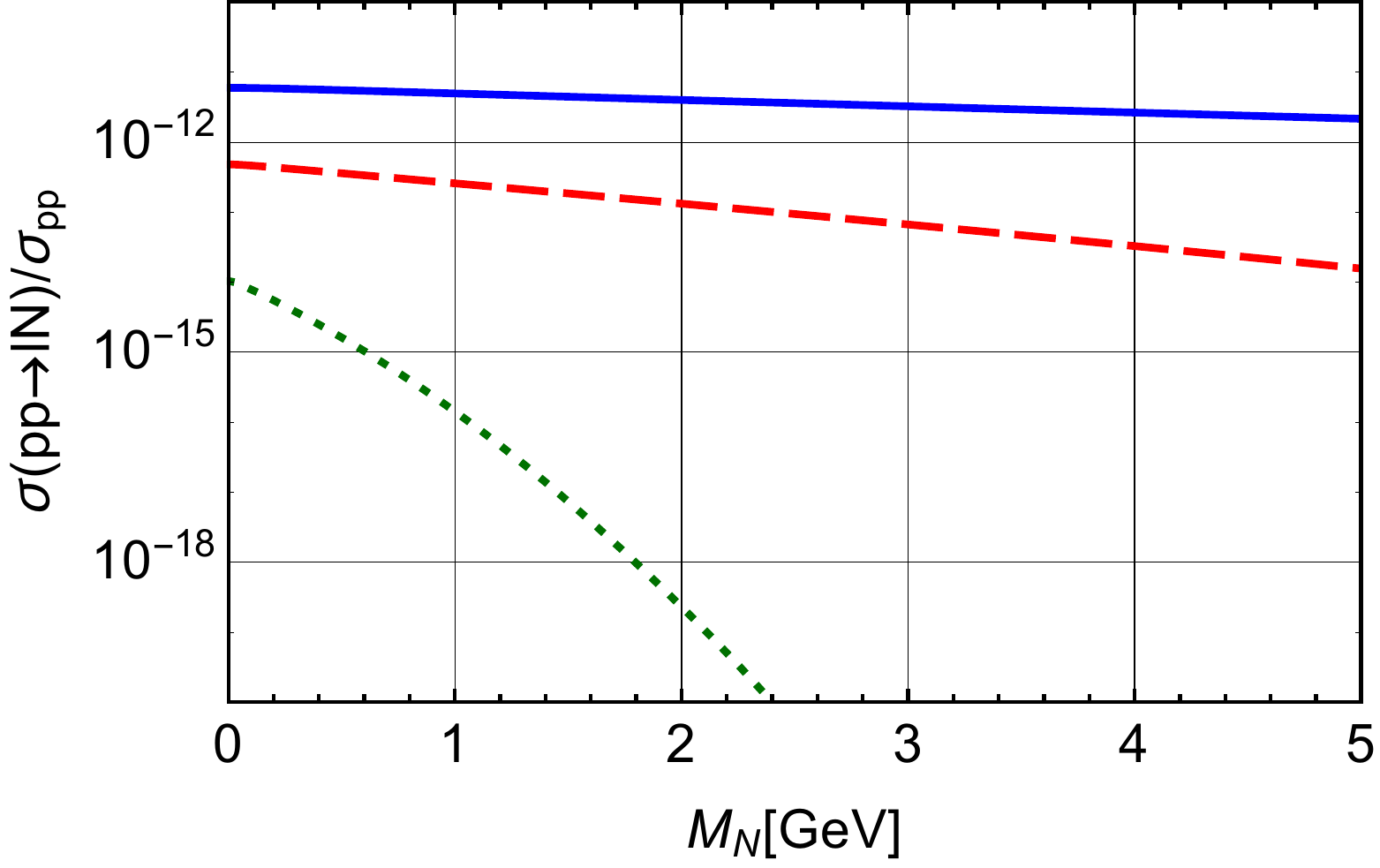}
\caption{Integral~\protect\eqref{eq:9} as a function of HNLs mass,
  neglecting lepton mass (left panel) and Probability of HNL
  production in $p$-$p$ collision for $|U_{\ell}|=1$ (right panel) for
  $\sqrt s = 100$~GeV (blue line), $\sqrt s = 28$~GeV (red dashed
  line) and $\sqrt s = 4$~GeV (green dotted line). The suppression of
  the integral as compared to $M_N=0$ case is due to PDFs being small
  at $x \sim 1$ and condition $x_1 x_2 s > M_N^2$. Total $p$-$p$
  cross section is taken from~\cite{Olive:2016xmw}.}
\label{fig:pdf}
\end{figure}

Let us start with the process a) in 
Fig.~\ref{fig:HNL_direct_production}. The cross section at the parton level is~\cite{Pilaftsis:1991ug,Datta:1993nm}
\begin{equation}
  \label{eq:7}
  \sigma(\bar q q' \to N \ell) = \frac{G_F^2 |V_{qq'}|^2 |U_{\ell}|^2 s_{\bar q q'}}{6 N_c\pi}\left(1 -\frac{3 M_N^2}{2 s_{\bar q q'}} + \frac{M_N^6}{2 s_{\bar q q'}^3} \right), \quad s_{\bar q q'} > M_N^2
\end{equation}
where $V_{qq'}$ is an element of the CKM matrix, $N_c=3$ is a number of colors and the centre-of-mass energy of the system $\bar q q'$ is given by
\begin{equation}
  \label{eq:8}
  s_{\bar q q'} = s x_1 x_2
\end{equation}
where $x_1$ and $x_2$ are fractions of the total proton's momentum
carried by the quark $q'$ and anti-quark $\bar q$ respectively. The
total cross section therefore is written as
\begin{multline}
  \label{eq:9}
  \sigma(\bar q q'\to N \ell) 
  = 2\sum_{\bar q,q'}\frac{G_F^2 |V_{qq'}|^2 |U_{\ell}|^2 s}{6 N_c\pi}
  \\ \times\int \frac{dx_1}{x_1}\, x_1^2 f_{\bar{q}}(x_1, s_{\bar q q'}) \int \frac{dx_2}{x_2} \, x_2^2 f_{q'}(x_2, s_{\bar q q'})
  \left( 1 -\frac{3 M_N^2}{2 s x_1 x_2} + \frac{M_N^6}{2 s^3x_1^3x_2^3} \right) \\
\equiv
  \frac{G_F^2 |V_{qq'}|^2 |U_{\ell}|^2 s}{6 N_c\pi} \,S(\sqrt{s},M_N)
\end{multline}
where $f_q(x, Q^2)$ is parton distribution function (PDF).
The corresponding integral $S(\sqrt s,M_N)$ as a function of $M_N$ and the production probability for this channel are shown in Fig.~\ref{fig:pdf}. For numerical estimates we have used LHAPDF package~\cite{Buckley:2014ana} with CT10NLO pdf set~\cite{Lai:2010vv}. 

\begin{figure}[!t]
  \centering
  \includegraphics[width=0.7\textwidth]{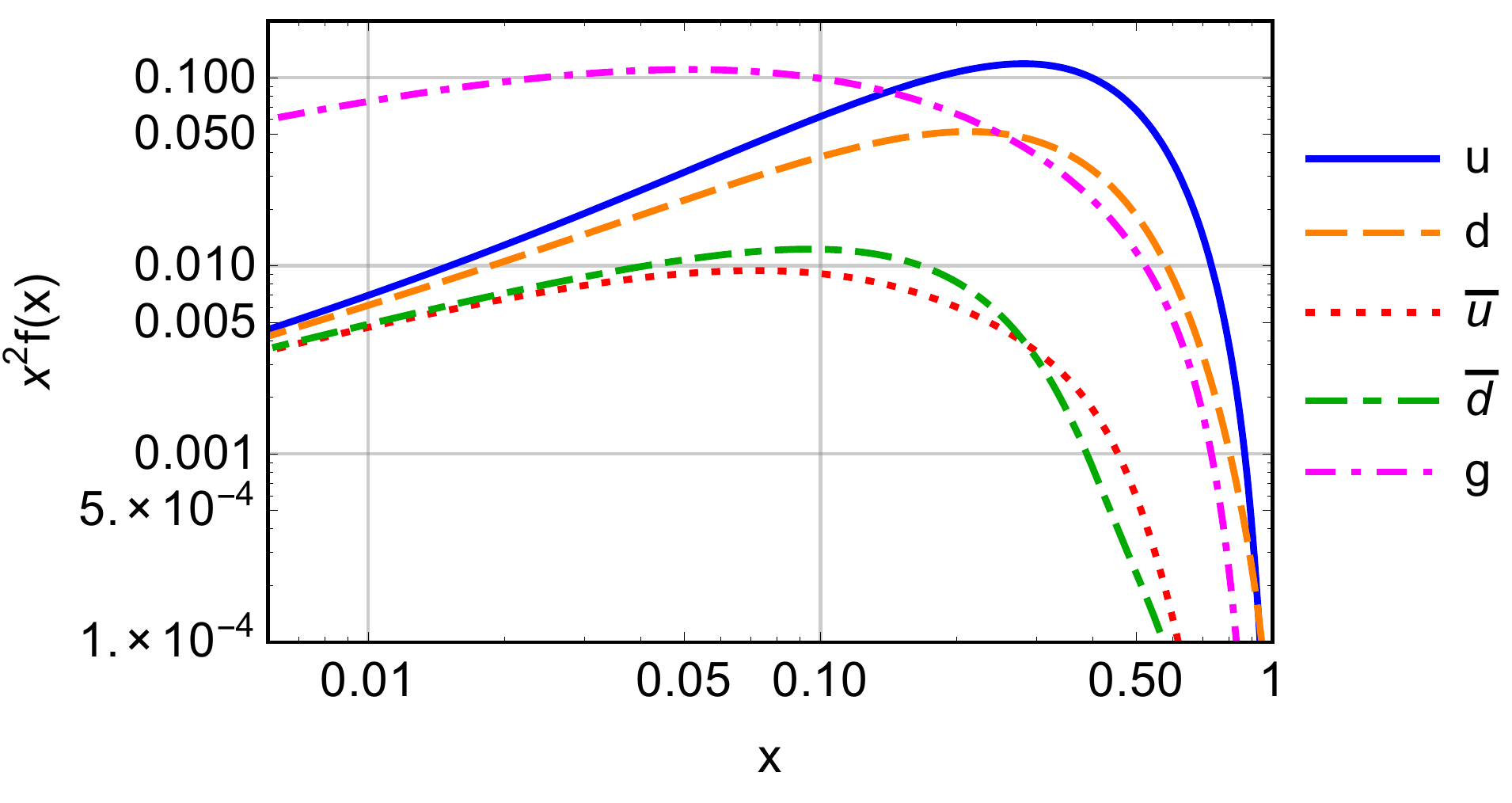}
  \caption[PDFs]{Combination $x^2 f(x)$ used in Eq.~\eqref{eq:9} for quark and
    gluon PDFs (for $\sqrt s \sim 30$~GeV). The functions peak at small values of $x$ and
    therefore a probability of the centre-of-mass energy of the parton pair
    close to $\sqrt s$ is small.}
  \label{fig:PDFs}
\end{figure}
This can be roughly understood as follows: PDFs peak at $x \ll 1$ (see Fig.~\ref{fig:PDFs})
and therefore the probability that the center-of-mass energy of a parton pair
exceeds the HNL mass, $\sqrt {s_{parton}} \gg M_N$, is small. On the other
hand, the 
probability of a flavour meson to decay to HNL (for $|U|^2 \sim 1$) is of the order of few \% and
therefore ``wins'' over the direct production, especially at the fixed-target
experiments where beam energies do not exceed hundreds of GeV. In case of the quark-gluon initial state (process c) in
Fig.\,\ref{fig:HNL_direct_production}) the similar considerations also work and the resulting cross section is also small, with an additional suppression due to the 3-body final state.
\emph{We see that the direct production channel is strongly suppressed in comparison with the production from mesons for HNLs with masses $M_N \lesssim \unit[5]{GeV}$.}

\subsection{Coherent proton-nucleus scattering}
\label{sec:coherent-pZ-scattering}

\begin{figure}[t!]
  \centering
  \includegraphics[width=0.9\linewidth]{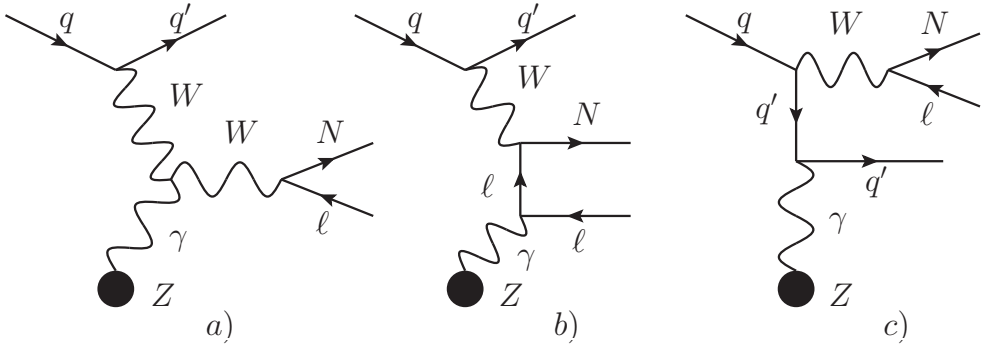}
  \caption{Possible Feynman diagrams for the HNL production in the proton coherent scattering off the nuclei.}
  \label{fig:HNLnuclei}
\end{figure}
The coherent scattering of a proton off the nuclei as a whole could be an
effective way of producing new particles in fixed target experiments.
There are two reasons for this.
First, parton scattering in the electromagnetic field of the nuclei is
proportional to $Z^2$ (where $Z$ is the nuclei charge) which can reach a factor $10^3$ enhancement for heavy nuclei.
Second, the centre of mass energy of proton-nucleus system is higher  than for the proton-proton scattering.
The coherent production of the HNLs will be discussed in the forthcoming
paper~\cite{Maxim}.
Here we announce the main result: the HNL coherent production channel is
subdominant to the meson decay for all HNL masses and mixing angles (for HNL
masses below 5~GeV).
In case of \ship on expects less than 1 HNL \emph{produced} via coherent
scattering for $10^{20}$ PoT.

\subsection{Summary}

In summary, production of HNL in proton fixed target experiments occurs
predominantly via (semi)leptonic decays of the lightest $c$- and $b$- mesons (Figs.~\ref{fig:Dbranching}, \ref{fig:Bbranching}). The production from heavier mesons is suppressed by the strong force mediated SM decays, while production from baryons is kinematically suppressed. 
Other production channels are subdominant for all masses $\unit[0.5]{GeV} \le
M_N \le \unit[5]{GeV}$ as discussed in Sections~\ref{sec:production-in-pp-collision}--\ref{sec:coherent-pZ-scattering}.


\section{HNL decay modes}
\label{sec:decays}

All HNL decays are mediated by charged current or neutral current
interactions~\eqref{eq:10}. In this Section we systematically revisit the
most relevant decay channels. Most of the results for sufficiently light HNLs
exist in the 
literature~\cite{Johnson:1997cj,Gribanov:2001vv,Gorbunov:2007ak,Atre:2009rg,Helo:2010cw,Cvetic:2016fbv}. 
For a few modes there are discrepancies by factors of few between different works, we comment on these
discrepancies in due course.

All the results presented below \emph{do not take into account charge
  conjugated channels} which are possible for the Majorana HNL; to account
for the Majorana nature one should multiply by 2 all the decay widths. The
branching ratios are the same for Majorana case and for the case considered here.

\subsection{3-body basic channels}

\begin{figure}[t]
  \centering
  \includegraphics[width=0.65\textwidth]{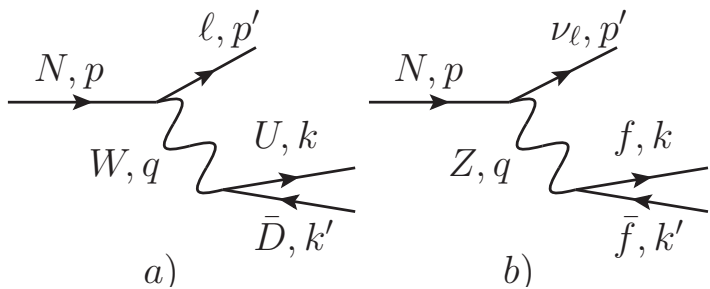}
  \caption{Diagram for the HNL decays mediated by charged a) and
    neutral b) currents.}
  \label{fig:Nbasicdecays}
\end{figure}

Two basic diagrams, presented in the Fig.~\ref{fig:Nbasicdecays},
contribute to all decays. For the charged current-mediated decay
(Fig.~\ref{fig:Nbasicdecays}(a)) the final particles $(U,D)$ could be
either a lepton pair $(\nu_{\alpha},\ell_{\alpha})$ or a pair of up
and down quarks $(u_{i},d_{j})$. For the neutral current-mediated
decay $f$ is any fermion. The tree-level decay width into free quarks,
while unphysical by itself for the interesting mass range, is important in estimates of the full
hadronic width at $M_N\gg \Lambda_{\text{QCD}}$, see
Section~\ref{sec:qcd} below.

For the decays $N\to \nu_{\alpha} \ell_{\alpha}^- \ell_{\alpha}^+$ and $N\to \nu_{\alpha} \nu_{\alpha} \bar{\nu}_{\alpha}$ both diagrams contribute, which leads to the interference (see Section~\ref{sec:Nqq-nc}).

\subsubsection{Charged current-mediated decays}
\label{sec:Nqq-cc}

\begin{figure}[!t]
  \centering
  \includegraphics[width=0.75\textwidth]{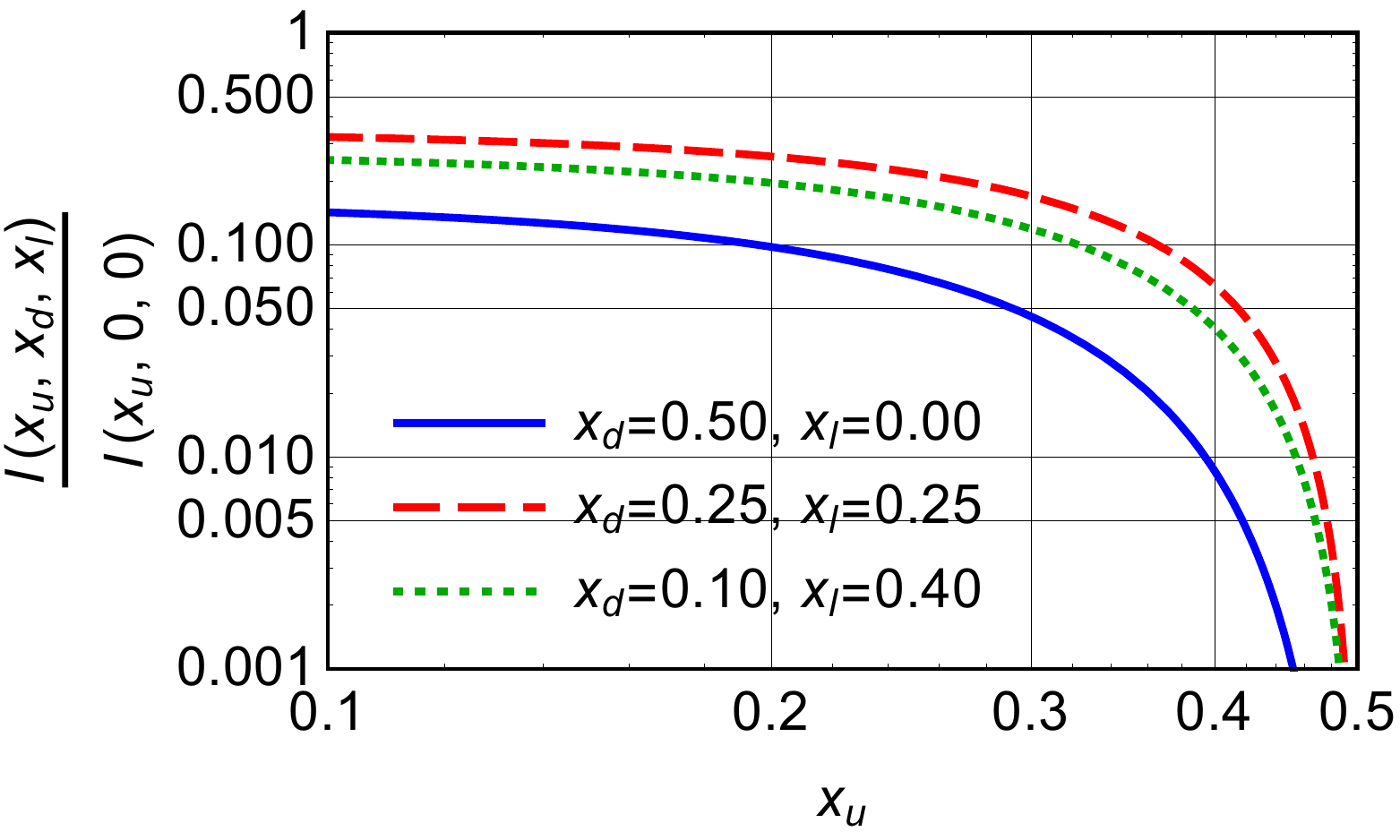}
  \caption{Function $I(x_u,x_d,x_l)/I(x_u,0,0)$ for several choices of $x_d$ and $x_l$ (see Eq.~\eqref{eq:38} for $I(x_u,x_d,x_l)$ definition).}
  \label{fig:Xudl}
\end{figure}

The general formula for the charged current-mediated processes $N\to
\ell_{\alpha}^- \nu_{\beta} \ell_{\beta}^+$, $\alpha\ne\beta$, and $N\to
\ell_{\alpha} u_i \bar{d}_j$ is~\cite{Shrock:1981wq,Gribanov:2001vv,Atre:2009rg,Helo:2010cw}
\begin{equation}
  \Gamma (N \to \ell^-_{\alpha} U \bar{D}) = N_W \frac{G_F^2 M_N^5}{192 \pi^3}
  |U_{\alpha}|^2  I(x_u,x_d,x_l)
  \label{eq:Glud}
\end{equation}
where $x_l = \dfrac{m_{\ell_{\alpha}}}{M_N}$,
$x_u = \dfrac{m_{U}}{M_N}$, $x_d = \dfrac{m_{D}}{M_N}$. The factor $N_W = 1$ for the case of the final leptons and
$N_W=N_c |V_{ij}|^2$ in the case of the final quarks, where $N_c = 3$ is the number of colors, and $V_{i j}$ is the corresponding matrix element of the CKM matrix. The function $I(x_u,x_d,x_l)$ that describes corrections due to finite masses of final-state fermions is given by
\begin{equation}
  \label{eq:38}
  I(x_u,x_d,x_l) \equiv 12\int\limits_{(x_d +
    x_l)^2}^{(1-x_u)^2}
  \frac{dx}{x} \left( x - x^2_{l} - x^2_d \right)\left( 1 + x_u^2 - x \right) 
  \sqrt{\lambda(x, x_l^2, x_d^2) \lambda(1, x, x_u^2)},
\end{equation}
where $\lambda(a,b,c)$ is given by Eq.~\eqref{eq:37}.

Several properties of the function~\eqref{eq:38} are in order:
\begin{enumerate}
\item $I(0,0,0) = 1$
\item Function $I(a,b,c)$ is symmetric under \emph{any} permutation of its arguments $a,b,c$.\footnote{This property is non-obvious but can be verified by the direct computation.}
\item  In the case of mass hierarchy $m_a, m_b \ll m_c$ (where $a,b,c$ are leptons
  and/or quarks in some order) one can use approximate result
  \begin{equation}
    \label{eq:39}
    I(x,0,0) = (1 - 8 x^2 + 8 x^6 - x^8 - 12 x^4 \text{log} x^2 )
\end{equation}
where $x = \dfrac{m_c}{M_N}$
\item The ratio $I(x_u,x_d,x_l)/I(x_u,0,0)$ for several choices of
  $x_d,x_l$ is plotted in Fig.~\ref{fig:Xudl}. It decreases with each
  argument.
\end{enumerate}

\subsubsection{Decays mediated by neutral current interaction and the interference case}
\label{sec:Nqq-nc}

Decay width for neutral current-mediated decay $N \to \nu_{\alpha}
f \bar{f}$ depends on the type of the final fermion.
For charged lepton pair $l_\beta\bar l_\beta$ the results are different for the case $\alpha\ne\beta$ and $\alpha=\beta$, because of the existence of the  charge current mediated diagrams in the latter case. Nevertheless, the decay width can be written in the unified way,
\begin{align}
  \nonumber
  \Gamma (N \to \nu_{\alpha} f \bar{f}) &= N_Z \frac{G_F^2 M_N^5}{192 \pi^3} 
  \cdot |U_{\alpha}|^2 \cdot \bigg[ C_1^{f} \bigg((1 - 14 x^2 - 2 x^4 - 12 x^6) 
  \sqrt{1 - 4 x^2} + \\
  \nonumber
  & + 12 x^4 (x^4 - 1)L(x) \bigg) + 4 C_2^{f} \bigg(x^2 (2 + 10 x^2 - 12 x^4) 
  \sqrt{1 - 4 x^2} + \\ 
  & + 6 x^4 (1 - 2 x^2 + 2 x^4) L(x) \bigg) \bigg], 
  \label{eq:Gvff}
\end{align}
where $x = \dfrac{m_f}{M_N}$,
$L(x) = \log \bigg[ \dfrac{1 - 3 x^2 - (1 - x^2) \sqrt{1 - 4 x^2}}{x^2
(1 + \sqrt{1 - 4 x^2})} \bigg]$ and $N_Z = 1$ for the case of leptons in the final state or $N_Z = N_c$ for the case of quarks. 
The values of $C_1^f$ and $C_2^f$ are given in the
Table~\ref{tab:C12f}. This result agrees with~\cite{Gorbunov:2007ak, Atre:2009rg, Helo:2010cw}.

\begin{table}[h]
\centering
\begin{tabular}{|c|c|c|}
\hline
$f$                            & $C_1^f$ & $C_2^f$ 
\\ \hline
$u$, $c$, $t$                            & 
$\frac{1}{4} \Big(1 - \frac{8}{3}\sin^2 \theta_W +\frac{32}{9}\sin^4\theta_W \Big)$       &
$\frac{1}{3} \sin^2 \theta_W\Big(\frac{4}{3} \sin^2 \theta_W - 1\Big)$
\\ \hline
$d$, $s$, $b$                            &
$\frac{1}{4} \Big(1 - \frac{4}{3}\sin^2 \theta_W +\frac{8}{9}\sin^4\theta_W \Big)$       &
$\frac{1}{6} \sin^2 \theta_W\Big(\frac{2}{3} \sin^2 \theta_W - 1\Big)$
\\ \hline
$\ell_\beta$, $\beta\ne\alpha$ &
$\frac{1}{4}\Big(1 - 4\sin^2 \theta_W + 8\sin^4\theta_W \Big)$  &
$\frac{1}{2} \sin^2 \theta_W \Big(2 \sin^2 \theta_W - 1 \Big)$
\\ \hline
$\ell_\beta$, $\beta=\alpha$   &
$\frac{1}{4}\Big(1 + 4\sin^2 \theta_W + 8\sin^4\theta_W \Big)$  &
$\frac{1}{2} \sin^2 \theta_W \Big(2 \sin^2 \theta_W + 1 \Big)$      
\\ \hline
\end{tabular}
\caption{Coefficients $C_1$ and $C_2$ for the neutral current-mediated decay width.}
\label{tab:C12f}
\end{table}

In the case of pure neutrino final state only neutral currents
contribute and the decays width reads 
\begin{equation}
	\Gamma (N \to \nu_{\alpha} \nu_{\beta} \bar{\nu}_{\beta}) =
	(1 + \delta_{\alpha\beta}) \frac{G_F^2 M_N^5}{768 \pi^3} |U_{\alpha}|^2.
	\label{eq:Gvvv}
\end{equation}

\subsection{Decay into hadrons}
\label{sec:decay-into-hadrons}

In this Section we consider hadronic final states for $M_N$ both below and above $\Lambda_{\rm QCD}$ scale and discuss the range of validity of our results.

\subsubsection{Single meson in the final state}
\label{sec:single-meson}

At $M_N \lesssim \Lambda_{\rm QCD}$ the quark pair predominantly binds into a single meson.
There are charged current- and neutral current-mediated processes with a meson in the final state: $N\to \ell_{\alpha} h_{P/V}^+$ and $N\to \nu_{\alpha} h_{P/V}^0$, where $h_{P}^+$ ($h_P^0$) are charged (neutral) pseudoscalar mesons and $h_V^+$ ($h_V^0$) are charged (neutral) vector mesons.
In formulas below $x_h \equiv m_h / M_N$, $x_\ell = m_\ell / M_N$, $f_h$ and $g_h$ are the corresponding meson decay constants (see Appendix~\ref{sec:phen-const}), $\theta_W$ is a Weinberg angle and the function $\lambda$ is given by eq.~(\ref{eq:37}).
The details of the calculations are given in the Appendix~\ref{sec:hnl-decaying-meson}.

The decay width to the charged pseudo-scalar mesons ($\pi^\pm,K^\pm, D^\pm, D_s, B^\pm, B_c$) is given by
\begin{align}
	\Gamma(N\to \ell_{\alpha}^- h_{P}^+) &= 
	\frac{G_F^2 f_h^2 |V_{UD}|^2 |U_{\alpha}|^2 M_N^3}{16\pi} 
	\left[ \left( 1 - x_\ell^2 \right)^2 - 
	x_h^2(1 + x_\ell^2) \right] 
	\sqrt{\lambda(1, x_h^2, x_\ell^2)},\label{eq:17}
\end{align}
in full agreement with the literature~\cite{Gorbunov:2007ak,Atre:2009rg,Helo:2010cw}.

The decay width to the pseudo-scalar neutral meson ($\pi^0,\eta,\eta',\eta_c$) is given by
\begin{align}
  \Gamma(N\to \nu_{\alpha} h_{P}^0) &=
  \frac{G_F^2 f_h^2 M_N^3}{32\pi} |U_{\alpha}|^2 
  \left( 1 - x_h^2 \right)^2
    \label{eq:29}
\end{align}
Our answer agrees with~\cite{Gorbunov:2007ak}, but
is twice larger than \cite{Atre:2009rg,Helo:2010cw}.
The source of the difference is
unknown.\footnote{This cannot be due to the Majorana or Dirac nature
  of HNL, because 
  the same discrepancy would then appear in Eq.~\eqref{eq:17}.}

The HNL decay width into charged vector mesons ($\rho^{\pm}$, $a_1^{\pm}$, $D^{\pm*}$, $D^{\pm*}_s$) is given by
\begin{align}
  \Gamma(N\to \ell_{\alpha}^- h_{V}^+) &= \frac{G_F^2 g_h^2 |V_{UD}|^2 |U_{\alpha}|^2 M_N^3}{16\pi m_h^2}  
	\left(
	\left(1 - x_\ell^2\right)^2 + x_h^2 \left(1 + x_\ell^2\right) - 2 x_h^4
	\right)
	\sqrt{\lambda(1, x_h^2, x_\ell^2)}
	\label{eq:40}
\end{align}
that agrees with the literature~\cite{Gorbunov:2007ak,Atre:2009rg,Helo:2010cw}.

\emph{However}, there is a disagreement regarding the numerical value of the meson constant $g_\rho$ between~\cite{Gorbunov:2007ak} and~\cite{Atre:2009rg,Helo:2010cw}.
We extract the value of this constant from the decay $\tau \to \nu_\tau \rho$ and obtain the result that numerically agrees with the latter works, see discussion in Appendix~\ref{sec:rho-decay-constant}.

For the decay into neutral vector meson ($\rho^0$, $a_1^0$, $\omega$,
$\phi$, $J/\psi$) we found that the result depends on the quark content of meson. To take it into account we introduce dimensionless $\kappa_h$ factor to the meson decay constant~\eqref{eq:vectorgh0}.
The decay width is given by
\begin{equation}
 \Gamma(N\to \nu_{\alpha} h_V^0) = \frac{G_F^2 \kappa_h^2 g_h^2 |U_{\alpha}|^2  M_N^3}{32\pi m_h^2}
 \left(1 + 2 x_h^2\right) \left(1 - x_h^2\right)^2.
 \label{eq:41}
\end{equation}
Our result for $\rho^0$ and results in~\cite{Gorbunov:2007ak} and~\cite{Atre:2009rg} are all different.
The source of the difference is unknown.
For decays into $\omega$, $\phi$ and $J/\psi$ mesons we agree with~\cite{Atre:2009rg}.
The result for the $a_1^0$ meson appears for the first
time.\footnote{Refs.\,\cite{Atre:2009rg,Helo:2010cw} quote also two-body decays $N\to \nu_{\alpha} h_V^0$, $h_V^0=K^{*0},\bar K^{*0},D^{*0},\bar D^{*0}$, with the rate given by \eqref{eq:41} (with a different $\kappa$).
  This is not justified, since the weak neutral current does not couple to
 the corresponding vector meson $h_V^0$ at tree level.}

\begin{figure}[!t]
  \centering
  \includegraphics[width=0.5\textwidth]{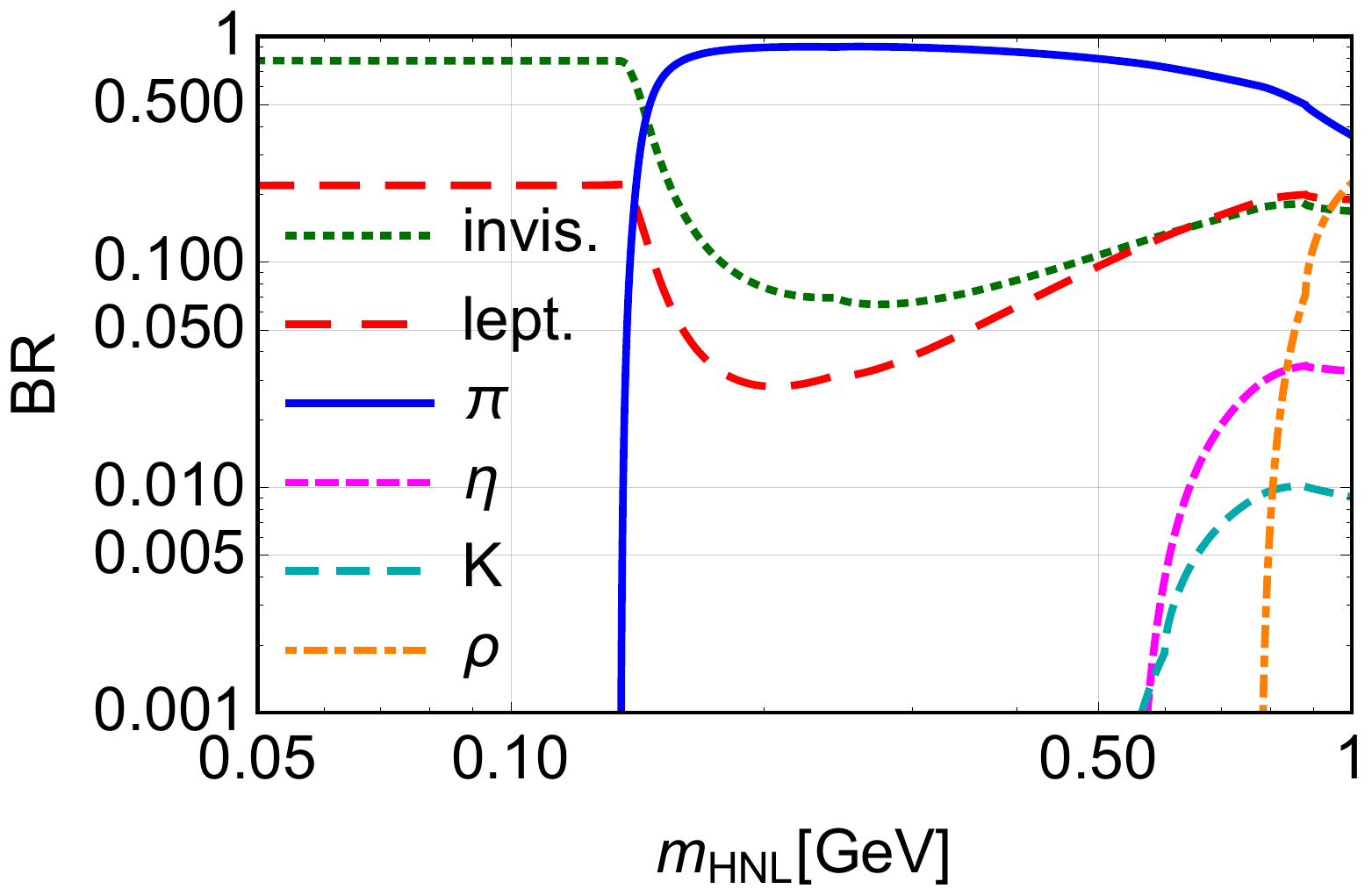}~  \includegraphics[width=0.5\textwidth]{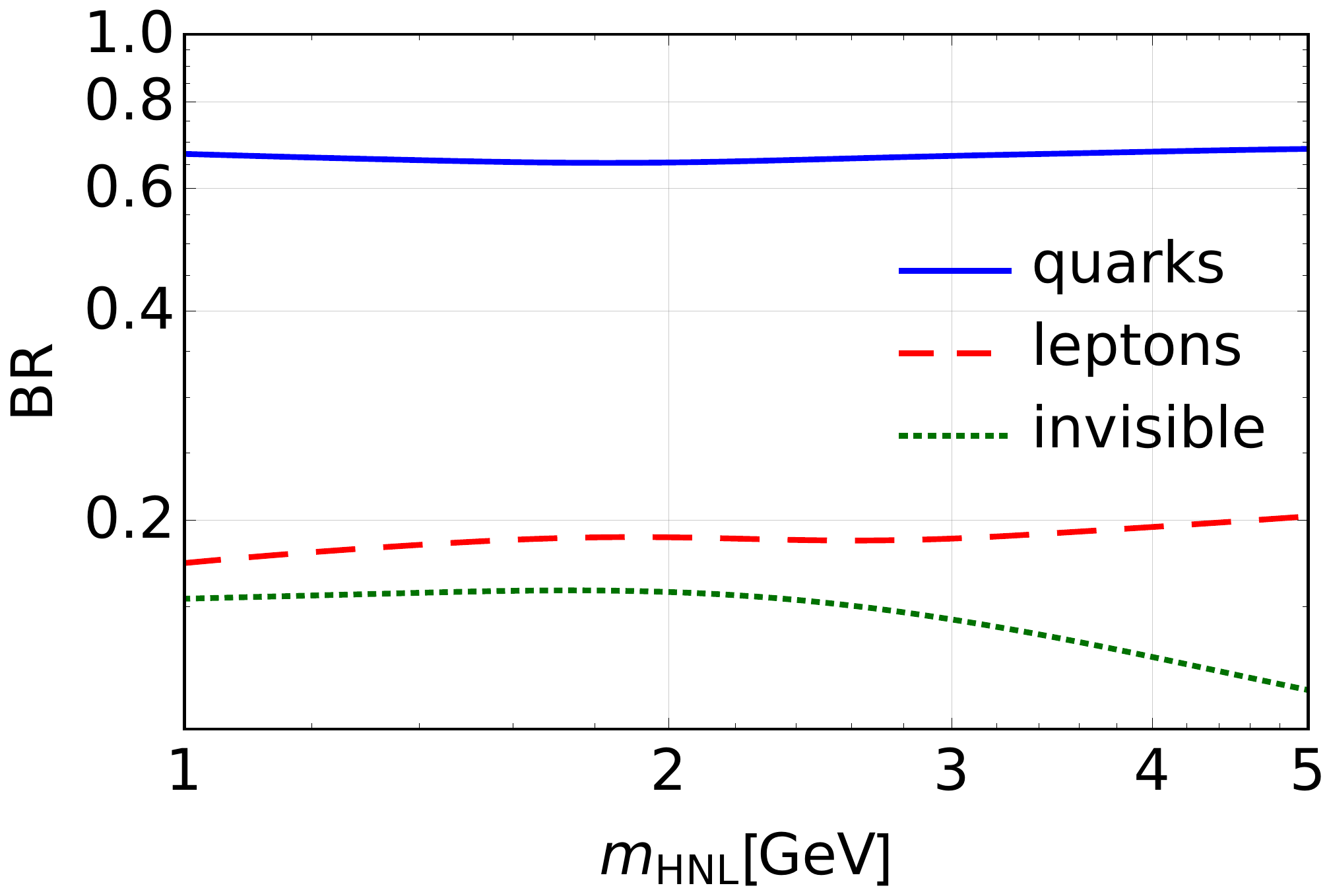}
  \caption{The branching ratios of the HNL for the mixing ratio
    $U_e:U_\mu:U_\tau=1:1:1$. \emph{Left panel:} region of masses
    below $1$~GeV; \emph{Right panel:} region of masses above $1$~GeV,
    for quarks the QCD corrections\,\eqref{eq:42}, \eqref{eq:tau-qcd}
    are taken into account.}
  \label{fig:BR1}
\end{figure}

The branching ratios for the one-meson and lepton channels below
$1$~GeV are given on the left panel of Fig.~\ref{fig:BR1}.

\subsubsection{Full hadronic width vs.\ decay into single meson final state}
\label{sec:qcd}

Decays into multi-hadron final states become kinematically accessible as soon as $M_N > 2 m_\pi$. 
To estimate their branching fractions and their contribution to the total decay width, we can compute the total hadronic decay width of HNLs, $\Gamma_{\rm had}$ and compare it with the combined width of all single-meson states, $\Gamma_{\rm 1\,meson}$. The total hadronic decay width can be estimated via decay width into quarks (Sections~\ref{sec:Nqq-cc}--\ref{sec:Nqq-nc}) times the additional loop corrections.

\begin{figure}[t]
  \centering
  \includegraphics[width=0.6\textwidth]{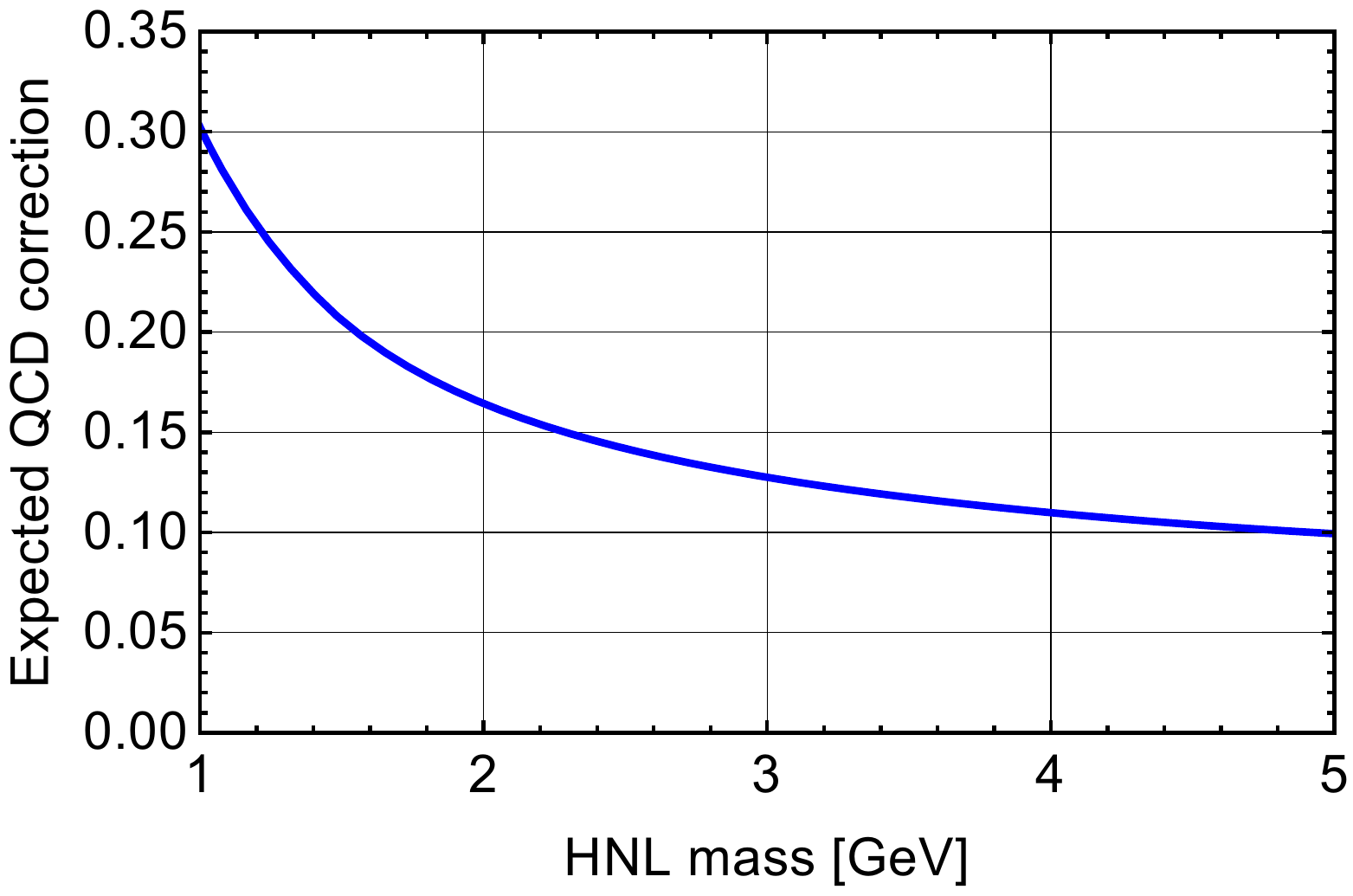}
  \caption{The estimate of the QCD corrections for the HNL decay into quark pairs, using the three-loop formula~\protect\eqref{eq:tau-qcd} for $\tau$-lepton.}
  \label{fig:QCDcorr}
\end{figure}

The QCD loop corrections to the tree-level decay into quarks have been estimated in case of $\tau$ lepton hadronic decays. In this case the tree level computation of the $\tau$ decay to two quarks plus neutrino underestimates  the full  hadronic decay width by $20\%$~\cite{Perl:1991gd, Braaten:1991qm, Gorishnii:1990vf}.  The loop corrections, $\Delta_{\rm QCD}$, defined via
\begin{equation}
  \label{eq:42}
  1 + \Delta_{\rm QCD} \equiv \frac{\Gamma(\tau\to \nu_{\tau}+\text{hadrons})}{\Gamma_{\text{tree}}(\tau\to \nu_{\tau}\bar{u}q)}
\end{equation}
have been computed up to three loops~\cite{Gorishnii:1990vf} and is given by:
\begin{equation}
  \label{eq:tau-qcd}
  \Delta_{\rm QCD} = \frac{\alpha_s}{\pi} + 5.2 \frac{\alpha_s^2}{\pi^2} + 26.4 \frac{\alpha_s^3}{\pi^3},
\end{equation}
where $\alpha_s = \alpha_s(m_{\tau})$.\footnote{Numerically this gives
  for $\tau$-lepton $\Delta_{\text{QCD}}\approx 0.18$, which is within
  a few \emph{per cent} of the experimental value
  $\Delta_{\text{Exp}}=0.21$. The extra difference comes from the QED
  corrections.} We use \eqref{eq:tau-qcd} with $\alpha_s =
\alpha_s(M_N)$ as an estimation for the QCD correction for the HNL
decay, for both charged and neutral current processes. We expect therefore that QCD correction to the HNL decay width into quarks is smaller than $30\%$ for $M_N \gtrsim 1$~GeV (Fig.~\ref{fig:QCDcorr}).

Full hadronic decay width dominates the HNL lifetime for masses $M_N \gtrsim
1$~GeV (see Fig.~\ref{fig:BR1}).
The latter is important to define the upper bound of
sensitivity for the experiments like SHiP or MATHUSLA (see Fig.~\ref{fig:HNLbounds}). This upper bound is defined by the requirements that HNLs can reach the detector.

\subsubsection{Multi-meson final states}
\label{sec:multi-meson-final}

\begin{figure}[t]
\centering
\includegraphics[width=0.48\textwidth]{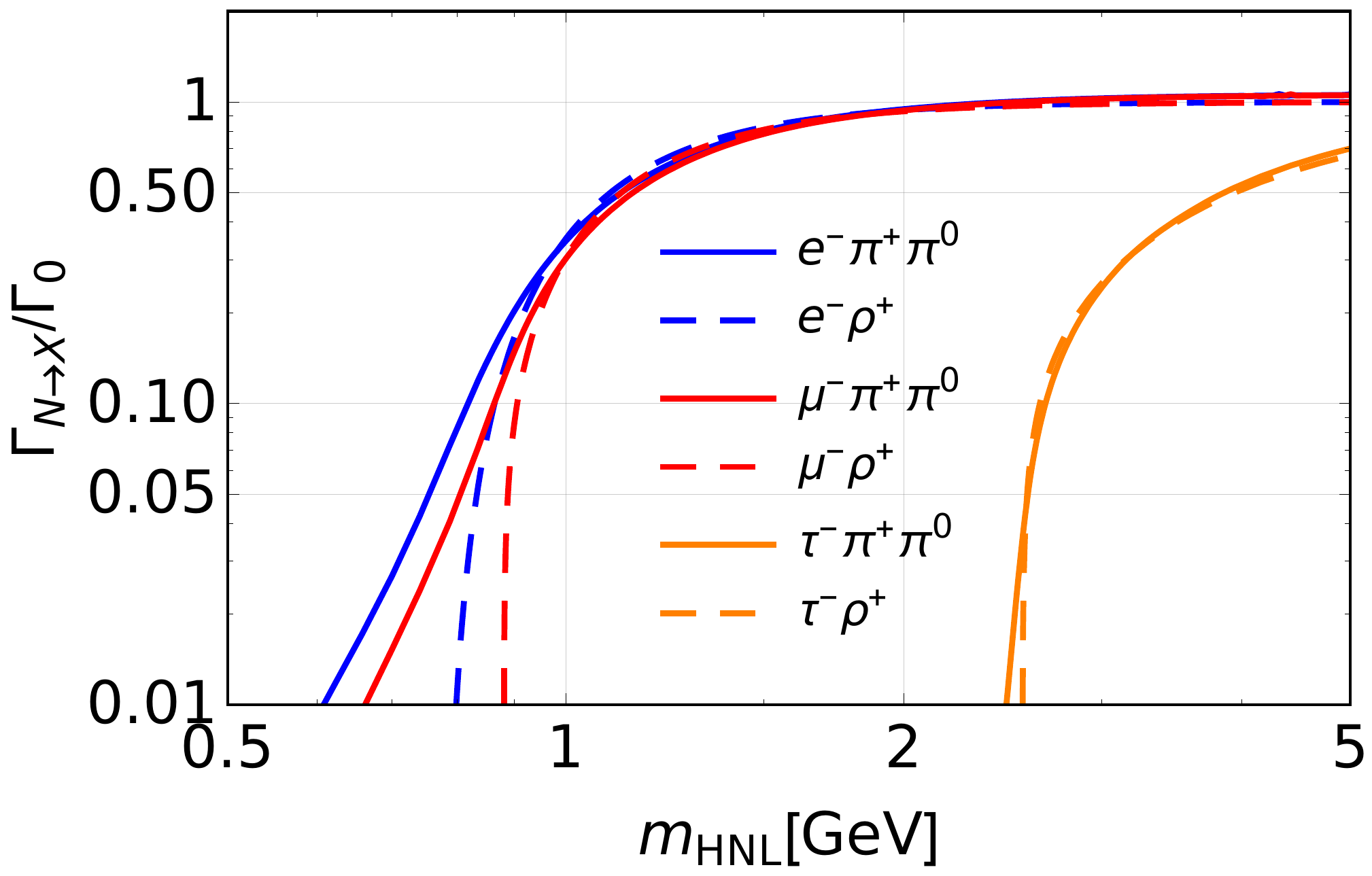}
~
\includegraphics[width=0.48\textwidth]{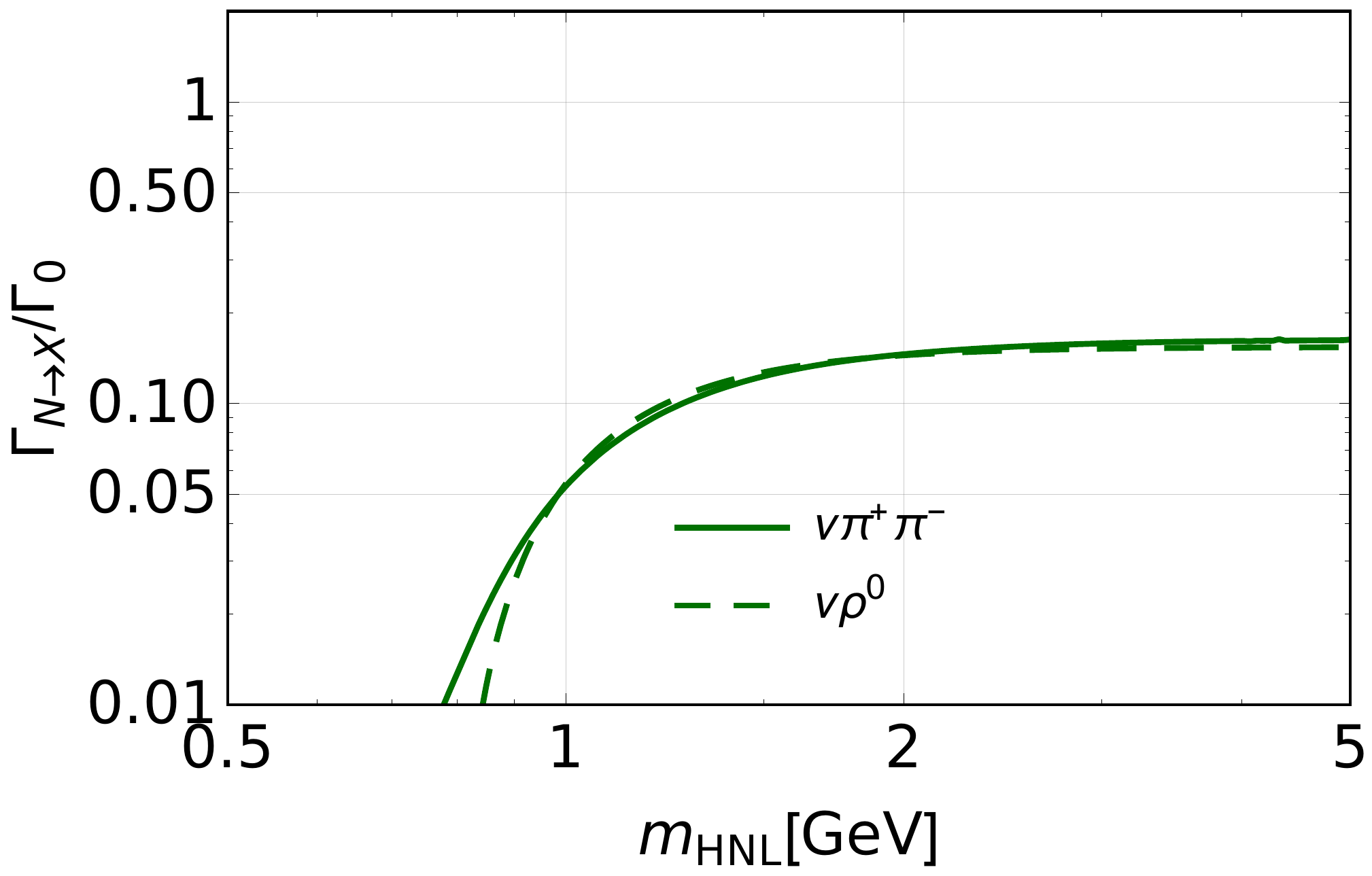}
\caption{\textit{Left panel:} Decay widths of charged current channels with $\rho$ or $2\pi$ divided by $\Gamma_0 = \frac{1}{16\pi} G_F^2 \frac{g_\rho^2}{m_\rho^2} |V_{ud}|^2 |U_{\alpha}|^2 M_N^3$, which is prefactor in Eq.~(\ref{eq:40}). \textit{Right panel:} The same for neutral current channels.}
  \label{fig:Grho_vs_G2pi}
\end{figure}

\begin{figure}[t]
  \centering
  \includegraphics[width=0.9\textwidth]{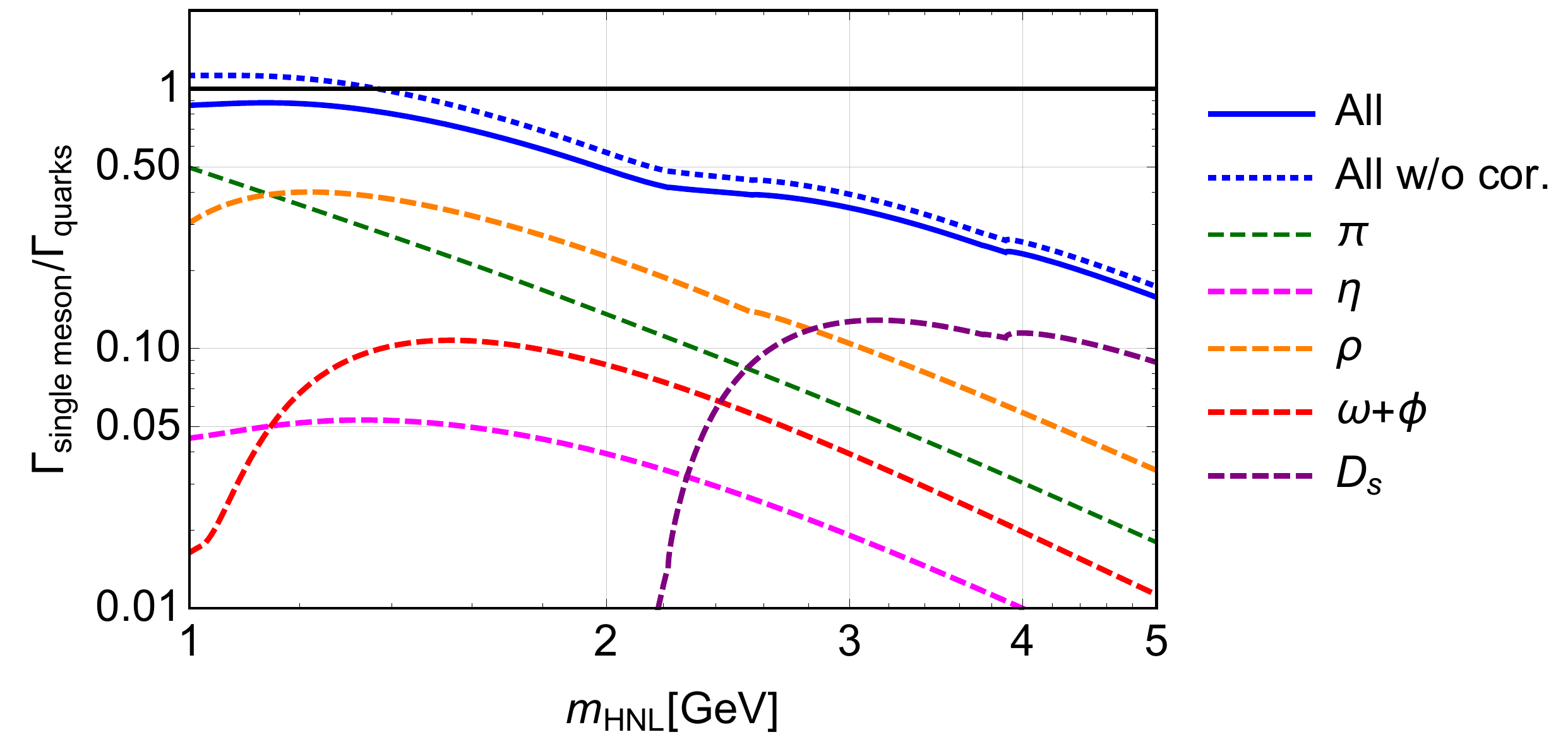}
  \caption{HNL decay widths into all relevant single meson channels,
    divided by the total decay width into quarks with QCD corrections,
    estimated as in~\protect\eqref{eq:tau-qcd} (all dashed lines). The
    blue solid line is the sum of the all mesons divided by decay
    width into quarks with QCD corrections, blue dotted line is the
    same but without QCD corrections.}
  \label{fig:GmGq}
\end{figure}

When discussing ``single-meson channels'' above, we have also included
there decays with the $\rho$-meson. By doing so, we have essentially
incorporated all the two-pion decays $N \to \pi^+\pi^0 \ell^-$ for
$M_N > m_\rho$.  Indeed, we have verified by direct computation of $N
\to \pi^+\pi^0 \ell^-$ that they coincide with $N \to \rho^+ \ell^-$
for all relevant masses (Fig.~\ref{fig:Grho_vs_G2pi}).  Of course the
decay channel to two pions is also open for $2m_\pi < M_N < m_\rho$,
but its contribution there is completely negligible and we ignore this
in what follows.

Figs.~\ref{fig:BR1} and~\ref{fig:GmGq} demonstrate that one-meson channels are definitely enough for all the
hadronic modes if sterile neutrino mass does not exceed 1 GeV. 
The ratio between the combined decay width into single-meson final
states ($\pi^\pm$, $\pi^0$, $\eta$, $\eta'$, $\rho^\pm$, $\rho^0$,
$\omega$+$\phi$, $D_s$) and
into quarks is shown in Fig.~\ref{fig:GmGq}.\footnote{We ignore
  CKM-suppressed decays into kaons as well as decays to heavy flavour
  meson ($D$,$B$).} One sees that the decay width into quarks is
larger for $M_N \gtrsim 2$~GeV, which means that multi-meson final
states are important in this region.
\begin{table}[!t]
    \centering
    \begin{tabular}[t]{|>{$}l<{$}|}
    \hline
    \text{3-body decays} \\
    \hline
        N\to \ell_\alpha^- \pi^+ \pi^0   \\
        N\to \nu_\alpha \pi^+\pi^-\\
        N\to \nu_\alpha \pi^0\pi^0\\
        N\to \ell_\alpha^- K^+ \bar K^0\\
        N\to \nu_\alpha K^+ K^-\\
        N\to \nu_\alpha K^0 \bar{K}^0 \\
        N\to \ell^-_\alpha K^+\pi^0\\
        N\to \ell^-_\alpha K^0 \pi^+\\
    \hline
    \end{tabular}~~~~
    \begin{tabular}[t]{|>{$}l<{$}|}
    \hline
    \text{4-body decays} \\
    \hline
        N\to \nu_\alpha  (3\pi)^0 \\
        N\to \ell_\alpha^- (3\pi)^+\\
        N\to \nu_\alpha (2\pi K)^0 \\
        N\to \ell_\alpha^- (2\pi K)^+\\
        N\to \nu_\alpha (2K \pi)^0 \\
        N\to \ell_\alpha^- (2K \pi)^+\\
        N\to \ell_\alpha^- (3K )^+\\
        N\to \nu_\alpha (3K )^0\\
    \hline
    \end{tabular}~~~~
    \begin{tabular}[t]{|>{$}l<{$}|r|}
    \hline
    \text{Branching ratios } & [\%]\\
    \tau^- \to \nu_\tau + X^- & \\
    \hline
    \tau \to \nu_\tau + \pi^- & 10.8\\
    \tau \to \nu_\tau + \pi^- \pi^0 & 25.5\\
    \tau \to \nu_\tau + \pi^0 \pi^- \pi^0& 9.2\\ 
    \tau \to \nu_\tau + \pi^- \pi^+ \pi^- & 9.0\\
    \tau \to \nu_\tau + \pi^-\pi^+\pi^-\pi^0 & 4.64\\
    \tau \to \nu_\tau + \pi^-\pi^0\pi^0\pi^0 & 1.04\\
    \tau \to \nu_\tau + 5\pi & $\mathcal{O}(1)$\\
    \tau \to \nu_\tau + K^-\text{ or } K^-\pi^0 & $\mathcal{O}(1)$\\
    \tau \to \nu_\tau + K^- K^0 &  $\mathcal{O}(0.1)$\\
    \tau \to \nu_\tau + K^- K^0 \pi^0 &  $\mathcal{O}(0.1)$\\
    \hline
    \end{tabular}

    \caption{Possible multi-meson decay channels of HNLs with $M_N >
      2m_\pi$ threshold. Right panel shows branching ratios of
      hadronic decays of $\tau$-lepton and demonstrates relative
      importance different hadronic 2, 3, 4 and 5 body channels. }
    \label{tab:multimeson}
\end{table}

The main expected 3- and 4-body decays channels of HNL and decay are
presented in Table~\ref{tab:multimeson}. In this table we also add
information about multimeson decays of $\tau$ because they give us
information about decay through charged current of the HNL of the same
mass as $\tau$-lepton. The main difference between HNL and $\tau$-lepton comes from the possibility of the HNL decay through the neutral current, which we discuss below.

As one observes, the main hadronic channels of the $\tau$ are
$n$-pions channels. Decay channel into 2 pions is the most probable,
but there is a large contribution from the 3 pions channels and still
appreciable contribution from the 4 pions ones. For
bigger masses the contribution from the channels with higher
multiplicity become more important as Fig.~\ref{fig:GmGq} demonstrates.

\begin{figure}[t]
 \centering
 \includegraphics[width=0.9\linewidth]{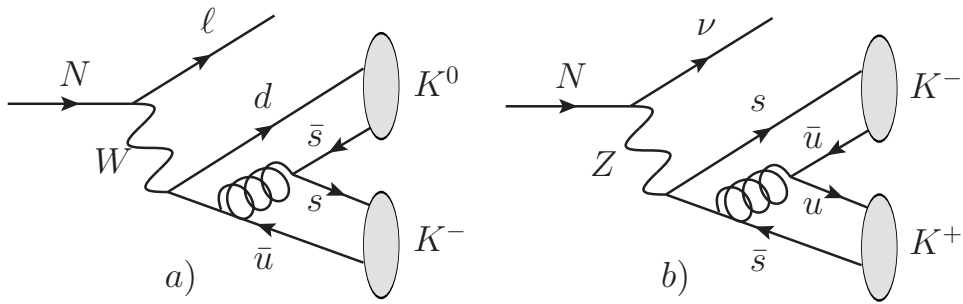}
 \caption{HNL decays into 2 kaons through charged a) and neutral b) currents.}
 \label{fig:Nto2K}
\end{figure}

The decay into kaons is suppressed for the $\tau$-lepton. For some channels like $\tau\to \nu_\tau K$ or $\tau\to \nu_\tau K \pi$ this suppression comes from the Cabibbo angle between $s$ and $u$ quarks. The same argument holds for HNL decays into lepton and $D$ meson, but not in $D_s$. The decays like $\tau\to \nu_\tau K^- K^0$ are not suppressed by CKM matrix and still are small. We think that this is because for such decays the probability of the QCD string fragmentation into strange quarks is much smaller than into $u$ and $d$ quarks for the given $\tau$-lepton mass (see diagram a) in Fig.~\ref{fig:Nto2K}). At higher masses the probability of such fragmentation should be higher, but still too small to take it into account. On the other hand, the HNL decay into two kaons can give a noticeable contribution, because of existence of the neutral current decay (see diagram b) in Fig.~\ref{fig:Nto2K}), for which the previous arguments do not apply.


\section{Summary}
\label{sec:summary}

In this paper we revise the phenomenology of the heavy neutral leptons,
including both their production and decays.
We concentrated on the HNL masses up to $\mathcal{O}(10)$~GeV.

The mechanisms of the HNL production are secondary decays of the hadrons produced in the initial collision (Section~\ref{sec:production-from-hadrons}), production in proton-nucleon collision (Section~\ref{sec:production-in-pp-collision}) and coherent scattering of the proton off nuclei (Section~\ref{sec:coherent-pZ-scattering}).
Of these mechanisms the production from the lightest flavored mesons dominate at all masses of interest.
Production from baryons is not efficient at any HNL mass (see discussion in Section~\ref{sec:prod-baryons}).
The main production channels above the kaon threshold are production from $D$ mesons for $M_N\lesssim 2$~GeV and production from the beauty mesons for $M_N \lesssim m_\Upsilon$.
For leptonic decays and two body semileptonic decays, the calculations are performed in Appendix~\ref{sec:production}.
Our results agree with~\cite{Gorbunov:2007ak}, for the case of the pseudoscalar and vector mesons we present the simplified version of the final formulas.
We additionally analyzed the HNL production in $B$ meson decays including multimeson final states, that were not previously discussed.
We estimate that contribution of the multimeson final state give not more than 20\% of production from $B$ mesons (Section~\ref{sec:multi-hadron-final}).

\begin{figure}[!t]
 \centering
 \includegraphics[width=0.47\textwidth]{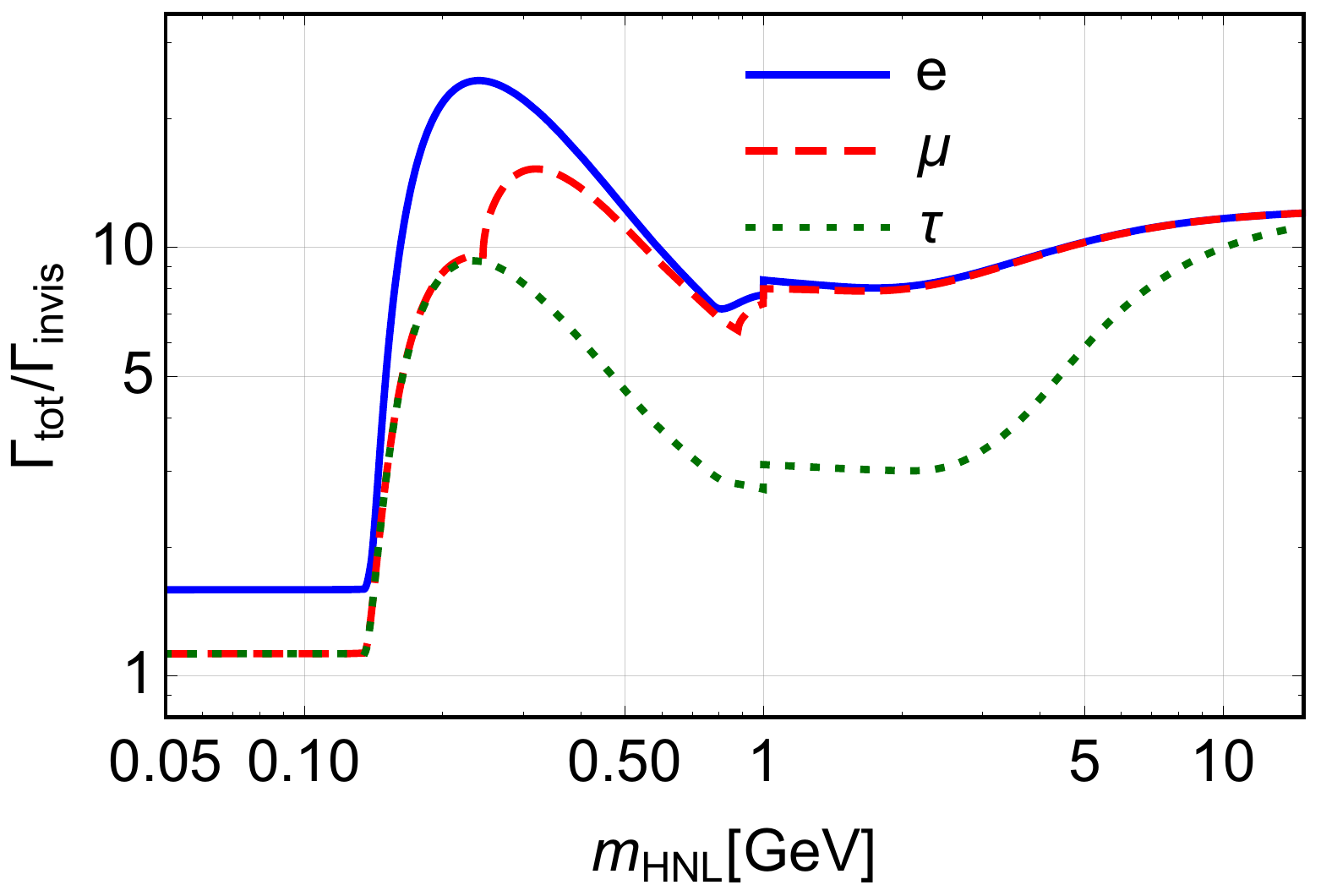}~\includegraphics[width=0.50\textwidth]{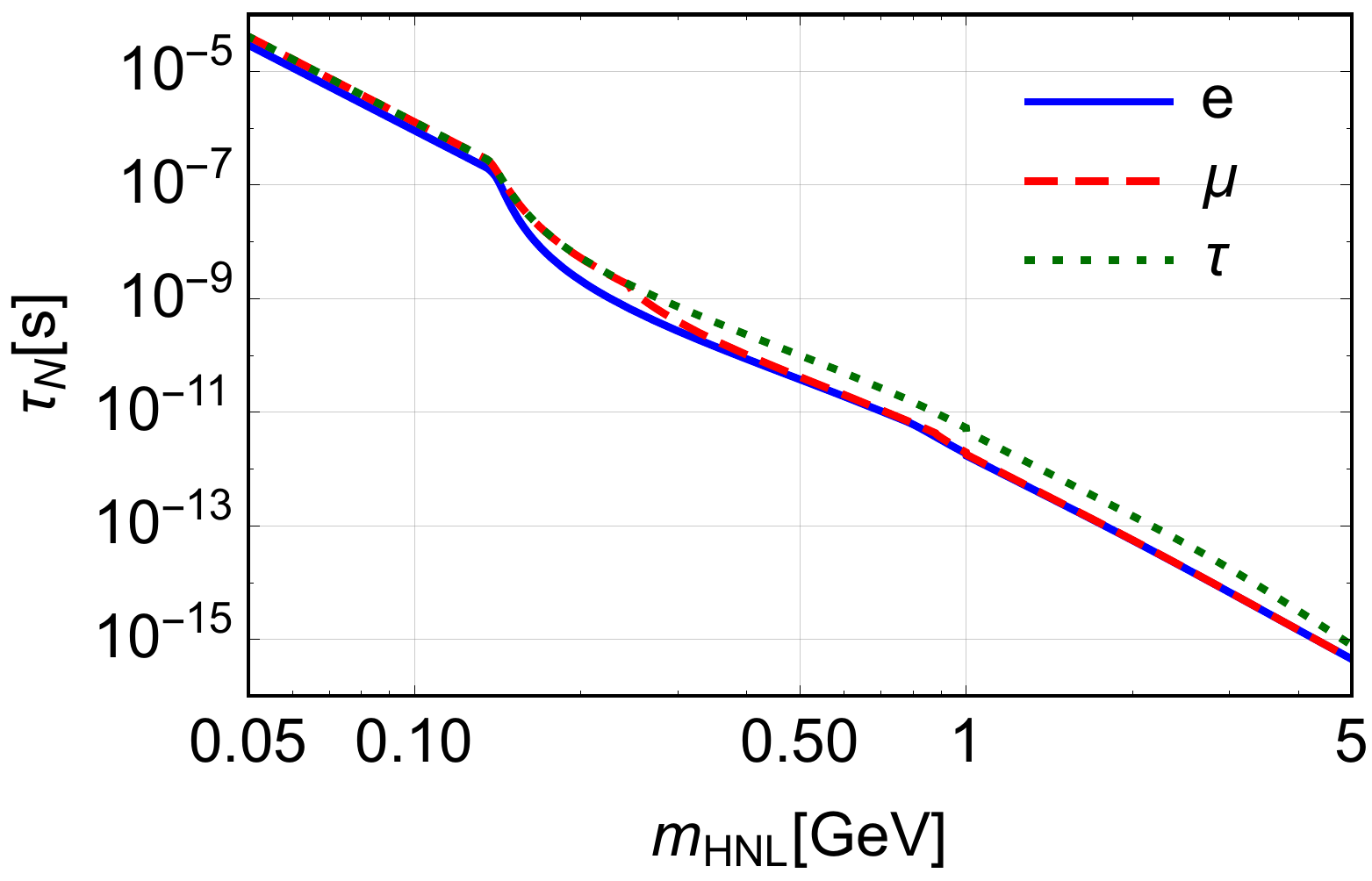}
 \caption{HNL decay width divided by the total decay width into active neutrinos Eq.~\eqref{eq:Gvvv} (left panel) and lifetime (right panel) as a function of HNL mass. Results are given for the pure $e$, $\mu$ and $\tau$ mixings, $U_{\alpha}=1$.
 }
 \label{fig:totalwidth}
\end{figure}

The HNL are unstable and decay into light SM particles which can be detected.
The HNL decay channels with branching ratio above $1\%$ in the region $M_N<5$~GeV are summarized in the Table~\ref{tab:decaychannels}.
For each channel we present the mass at which it opens, mass range where it is
relevant and maximal branching ratio.
The total decay width and the lifetime are summarized in Fig.~\ref{fig:totalwidth}.

All HNL decay channels can be divided into purely leptonic and semileptonic (hadronic)
ones. The decay widths into leptons are given by~\eqref{eq:Glud}, \eqref{eq:Gvff}, \eqref{eq:Gvvv} and are in full agreement with the literature~\cite{Gorbunov:2007ak,Atre:2009rg}.

For HNL masses above $m_\pi$ semileptonic decay channels quickly start to dominate, the hardonic branching ratio reaches $\sim 70\%$ at  $M_N \gtrsim 1$~GeV.
Single-meson final states (including decay into on-shell $\rho$ mesons) saturate hadronic decay width till about $1.5$~GeV (Fig.~\ref{fig:GmGq}).
In the HNL mass region $2-5$~GeV from 50\% to 80\% of the semileptonic decay width is saturated by multimeson states.
For completeness we summarize all relevant  hadronic form factors  in Appendices.

Our final results are directly suitable for sensitivity studies of
particle physics experiments (ranging from proton beam-dump to the LHC) aiming
at searches for heavy neutral leptons.

\subsubsection*{Acknowledgements}

For this project K.B., A.B.\ and O.R.\ have received funding from the European Research Council (ERC)
under the European Union's Horizon 2020 research and innovation programme (GA
694896) and from the Netherlands Science Foundation (NWO/OCW).

\begin{small}
\LTcapwidth=\textwidth
\begin{longtable}{|c|c|c|c|c|c|}
  \caption[Table of the relevant HNL decay channels]{\small
  Relevant HNL decay channels. Only channels with the branching
  ratio above $1\%$ covering the HNL mass range up to $5$~GeV are shown.
  The
  numbers are provided for 
  $|U_e|=|U_\mu|=|U_\tau|$.
  For neutral current channels (with $\nu_a$ in the final state) the sum over
  neutrino flavors is taken, otherwise the lepton flavor is shown explicitly.\newline
  \emph{Columns}: (1) the HNL decay channel.
  Notation
  $(n\pi)^a$ means a system of $n$ pions with the total charge $a$.
  (2)
  The HNL mass at which the channel opens;
  (3) The HNL mass
  starting from which the channel becomes relevant.
  For multimeson
  final states we provide our ``best-guess estimates'';
  (4) HNL mass above which the
  channel contributes less than $1\%$;
  ``---'' means that the channel is
  still relevant at $M_N=5$~GeV, ``?''
  means that we could not estimate the relevance of the channel;
  (5)
  The maximal branching ratio of the channel for $M_N<5$~GeV;
  (6) Reference to the formula for the decay
  width of the channel, if present in the text.
\label{tab:decaychannels}} \\
\hline
  Channel & Opens at& Relevant from & Relevant to & Max BR & Reference \\
  & [MeV]  &  [MeV]  &  [MeV] &  [\%] & in text\\
  \endfirsthead 
\caption[]{(continued)} \\
\hline
Channel & Open, MeV & Rel. from, MeV & Rel. to, MeV & Max BR, \% & Formula \endhead \hline
$N\to \nu_{\alpha} \nu_{\beta} \bar{\nu}_{\beta}$ 
	& $\sum m_{\nu}\approx 0$ & $\sum m_{\nu}\approx 0$ & --- & 100 & (\ref{eq:Gvvv}) \\ \hline
$N\to \nu_\alpha e^+e^-$ 
	& 1.02 & 1.29 & --- & 21.8 & (\ref{eq:Gvff}) \\ \hline
$N\to \nu_\alpha \pi^0$ 
	& 135 & 136 & 3630 & 57.3 & (\ref{eq:29}) \\ \hline
$N\to e^- \pi^+$ 
	& 140 & 141 & 3000 & 33.5 & (\ref{eq:17}) \\ \hline
$N\to \mu^-\pi^+$ 
	& 245 & 246 & 3000 & 19.7 & (\ref{eq:17}) \\ \hline
$N\to e^- \nu_\mu\mu^+$ 
	& 106 & 315 & --- & 5.15 & (\ref{eq:Glud}) \\ \hline
$N\to \mu^- \nu_e e^+$ 
	& 106 & 315 & --- & 5.15 & (\ref{eq:Glud}) \\ \hline
$N\to \nu_\alpha \mu^+ \mu^-$ 
	& 211 & 441 & --- & 4.21 & (\ref{eq:Gvff}) \\ \hline
$N\to \nu_{\alpha} \eta$ 
	& 548 & 641 & 2330 & 3.50 & (\ref{eq:29}) \\ \hline
$N\to e^-\pi^+\pi^0$ 
	& 275 & 666 & 4550 & 10.4 & (\ref{eq:N2lpipi}) \\ \hline
$N\to \nu_{\alpha}\pi^+\pi^-$ 
	& 279 & 750 & 3300 & 4.81 & (\ref{eq:N2vpipi}) \\ \hline
$N\to \mu^-\pi^+\pi^0$ 
	& 380 & 885 & 4600 & 10.2 & (\ref{eq:N2lpipi}) \\ \hline
$N\to \nu_{\alpha} \omega$ 
	& 783 & 997 & 1730 & 1.40 & \eqref{eq:41}
	\\ \hline
$N\to \nu_{\alpha} (3\pi)^0$ 
	&  $\gtrsim 405$  & $\gtrsim 1000$ & ? & ? & No \\ \hline
$N\to e^- (3\pi)^+ $ 
	&  $\gtrsim 410$  & $\gtrsim 1000$ & ? & ? & No \\ \hline
$N\to \nu_{\alpha} \eta'$ 
	& 958 & 1290 & 2400 & 1.86 & (\ref{eq:29}) \\ \hline
$N\to \nu_{\alpha} \phi$ 
	& 1019 & 1100 & 4270 & 5.90 & \eqref{eq:41} 
	\\ \hline
$N\to \mu^- (3\pi)^+ $ 
	& $\gtrsim 515$ & $\gtrsim 1100$ & ? & ? & No \\ \hline
$N\to \nu_{\alpha}K^+K^-$ 
	& 987 & $\gtrsim 1100$ & ? & ? & No \\ \hline
$N\to \nu_{\alpha} (4\pi)^0$ 
	& $\gtrsim 540$ & $\gtrsim 1200$ & ? & ? & No \\ \hline
$N\to e^- (4\pi)^+ $ 
	& $\gtrsim 545$ & $\gtrsim 1200$ & ? & ? & No \\ \hline
$N\to \mu^- (4\pi)^+ $ 
	& $\gtrsim 649$ & $\gtrsim 1300$ & ? & ? & No \\ \hline
$N\to \nu_{\alpha} (5\pi)^0$ 
	& $\gtrsim 675$  & $\gtrsim m_\tau\approx1780$ & ? & ? & No \\ \hline
$N\to e^- (5\pi)^+ $ 
	& $\gtrsim 680$  & $\gtrsim m_\tau\approx1780$ & ? & ? & No \\ \hline
$N\to \mu^- (5\pi)^+ $ 
	& $\gtrsim 785$ & $\gtrsim m_\tau\approx1780$ & ? & ? & No \\ \hline
$N\to e^- D_s^{*+}$ 
	& 2110 & 2350 & --- & 3.05 & (\ref{eq:40}) \\ \hline
$N\to \mu^- D_s^{*+}$ 
	& 2220 & 2370 & --- & 3.03 & (\ref{eq:40}) \\ \hline
$N\to e^- D_s^+$ 
	& 1970 & 2660 & 4180 & 1.23 & (\ref{eq:17}) \\ \hline
$N\to \mu^- D_s^+$ 
	& 2070 & 2680 & 4170 & 1.22 & (\ref{eq:17}) \\ \hline
$N\to \nu_{\alpha} \eta_c$ 
	& 2980 & 3940 & --- & 1.26 & (\ref{eq:29}) \\ \hline
$N\to \tau^- \nu_e e^+$
	& 1780 & 3980 & --- & 1.52 & (\ref{eq:Glud}) \\ \hline
$N\to e^- \nu_\tau \tau^+$
	& 1780 & 3980 & --- & 1.52 & (\ref{eq:Glud}) \\ \hline
$N\to \tau^- \nu_\mu \mu^+$ 
	& 1880 & 4000 & --- & 1.51 & (\ref{eq:Glud}) \\ \hline
$N\to \mu^- \nu_\tau \tau^+$ 
	& 1880 & 4000 & --- & 1.51 & (\ref{eq:Glud}) \\ \hline
\end{longtable}
\end{small}


\appendix

\section{HNL production from hadrons}
\label{sec:production}

The calculation of weak decays, involving hadrons is summarized 
in~\cite{Antonelli:2009ws}. 
In the absence of QED and QCD corrections the effective weak interaction Lagrangian at low energies can be written as 
\begin{equation}
  \label{eq:11}
  \mathcal{L}_\text{weak} = \mathcal{L}_{\text{cc}} + \mathcal{L}_{\text{nc}}
\end{equation}
where the \emph{charged current} terms have the form
\begin{equation}
  \label{eq:12}
  \mathcal{L}_{\text{cc}} = \frac{G_F}{\sqrt 2} \left|\sum_{U,D} V_{UD} J^{\mu,+}_{UD} + \sum_\ell J^{\mu,+}_\ell \right|^2,
\end{equation}
where 
\begin{align}
	J^{\mu,+}_{UD} = \bar{D}\gamma^{\mu} (1-\gamma^5) U, \\
	J^{\mu,+}_{\ell} = \bar{\ell}\gamma^{\mu} (1-\gamma^5) \nu_{\ell},
\end{align}
and $V_{UD}$ is the CKM element which corresponds to quark flavour transition in hadronic current. For \emph{neutral current} the interaction has the same form
\begin{equation}
  \label{eq:19}
  \mathcal{L}_{\text{nc}} = \frac{G_F}{\sqrt 2} \left( \sum_f J^{\mu,0}_f\right)^2,
\end{equation}
where summation goes over all fermions,
\begin{align}
	J^{\mu,0}_f = \bar{f} \gamma^\mu (v_f - a_f &\gamma^5) f, \\
	v_f = I_{3f} - 2 Q_f \sin^2\theta_W, &\quad a_f = I_{3f}
\end{align}
and $I_{3f}$ is the fermion isospin projection and $Q_f$ is its electric charge ($Q_e = -1$). In the following Sections we describe different processes with HNL and hadrons. 

\subsection{Leptonic decay of a pseudoscalar meson}
\label{sec:leptonic-decays}

Consider a decay of pseudoscalar meson $h$ into charged lepton $\ell$ and HNL:
\begin{equation}
  \label{eq:13}
  h \to \ell+ N,
\end{equation}
see left diagram in Fig.~\ref{fig:h2l_decay}. The corresponding matrix element is given by
\begin{equation}
  \label{eq:14}
  \CM = \frac{G_F}{\sqrt{2}} V_{UD} \bra{0}J^\mu_{UD}\ket{h}\bra{\ell,N}J_{\ell,\mu}\ket0,
\end{equation}
where the corresponding quark contents of meson $h$ is $\ket{h} = \ket{\bar U
  D}$. 
In order to fix the notations we remind that the charged meson coupling
constant, $f_h$, for a pseudoscalar meson constructed from up ($U$) and down
($D$) type quarks is defined as
\begin{equation}
  \label{eq:15}
  \bra{0}J^\mu_{UD}\ket{h} = 
  \bra{0} \bar U \gamma^\mu \gamma_5 D \ket{h}
  \equiv  i f_h p^\mu
\end{equation}
where $p_\mu$ is 4-momentum of the pseudo-scalar meson $h$.
The numerical values of the decay constants for different mesons are summarized in Table~\ref{tab:f_meson}.

After standard calculation one finds the decay width of this reaction
\begin{equation}
	\Gamma(h\to \ell_{\alpha} N) = \frac{G_F^2 f_h^2 m_h^3}{8\pi} 
	|V_{UD}|^2 |U_{\alpha}|^2 \left[ y_N^2 + y_\ell^2 - 
	\left( y_N^2 - y_\ell^2 \right)^2 \right] 
	\sqrt{\lambda(1, y_N^2, y_\ell^2)},
\end{equation}
where \mbox{$y_\ell = m_\ell / m_h$}, \mbox{$y_N = M_N / m_h$} and $\lambda$ is given by~\eqref{eq:37}.

\subsection{Semileptonic decay of a pseudoscalar meson}
\label{sec:decay-with-meson}

The process with pseudoscalar or vector meson $h'_{P/V}$ in the final state
\begin{equation}
  \label{eq:21}
  h \to h'_{P/V} + \ell + N,
\end{equation}
is mediated by the current that has $V-A$ form (see right diagram in Fig.~\ref{fig:h2l_decay}). Properties of the hadronic matrix element 
$\bra{h'_{P/V}}J^\mu_\text{hadron}\ket{h\vphantom{h'_{P/V}}}$ 
depend on the type of final meson $h'$~\cite{Gilman:1989uy}. In the case of pseudoscalar meson only vector part of the current plays role:
\begin{align}
  \bra{h'_P(p')}V_\mu\ket{h(p)} &= 
  f_+(q^2) (p+p')_\mu + f_-(q^2) q_\mu = \nonumber\\
  & = f_+(q^2) \left(p_\mu + p'_\mu - \frac{m_h^2 - m_{h'}^2}{q^2} q_\mu\right) + f_0(q^2) \frac{m_h^2 - m_{h'}^2}{q^2} q_\mu
  \label{eq:22}
\end{align}
where $q_\mu = (p - p')_\mu$ is a transferred momentum and
\begin{equation}
 f_0(q^2) \equiv f_+(q^2) + \frac{q^2}{m_h^2 - m_{h'}^2} f_-(q^2).
 \label{eq:f0def}
\end{equation}

For the case of a vector meson $h'_V$ in the final state both vector
and axial part of the current contribute. The standard parametrization
with form factors is
\begin{equation}
 \label{eq:24}
  \bra{h'_V(\epsilon,p'_\nu)}  V^\mu \ket{h(p_\nu)} = i g(q^2) \varepsilon^{\mu\nu\sigma\rho} 
  \epsilon^{*}_{\nu} (p+p')_{\sigma} (p-p')_{\rho},
\end{equation} 
\begin{equation}
  \label{eq:23}
  \bra{h'_V(\epsilon,p'_\nu)}  A_\mu \ket{h(p_\nu)} = 
  f(q^2) \epsilon_\mu^* + 
  a_+(q^2) (\epsilon^*\cdot p) (p+p')_\mu + 
  a_-(q^2) (\epsilon^*\cdot p) (p-p')_\mu,
\end{equation}
where $\epsilon_\mu$ is a polarization vector of the vector meson $h'_V$.

Using matrix elements~(\ref{eq:22}--\ref{eq:23}) it is straightforward to calculate decay widths of the reactions. In the case of pseudoscalar meson $h'_P$ we follow Ref.~\cite{Abada:2013aba} and decompose full decay width into 4 parts,
\begin{equation}
	\Gamma(h\to h'_P \ell_{\alpha} N) = \frac{G_F^2 m_h^5}{64 \pi^3} C_K^2
	|V_{UD}|^2 |U_\alpha|^2 \left(I_{P,1} + I_{P,2} + I_{P,3} + I_{P,4}\right),
\end{equation}
where $I_{P,1},I_{P,2}$ depend on $|f_+(q^2)|^2$, $I_{P,3}$ on $|f_0(q^2)|^2$ and $I_{P,4}$ on $\text{Re}\left(f_0(q^2)f_+^*(q^2)\right)$. It turns out that $I_{P,4} = 0$, the explicit expressions for others are
\begin{align}
	I_{P,1} &=
	\int\limits_{(y_{\ell}+y_N)^2}^{(1-y_{h'})^2}
	\frac{d\xi}{3\xi^3} |f_+(q^2)|^2 
	\Lambda^3(\xi),
	\label{eq:GP1} \\
	I_{P,2} &=
	\int\limits_{(y_{\ell}+y_N)^2}^{(1-y_{h'})^2}
	\frac{d\xi}{2\xi^3} |f_+(q^2)|^2
	\Lambda(\xi) G_-(\xi)
	\lambda(1,y_{h'}^2,\xi),
	\label{eq:GP2} \\
	I_{P,3} &= 
	\int\limits_{(y_{\ell}+y_N)^2}^{(1-y_{h'})^2}
	\frac{d\xi}{2\xi^3} |f_0(q^2)|^2 
	\Lambda(\xi) G_-(\xi) (1 - y_{h'}^2)^2,
	\label{eq:GP3}
\end{align}
where  
\begin{align}
 \Lambda(\xi) &= \lambda^{1/2}(1,y_{h'}^2,\xi) \lambda^{1/2}(\xi,y_{N}^2,y_{\ell}^2),
 \\
 G_-(\xi) &= \xi \left(y_{N}^2+y_{\ell}^2\right) 
 -\left(y_{N}^2-y_{\ell}^2\right)^2,
\end{align}
$y_i = \dfrac{m_i}{m_h}$, $\xi = \dfrac{q^2}{m_h^2}$ and function $\lambda(a,b,c)$ is given by~(\ref{eq:37}). $C_K$ is a Clebsh-Gordan coefficient, see for example~\cite[(14)]{Blucher:2005dc} and~\cite[(2.1)]{Antonelli:2008jg}, $C_K = 1/\sqrt{2}$ for decays into $\pi^0$ and $C_K = 1$ for all other cases.

For the decay into vector meson the expression is more bulky,
\begin{align}
	\Gamma(h\to h'_V \ell_{\alpha} N) &= \frac{G_F^2 m_h^7}{64 \pi^3 m_{h'}^2}
	C_K^2 |V_{UD}|^2 |U_\alpha|^2 \Big(I_{V,g^2} + I_{V,f^2} + I_{V,a_+^2} + I_{V,a_-^2} 
	+ \nonumber \\ &+ 
	I_{V,gf} + I_{V,g a_+} + I_{V,g a_-} + I_{V,f a_+} + I_{V,f a_-} + I_{V,a_+ a_-}\Big),
\end{align}
where $I_{V, F G}$ are parts of the decay width that depend on the $F
G$ form factors combination\footnote{In this computation we take all
  form factors as real-valued functions.} and $C_K$ is a Clebsh-Gordan coefficient, $C_K = 1/\sqrt{2}$ for decays into $\rho^0$ and $C_K = 1$ for all other cases in this paper. It turns out that $I_{V, g f} = I_{V, g a_+}= I_{V, g a_-} = 0$, the other terms are given by
\begin{align}
	I_{V,g^2} &= 
	\frac{m_h^2 y_{h'}^2}{3}
	\!\!\!\!
	\int\limits_{(y_{\ell}+y_N)^2}^{(1-y_{h'})^2}
	\frac{d\xi}{\xi^2} g^2(q^2) 
	\Lambda(\xi) F(\xi) 
	\left(
	2 \xi^2 - G_+(\xi)
	\right),
\end{align}
\begin{align}
I_{V,f^2} &= 
	\frac{1}{24 m_h^2}
	\int\limits_{(y_{\ell}+y_N)^2}^{(1-y_{h'})^2}
	\frac{d\xi}{\xi^3} f^2(q^2) 
	\Lambda(\xi) 
	\times \nonumber \\ & \times
	\left(3 F(\xi)
   \left[\xi^2 - \left(y_{\ell}^2-y_N^2\right)^2\right] -\Lambda^2(\xi)
   +12 y_{h'}^2 \xi \left[2 \xi^2 - G_+(\xi)\right]
   \right),
\end{align}
\begin{align}
 I_{V,a_+^2} &= 
	\frac{m_h^2}{24}
	\int\limits_{(y_{\ell}+y_N)^2}^{(1-y_{h'})^2}
	\frac{d\xi}{\xi^3} a_+^2(q^2) 
	\Lambda(\xi) F(\xi)
   \left( F(\xi) \left[
   2\xi^2 - G_+(\xi)
   \right]
   + 3 G_-(\xi) \left[1-y_{h'}^2\right]^2 
   \right),
\end{align}
\begin{align}
I_{V,a_-^2} &= 
	\frac{m_h^2}{8}
	\int\limits_{(y_{\ell}+y_N)^2}^{(1-y_{h'})^2}
	\frac{d\xi}{\xi} a_-^2(q^2) 
	\Lambda(\xi) F(\xi) G_-(\xi),
    \\
	I_{V,f a_+} &= 
	\frac{1}{12}
	\int\limits_{(y_{\ell}+y_N)^2}^{(1-y_{h'})^2}
	\frac{d\xi}{\xi^3} f(q^2)a_+(q^2) 
	\Lambda(\xi)
	\times \nonumber \\ & \times
	\left(
	3 \xi F(\xi) G_-(\xi) +\left(1-\xi-y_{h'}^2\right) \left[3 F(\xi)
   \left(\xi^2 - \left(y_l^2-y_N^2\right)^2\right)-\Lambda^2(\xi)
   \right]
   \right),
    \\
	I_{V,f a_-} &= 
	\frac{1}{4}
	\int\limits_{(y_{\ell}+y_N)^2}^{(1-y_{h'})^2}
	\frac{d\xi}{\xi^2} f(q^2)a_-(q^2) 
	\Lambda(\xi) F(\xi) G_-(\xi),
    \\
	I_{V,a_+ a_-} &= 
	\frac{m_h^2}{4}
	\int\limits_{(y_{\ell}+y_N)^2}^{(1-y_{h'})^2}
	\frac{d\xi}{\xi^2} a_+(q^2)a_-(q^2) 
	\Lambda(\xi) F(\xi) G_-(\xi) \left(1 - y_{h'}^2\right),
\end{align}
where the notation is the same as in Eqs.~(\ref{eq:GP1}-\ref{eq:GP3}) and
\begin{align}
 F(\xi) &= (1-\xi)^2 - 2 y_{h'}^2(1+\xi) 
 + y_{h'}^4,
 \\
 G_+(\xi) &= \xi \left(y_{N}^2+y_{\ell}^2\right) 
 + \left(y_{N}^2-y_{\ell}^2\right)^2.
\end{align}

\section{HNL decays into hadronic states}

\subsection{Connection between matrix elements of the unflavoured mesons}
\label{sec:connection-between-matrix-elements-of-mesons}

\subsubsection{G-symmetry}

An important symmetry of the low-energy theory of strong interactions is the so-called \emph{$G$-symmetry} which is a combination of the charge conjugation $\hat{C}$ and rotation of $180^\circ$ around the $y$ axis in the isotopic space $\hat{R}_y$.\footnote{The latter corresponds to the interchange of $u$ and $d$ quarks with an additional phase, see Eq.~\eqref{eq:44} below.} The operation of charge conjugation acts on bilinear combinations of fermions $f_1$, $f_2$ as follows:
\begin{align}
	\hat{C} \bar{f}_1 f_2 &= \bar{f}_2 f_1, \\
	\hat{C} \bar{f}_1\gamma_5 f_2 &= \bar{f}_2 \gamma_5 f_1, \\
	\hat{C} \bar{f}_1 \gamma_\mu f_2 &= - \bar{f}_2 \gamma_\mu f_1, \\
	\hat{C} \bar{f}_1 \gamma_\mu \gamma_5 f_2 &= \bar{f}_2 \gamma_\mu \gamma_5 f_1.
\end{align}
$\hat{R}_y$ acts on the isospin doublet as
\begin{equation}
  \label{eq:44}
	\hat{R}_y 
	\begin{pmatrix}
	 u \\ d
	\end{pmatrix}
	=
	\begin{pmatrix}
	 d \\ -u
	\end{pmatrix}.
\end{equation}

Acting on pion states, that are pseudoscalar isovectors, one gets
\begin{align}
	\hat{G} \ket{\pi^+} &= 
	\hat{R}_y \hat{C} \ket{\bar{d}\gamma_5 u} = 
	\hat{R}_y \ket{\bar{u}\gamma_5 d} = 
	- \ket{\bar{d}\gamma_5 u} = - \ket{\pi^+}, \\
	\hat{G} \ket{\pi^0} &= 
	\hat{R}_y \hat{C} \frac{1}{\sqrt{2}}\ket{\bar{u}\gamma_5 u-\bar{d}\gamma_5 d} = 
	\hat{R}_y \frac{1}{\sqrt{2}}\ket{\bar{u}\gamma_5 u-\bar{d}\gamma_5 d} = \nonumber \\ 
	&=  - \frac{1}{\sqrt{2}}\ket{\bar{u}\gamma_5 u-\bar{d}\gamma_5 d} =  
	- \ket{\pi^0},
\end{align}
so any pion is an odd state under $G$-symmetry. As a consequence,
for the system of $n$ pions
\begin{equation}
	\hat{G} \ket{n \pi} = (-1)^n \ket{n \pi}.
\end{equation}
For $\rho$ mesons, which are vector isovectors, $G$-parity is positive,
\begin{align}
	\hat{G} \ket{\rho^+} &= 
	\hat{R}_y \hat{C} \ket{\bar{d}\gamma_{\mu}u} = 
	- \hat{R}_y \ket{\bar{u}\gamma_{\mu} d} = 
	\ket{\bar{d}\gamma_{\mu}u} = \ket{\rho^+}, \\
	\hat{G} \ket{\rho^0} &= 
	\hat{R}_y \hat{C} \frac{1}{\sqrt{2}}\ket{\bar{u}\gamma_{\mu}u-\bar{d}\gamma_{\mu}d} = 
	- \hat{R}_y \frac{1}{\sqrt{2}}\ket{\bar{u}\gamma_{\mu}u-\bar{d}\gamma_{\mu}d} 
	= \nonumber \\ 
	&= \frac{1}{\sqrt{2}}\ket{\bar{u}\gamma_{\mu}u-\bar{d}\gamma_{\mu}d} = \ket{\rho^0},
\end{align}
while for $a_1$ mesons,  which are pseudovector isovectors, $G$-parity is negative,
\begin{align}
	\hat{G} \ket{a_1^+} &= 
	\hat{R}_y \hat{C} \ket{\bar{d}\gamma_{\mu}\gamma_5 u} = 
	\hat{R}_y \ket{\bar{u}\gamma_{\mu}\gamma_5 d} = 
	- \ket{\bar{d}\gamma_{\mu}\gamma_5 u} = - \ket{a_1^+}, \\
	\hat{G} \ket{a_1^0} &= 
	\hat{R}_y \hat{C} \frac{1}{\sqrt{2}}\ket{\bar{u}\gamma_{\mu}\gamma_5 u-\bar{d}\gamma_{\mu}\gamma_5 d} = 
	\hat{R}_y \frac{1}{\sqrt{2}}\ket{\bar{u}\gamma_{\mu} \gamma_5 u-\bar{d}\gamma_{\mu}\gamma_5 d} = \nonumber \\
	&=  
	- \frac{1}{\sqrt{2}}\ket{\bar{u}\gamma_{\mu}\gamma_5 u-\bar{d}\gamma_{\mu}\gamma_5 d} =  
	-\ket{a_1^0},
\end{align}

\subsubsection{Classification of currents}

Unflavoured quarks system interacts with electromagnetic field, $W$- and $Z$-bosons through currents
\begin{align}
	J_\mu^{\text{EM}} &= \frac{2}{3}\bar{u}\gamma_\mu u - \frac{1}{3}\bar{d}\gamma_\mu d, 
	\label{eq:JEM}\\
	J^W_\mu &= \bar u \gamma_\mu (1 - \gamma_5) d, 
	\label{eq:JW}\\
	J^Z_\mu &= \bar{u}\gamma_\mu (v_u - a_u \gamma_5) u + \bar{d}\gamma_\mu (v_d - a_d \gamma_5) d,
	\label{eq:JZ}
\end{align}
where 
\begin{align}
	v_u &= \frac{1}{2} - \frac{4}{3} \sin^2\theta_W, 
	\qquad a_u = \frac{1}{2}, \\
	v_d &= -\frac{1}{2} + \frac{2}{3} \sin^2\theta_W, 
	\qquad a_d = -\frac{1}{2}.
\end{align}

\begin{table}[t]
\centering
\begin{tabular}{|l|c|c|c|c|}
\hline
Current    & $j_{\mu}^{V,s}$ & $j_{\mu}^{V,+/0/-}$ & 
$j_{\mu}^{A,s}$ & $j_{\mu}^{A,+/0/-}$ \\ \hline
$G$-parity & $-$ & $+$ & $+$ & $-$ \\ \hline
\end{tabular}
\caption{Properties of axial and vector currents under $G$-symmetry.}
\label{tab:currentsG}
\end{table}

To divide currents~(\ref{eq:JEM})-(\ref{eq:JZ}) into G-odd and G-even parts let us introduce isoscalar and isovector vector currents
\begin{align}
	j_{\mu}^{V,s} &= \frac{1}{\sqrt{2}} 
	(\bar{u}\gamma_\mu u + \bar{d}\gamma_\mu d), 
	\label{eq:jVs}\\
	j_{\mu}^{V,0} &= \frac{1}{\sqrt{2}} 
	(\bar{u}\gamma_\mu u - \bar{d}\gamma_\mu d),\label{eq:45} \\
	j_{\mu}^{V,+} &= \bar d \gamma_\mu u\,, 
	\qquad j_{\mu}^{V,-} = \bar u \gamma_\mu d\,, \label{eq:46}
\end{align}
and isoscalar and isovector axial currents
\begin{align}
	j_{\mu}^{A,s} &= \frac{1}{\sqrt{2}} 
    (\bar{u}\gamma_\mu \gamma_5 u + \bar{d}\gamma_\mu \gamma_5 d), \\
	j_{\mu}^{A,0} &= \frac{1}{\sqrt{2}} 
	(\bar{u}\gamma_\mu \gamma_5 u - \bar{d}\gamma_\mu \gamma_5 d), \\
	j_{\mu}^{A,+} &= \bar d \gamma_\mu \gamma_5 u\,,
	\qquad j_{\mu}^{A,-} = \bar u \gamma_\mu \gamma_5 d\,.
	\label{eq:jApm}
\end{align}
Currents~(\ref{eq:jVs})-(\ref{eq:jApm}) have certain $G$-parity
presented in Table~\ref{tab:currentsG}. Using these currents one can rewrite physical currents as
\begin{align}
	J_\mu^{\text{EM}} &= \frac{1}{\sqrt{2}} j_{\mu}^{V,0} + \frac{1}{3\sqrt{2}} j_{\mu}^{V,s}, 
	\label{eq:JEMdivided}\\
	J^W_\mu &= j_{\mu}^{V,-} - j_{\mu}^{A,-}, 
	\label{eq:JWdivided}\\
	J^Z_\mu &= \frac{1-2\sin^2\theta_W}{\sqrt{2}} j_{\mu}^{V,0} - 
	\frac{\sqrt{2}\sin^2\theta_W}{3} j_{\mu}^{V,s} - 
	\frac{1}{\sqrt{2}} j_{\mu}^{A,0}.
	\label{eq:JZdivided}
\end{align}

\subsubsection{Connection between matrix elements}


$G$-even part of currents~(\ref{eq:JEMdivided})-(\ref{eq:JZdivided})
belongs to one isovector family, therefore there is an approximate connection between matrix elements for the system of even number of pions or $\rho$-meson,
\begin{equation}
	\bra{0} J_\mu^{\text{EM}} \ket{2n \pi/\rho} \approx 
	\frac{1}{\sqrt{2}} \bra{0} J_\mu^{W} \ket{2n \pi/\rho} \approx 
	\frac{1}{1-2\sin^2\theta_W} \bra{0} J_\mu^{Z} \ket{2n \pi/\rho}.
	\label{eq:relation4even}
\end{equation}
The special case to mention here is $\ket{2\pi^0}$ state. In $V \pi^0
\pi^0$ vertex, where $V=\gamma/Z$, system of 2 pions should have total
angular momentum $J=1$. Pions are spinless particles, so their
coordinate wavefuction has negative parity, which is forbidden by the Bose--Einstein statistics. Therefore
\begin{equation}
	\bra{0} J_\mu^{\text{EM}} \ket{2\pi^0} = \bra{0} J_\mu^{Z} \ket{2\pi^0} = 0.
	\label{eq:2pi0prohibition}
\end{equation}
This result is equivalent to the prohibition of the $\rho^0\to 2\pi^0$ decay.

$G$-odd parts of the
currents~(\ref{eq:JEMdivided})-(\ref{eq:JZdivided}), see Table~\ref{tab:currentsG}, belong to one isoscalar and one isovector families, so there is only one relation between matrix elements for the system of odd number of pions or for $a_1$-mesons,
\begin{equation}
	\frac{1}{\sqrt{2}}\bra{0} J_\mu^{W} \ket{(2n+1) \pi/a_1} \approx
	\bra{0} J_\mu^{Z} \ket{(2n+1) \pi/a_1} +
	2\sin^2\theta_W \bra{0} J_\mu^{\text{EM}} \ket{(2n+1) \pi/a_1}.
	\label{eq:relation4odd}
\end{equation}
The last formula can be simplified in the case of the one-pion or $a_1$ state. The direct interaction between photon and $\pi^0$ is forbidden because of the $C$ symmetry, while photon-to-$a_1$ interaction violates both $P$ and $C$ symmetry. Therefore, the matrix element $\bra{0} J_\mu^{\text{EM}} \ket{\pi/a_1} = 0$ and
\begin{equation}
	\frac{1}{\sqrt{2}}\bra{0} J_\mu^{W} \ket{\pi/a_1} \approx
	\bra{0} J_\mu^{Z} \ket{\pi/a_1}.
	\label{eq:relationpia1}
\end{equation}

All the approximate relations discussed above hold up to isospin violating terms of order \mbox{$(m_{\pi^+} - m_{\pi^0})/m_\pi\sim 3.4\%$}.

\subsection{HNL decays to a meson and a lepton}
\label{sec:hnl-decaying-meson}

There are 4 types of this decay: $N\to \ell_{\alpha} + h_{P/V}$ and $N\to \nu_{\alpha} + h_{P/V}$, where $h_P$ and $h_V$ are pseudoscalar and vector mesons respectively. Reaction $N\to \ell_{\alpha} + h_P$ is closely related to the process calculated in 
Section~\ref{sec:leptonic-decays}. It utilizes the same matrix element and differs only by
kinematics. Using the same notation, the decay width is
\begin{equation}
	\Gamma(N\to \ell_{\alpha} h_P) = \frac{G_F^2 f_h^2 M_N^3}{16\pi} 
	|V_{UD}|^2 |U_{\alpha}|^2 \left[ \left( 1 - x_\ell^2 \right)^2 - 
	x_h^2(1 + x_\ell^2) \right] 
	\sqrt{\lambda(1, x_h^2, x_\ell^2)},
\end{equation}
where \mbox{$x_h = m_h / M_N$}, \mbox{$x_\ell = m_\ell / M_N$} and function
$\lambda$ is given by eq.~(\ref{eq:37}).

In case of the neutral current-mediated decay $N \to \nu_{\alpha} +
h_P$ the hadronic matrix element reads (see Section \ref{sec:etaetaprime} for details) 
\begin{equation}
  \bra{0}J_\mu^{Z}\ket{h_P^0} \equiv - i \frac{f_h}{\sqrt{2}} p_\mu,
  \label{eq:fhneutralscalar}
\end{equation}
where $p_\mu$ is the 4-momentum of the pseudo-scalar meson $h$, $J_\mu^{Z}$ current is given by Eq.~\eqref{eq:JZ}.
The decay width is
\begin{equation}
	\Gamma(N\to \nu_{\alpha} h_P) = \frac{G_F^2 f_h^2 M_N^3}{32\pi} 
	|U_{\alpha}|^2 \left( 1 - x_h^2 \right)^2,
\end{equation}
where \mbox{$x_h = m_h / M_N$} and $f_h$ are neutral meson decay constants presented in the right part of Table~\ref{tab:f_meson}.

Consider the process $N\to \ell_{\alpha} + h_{V}$. For the vector meson the hadronic matrix element of the charged current is defined as
\begin{equation}
  \bra{0}J^\mu_{UD}\ket{h_V} \equiv  i g_h \varepsilon^\mu(p),
  \label{eq:vectorgh}
\end{equation}
where $\varepsilon^{\mu}(p)$ is polarization vector of the meson and $g_h$ is the vector meson decay constant. The values of the $g_h$ are given in Table~\ref{tab:g_meson}. Using previous notations, the decay width of this process is
\begin{equation}
  \Gamma(N\to \ell_{\alpha}^- h_{V}^+) = \frac{G_F^2 g_h^2 |V_{UD}|^2 |U_{\alpha}|^2 M_N^3}{16\pi m_h^2}  
	\left(
	\left(1 - x_\ell^2\right)^2 + x_h^2 \left(1 + x_\ell^2\right) - 2 x_h^4
	\right)
	\sqrt{\lambda(1, x_h^2, x_\ell^2)}.
\end{equation}

Finally, to calculate HNL decay into neutral vector meson $N\to \nu_{\alpha} + h_{V}$ we define the hadronic matrix element as
\begin{equation}
  \bra{0}J_\mu^{Z}\ket{h_V^0} \equiv  i \frac{\kappa_h g_h}{\sqrt{2}} \varepsilon^\mu(p),
  \label{eq:vectorgh0}
\end{equation}
where $g_h$ is the vector meson decay constant and $\kappa_h$ is
dimensionless correction factor, their values are given in
Table~\ref{tab:g_meson}. For the decay width one obtains
\begin{equation}
 \Gamma(N\to \nu_{\alpha} h_V) = \frac{G_F^2 \kappa_h^2 g_\rho^2 |U_{\alpha}|^2  M_N^3}{32\pi m_h^2}
 \left(1 + 2 x_h^2\right) \left(1 - x_h^2\right)^2.
\end{equation}

\subsection{HNL decays to a lepton and two pions}
\label{sec:hnl-two-pion-decays}

\begin{figure}[t]
  \centering
  \includegraphics[width=0.45\textwidth]{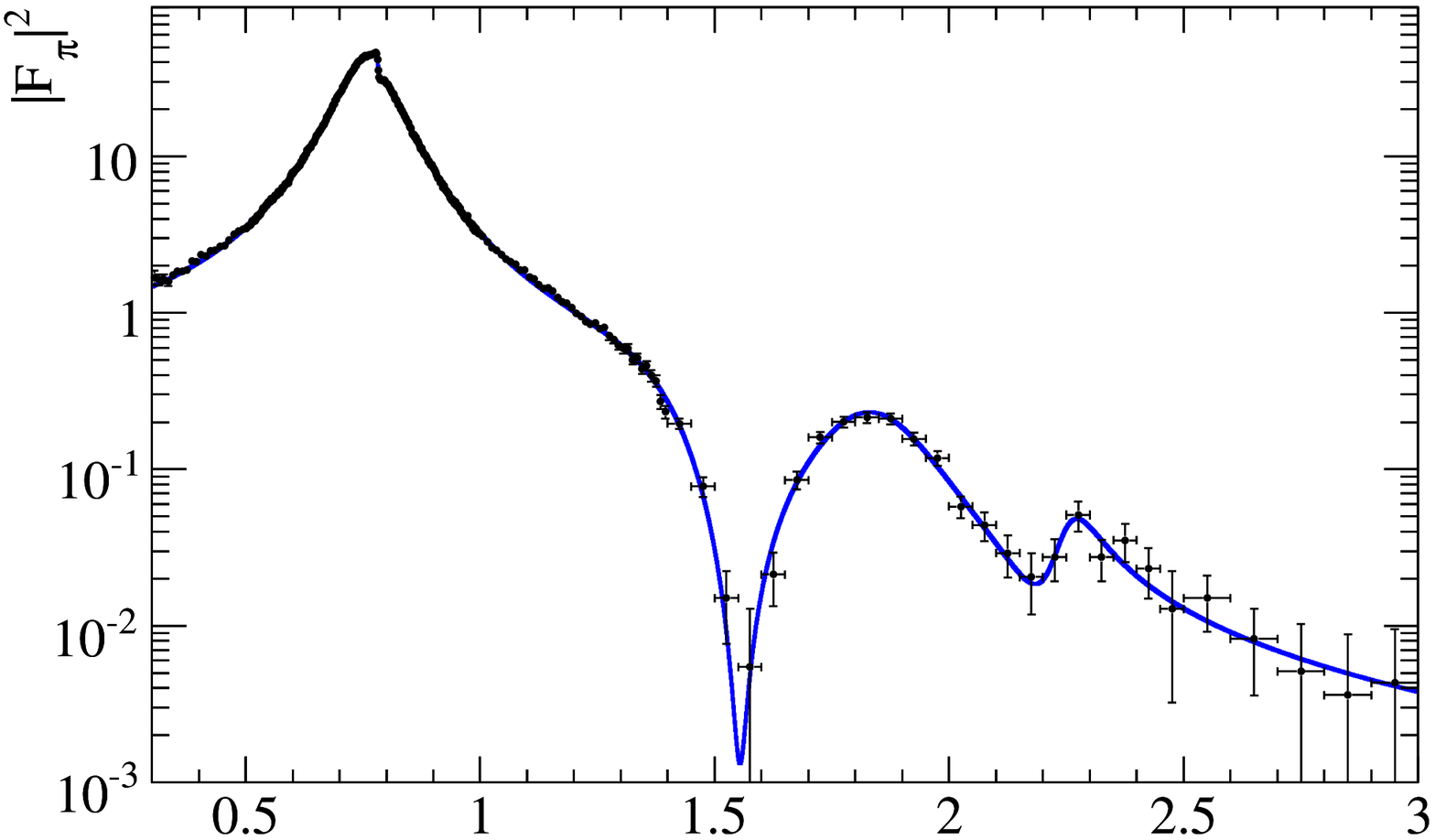}~
  \includegraphics[width=0.45\textwidth]{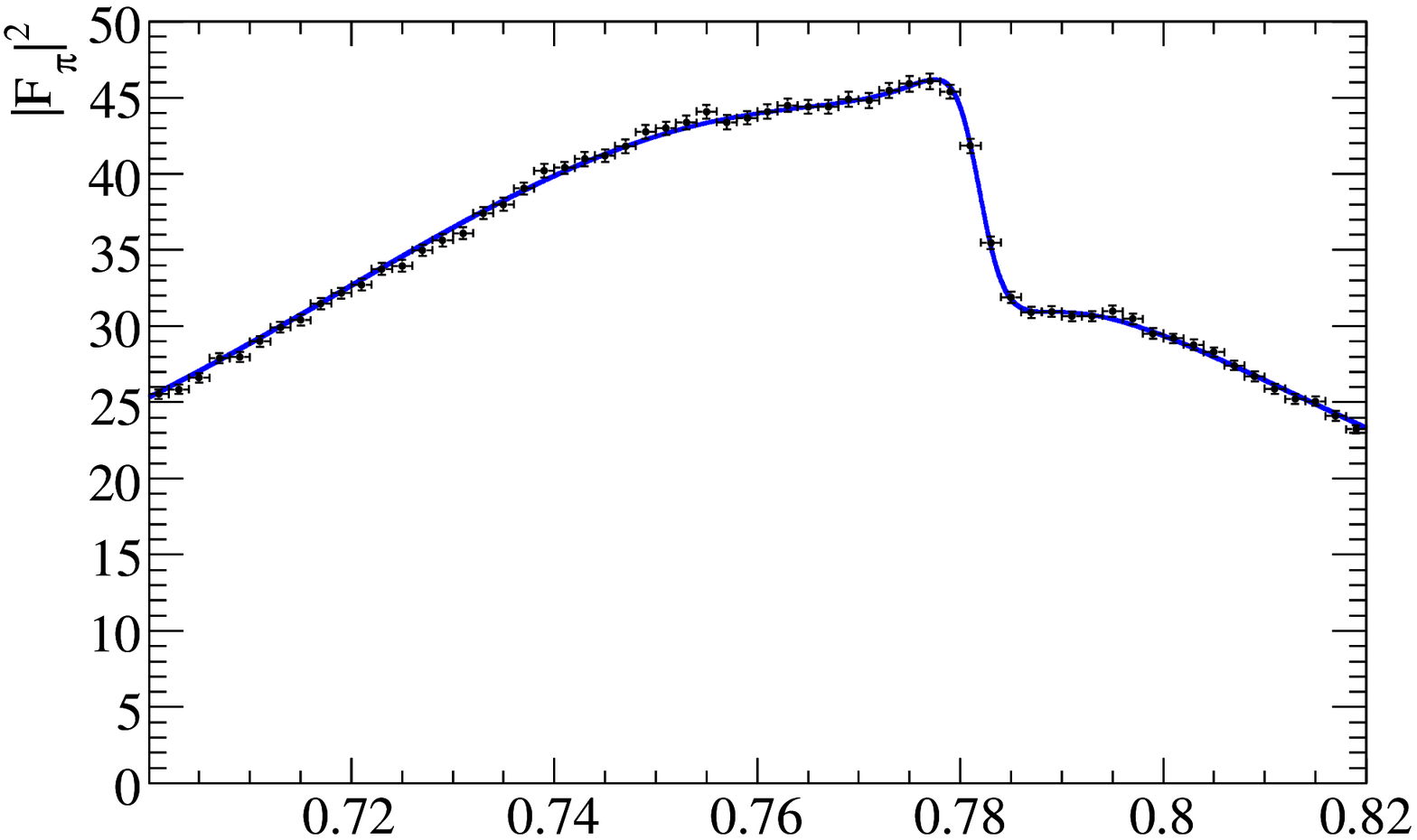}
  \put(-315,0){\scriptsize {$\bm{\sqrt{s}}$ \textbf{[GeV]}}}
  \put(-110,0){\scriptsize {$\bm{\sqrt{s}}$ \textbf{[GeV]}}}
  \caption{Pion form factor squared, $|F_\pi|^2$. \textit{Left:} Fit
    to the BaBaR data~\cite{Lees:2012cj} using   the vector-dominance
    model (blue line). \textit{Right:} Zoom to the energies around
    $\sqrt s \simeq m_\rho$.}
  \label{fig:BaBaR}
\end{figure}

\begin{figure}[t]
  \centering
  \includegraphics[width=0.65\textwidth]{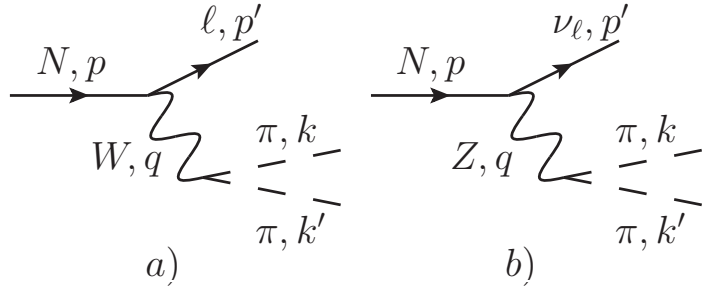}
  \caption{Diagram for the HNL decay into 2 pions.}
  \label{fig:Nto2pi}
\end{figure}

For the case of 2 pions the matrix element of the axial current is
equal to zero, so the general expression for matrix element is (c.f. \eqref{eq:22})
\begin{equation}
	\bra{\pi(p')}J_\mu \ket{\pi(p)} =
  f_+(q^2) (p+p')_\mu + f_-(q^2) q_\mu,
\end{equation}
where $J_\mu$ is one of the currents~(\ref{eq:JEM})-(\ref{eq:JZ}) and
$q_{\mu} = (p-p')_\mu$. Because of isospin
symmetry~(\ref{eq:relation4even}) the form factors are related as
\begin{equation}
	f_{\pm}^{\text{EM}} \approx \frac{1}{\sqrt{2}} f_{\pm}^W \approx \frac{1}{1 - 2 \sin^2 \theta_W} f_{\pm}^Z
\end{equation}
Electromagnetic current conservation $q_\mu J^\mu=0$ implies
$f_-^{\text{EM}}(q^2)=0$. Therefore all the matrix elements could be
expressed via only one form factor, called \emph{pion electromagnetic form factor},
\begin{equation}
	\bra{\pi(p')}J_\mu^{\text{EM}} \ket{\pi(p)} =
  F_\pi(q^2) (p+p')_\mu.
\end{equation}

Pion electromagnetic form factor is related to the cross section of reaction $e^+ e^-\to 2\pi$ as
\begin{equation}
  \sigma(e^+e^- \to 2\pi) = \frac{\pi \alpha_{\text{EM}}^2}{3 s}\beta_{\pi}^3(s) |F_\pi(s)|^2,
\end{equation}
where $\beta_{\pi}(s) = \sqrt{1 - 4 m_{\pi}^2/s}$, so it is
well-measured experimentally. There are a lot of data on electromagnetic form factor~\cite{Barkov:1985ac, Akhmetshin:2006bx,Ambrosino:2010bv,Babusci:2012rp,Lees:2012cj,Ablikim:2015orh}, which agree with each other. Good description of the data is given by the vector-dominance model (VDM), see Fig.~\ref{fig:BaBaR}~\cite{Lees:2012cj} and Appendix~\ref{app:VDM} for model description.

Using matrix elements described above it is easy to find the decay widths of $N\to \ell \pi^0 \pi^+$ and 
$N\to \nu_{\ell} \pi^+ \pi^-$ (see Feynman diagrams in Fig.~\ref{fig:Nto2pi}),
\begin{align}
    \Gamma(N \to \ell_\alpha \pi^0 \pi^+) &= 
    \frac{G_F^2 M_N^3}{384\pi^3} 
    |V_{ud}|^2 |U_{\alpha}|^2
    \!\!\!\!
    \int\limits_{4 m_\pi^2}^{(M_N - m_\ell)^2}
    \left(
    (1-x_\ell^2)^2 + \frac{q^2}{M_N^2}(1 + x_\ell^2) - 2\frac{q^4}{M_N^4}
    \right)
    \times \nonumber \\ &\qquad\qquad\qquad\times
    \lambda^{1/2}\left(1, \frac{q^2}{M_N^2}, x_\ell^2 \right)
    \beta_{\pi}^3(q^2) |F_\pi(q^2)|^2 dq^2,
    \label{eq:N2lpipi}
    \\
    \Gamma(N \to \nu_\alpha \pi^+ \pi^-) &=
    \frac{G_F^2 M_N^3}{768\pi^3} 
    |U_{\alpha}|^2 (1 - 2 \sin^2\theta_W)^2
    \times \nonumber \\ &\qquad\times
    \int\limits_{4 m_\pi^2}^{M_N^2}
    \left(1-\frac{q^2}{M_N^2}\right)^2
    \left(1+\frac{2q^2}{M_N^2}\right)
    \beta_{\pi}^3(q^2) |F_\pi(q^2)|^2 dq^2,
    \label{eq:N2vpipi}
\end{align}
where
$x_\ell = \dfrac{m_\ell}{M_N}$ and the function $\lambda$ is given
by~(\ref{eq:37}). The decay width $\Gamma(N\to \nu_{\alpha} \pi^0 \pi^0) = 0$ because of Eq.~\eqref{eq:2pi0prohibition}.

Using VDM model, formula~(\ref{eq:N2lpipi}) and lifetime of the $\tau$-lepton we have calculated the branching ratio $\text{BR}(\tau \to \nu_{\tau} \pi^-\pi^0) = 25.2\%$ which is close to the experimental value $25.5\%$. 

The decay into 2 pions is significantly enhanced by the
$\rho$-resonance. It turns out, that this is the dominant channel, see
Fig.~\ref{fig:Grho_vs_G2pi} with comparison of the decay width of
HNL into 2 pions and into $\rho$-meson. Therefore, one can replace the
decay into 2 pions with 2-body decay into $\rho$-meson.

%
%

\section{Phenomenological parameters}

In this Section we summarize parameters used in this work. Values of
the CKM matrix elements are given in Table~\ref{tab:CKM}.

\begin{table}[!ht]
  \centering
  \begin{tabular}[c]{|>{$}c<{$}|>{$}c<{$}|>{$}c<{$}|>{$}c<{$}|>{$}c<{$}|>{$}c<{$}|}
    \hline
    V_{ud}  & V_{us} & V_{ub} & V_{cd} & V_{cs} & V_{cb}
    \\ \hline
    0.974 & 0.225 & 0.00409 & 0.220 & 0.995 & 0.0405
    \\ \hline
  \end{tabular}
  \caption[CKM matrix elements]{CKM matrix elements~\cite{Olive:2016xmw} adopted in this work.}
  \label{tab:CKM}
\end{table}

\subsection{Meson decay constants}
\label{sec:phen-const}

The decay constants for charged pseudoscalar mesons are defined by
Eq.~\eqref{eq:15}, Values of $f_h$ (Table~\ref{tab:f_meson})
are measured experimentally and/or obtained by lattice
calculations~\cite{Rosner:2015wva}.

Meson decay constants for the mesons with the same-flavour quarks are defined by Eq.~\eqref{eq:fhneutralscalar}.
There is a discrepancy regarding their values in the literature, therefore we have computed them directly (see Appendix~\ref{sec:etaetaprime}).
The results of these computations are given in the right column of Table~\ref{tab:f_meson}.
The meson decay constants for neutral mesons consisted of quarks of different flavors (such as $K^0$, $D^0$, $B^0$, $B_s$) are not needed in computing HNL production or decay, we do not provide them here.

For vector charged mesons the decay constants $g_h$ are defined by
Eq.~\eqref{eq:vectorgh}. In the literature they often appear as $f_h$,
connected to our prescription by mass of the meson $g_h = f_h
m_h$. Their values are presented in Table~\ref{tab:g_meson}. 
For vector neutral mesons the decay constants $g_h$ and dimensionless
factors $\kappa_h$ are defined by Eq.~\eqref{eq:vectorgh0}. Their
values are presented in Table~\ref{tab:g_meson} as well.

\begin{table}[!h]
  \centering
  \begin{tabular}[t]{|>{$}l<{$}|>{$}r<{$}|}
    \hline
    f_{\pi^+} & \unit[130.2]{MeV}~\text{\cite{Rosner:2015wva}} \\
    f_{K^+} & \unit[155.6]{MeV}~\text{\cite{Rosner:2015wva}} \\
    f_{D^+} & \unit[212]{MeV}~\text{\cite{Rosner:2015wva}} \\
    f_{D_s} & \unit[249]{MeV}~\text{\cite{Rosner:2015wva}} \\    
    f_{B^+} & \unit[187]{MeV}~\text{\cite{Rosner:2015wva}} \\
    f_{B_c} & \unit[434]{MeV}~\text{\cite{Colquhoun:2015oha}}\\ 
    \hline    
  \end{tabular}
  ~ 
  \begin{tabular}[t]{|>{$}l<{$}|r|}
    \hline
    f_{\pi^0} & $\unit[130.2]{MeV}$\,\footnote{It should be equal to $f_{\pi^+}$, according to Eq.~\eqref{eq:relationpia1}.} \\ 
    f_{\eta} & $\unit[81.7]{MeV}$\,\footnote{See discussion in Section~\ref{sec:etaetaprime}.\label{foot:eta}}  \\ 
    f_{\eta'} & $-\unit[94.7]{MeV}^{\,\text{\ref{foot:eta}}}$  \\
    f_{\eta_c} & $\unit[237]{MeV}$\,\footnote{See discussion in Section~\ref{sec:fhetac}.}
    \\
    \hline    
  \end{tabular}
  \caption{Decay constants of pseudoscalar charged mesons (left table) and
    pseudoscalar neutral mesons (right table). }
  \label{tab:f_meson}
\end{table}

\begin{table}[!h]
  \centering
  \begin{tabular}[c]{|>{$}l<{$}|r|}
    \hline
    g_{\rho^+} & $\unit[0.162]{GeV^2}$~\cite{Ebert:2006hj}\footnote{See discussion in the section~\ref{sec:rho-decay-constant}.} \\
    g_{D^{+*}} & $\unit[0.535]{GeV^2}$~\cite{Dhiman:2017urn}\\ 
    g_{D_s^{+*}} & $\unit[0.650]{GeV^2}$~\cite{Dhiman:2017urn}\\
    \hline    
  \end{tabular}
  ~ 
  \begin{tabular}[c]{|>{$}c<{$}|c|c|}
    \hline
    h & $g_h~[\unit{GeV^2}]$ & $\kappa_h$ \\
    \hline
    \rho^0 & $0.162$ & $1-2\sin^2\theta_W$\footnote{See Eq.~\eqref{eq:relation4even}.} \\
    \omega & $0.153$~\cite{Atre:2009rg} & $\frac{4}{3}\sin^2\theta_W$ \\
    \phi & $0.234$~\cite{Ebert:2006hj} & 
    $\frac{4}{3}\sin^2\theta_W - 1$\\
    J/\psi & $1.29$~\cite{Becirevic:2013bsa} & 
    $1 - \frac{8}{3}\sin^2\theta_W$\\
    \hline    
  \end{tabular}
  \caption{Decay constants of vector charged mesons (left table) and
    vector neutral mesons (right table). Decay constants for
    $D_{(s)}^*$ mesons in~\cite{Dhiman:2017urn} show large theoretical
    uncertainty, we quote only the average value here.}
  \label{tab:g_meson}
\end{table}

\subsubsection[Decay constants of $\eta$ and $\eta'$ mesons]{Decay constants of $\eta$ and $\eta'$ mesons}
\label{sec:etaetaprime}

To describe HNL decays into $\eta$ and $\eta'$ mesons we need to know
the corresponding neutral current decay constants, that we define as \eqref{eq:fhneutralscalar}
\begin{equation*}
  \bra{0}J_\mu^{Z}\ket{h_P^0} \equiv - i \frac{f_h}{\sqrt{2}} p_\mu,
\end{equation*}
where where $p_\mu$ is the 4-momentum of the pseudo-scalar meson $h$, $J_\mu^{Z}$ current is given by Eq.~\eqref{eq:JZ}.
The choice of the additional factor $(-1/\sqrt{2})$ is introduced in order to obtain $f_{\pi^0} = f_{\pi^\pm}$ and $f_{\pi^0} > 0$, see discussion below. Taking into account that for pseudoscalar mesons only axial part of the current contributes to this matrix element we can write the matrix element as
\begin{equation}
  \bra{0}J_\mu^{Z}\ket{h_P^0} = 
  \bra{0}\bar{q}\gamma_{\mu}\gamma^5 \lambda^Z q\ket{h_P^0},
\end{equation}
where
\begin{equation}
 q =
 \begin{pmatrix}
  u \\
  d \\
  s
 \end{pmatrix},
 \qquad
 \lambda^Z = \frac{1}{2}
 \left(\begin{array}{rrr}
  -1 & 0 & 0 \\
  0  & \hphantom{-}1 & 0 \\
  0  & 0 & \hphantom{-}1
 \end{array}\right).
\end{equation}
The relevant decay constants are $f^0$ and $f^8$, they come from the
set of extracted from experiments decay constants defined as~\cite{Escribano:2015yup}
\begin{equation}
 \bra{0} J_{\mu}^a\ket{h} = i f_{h}^a p_{\mu},
\end{equation}
where $J_{\mu}^a = \bar{q}\gamma_\mu \gamma^5 \dfrac{\lambda^a}{\sqrt{2}} q$ with $\lambda^a$ being the Gell-Mann matrices for $a=1\dots 8$ and
\begin{equation}
 \lambda^0 = \sqrt{\frac{2}{3}}\left(\begin{array}{rrr}
  1 & 0 & 0 \\
  0 & \hphantom{-}1 & 0 \\
  0 & 0 & \hphantom{-}1
 \end{array}\right)
\end{equation}
The overall factor in $\lambda^0$ is chosen to obey normalization condition $\text{Tr}(\lambda^a \lambda^b) = 2 \delta^{ab}$.

Within the chiral perturbation theory
($\chi$PT)~(see~\cite{Scherer:2002tk} and references therein), the lightest mesons
corresponds to pseudogoldstone bosons $\phi^a$, that appear after
the spontaneous breaking of $U_L(3)\times U_R(3)$ symmetry to group
$U_V(3)$. States $\phi^a$ are orthogonal in the sense
\begin{equation}
 \bra{0} J_{\mu}^a\ket{\phi^b} = i f_{\phi^b}^a p_{\mu},\quad f_{\phi^b}^a = f_{b}^a\delta^{a b}
\end{equation}
where 
and $f_{b}^a$ are corresponding decay constants. 
Using
\begin{equation}
 \lambda^3 = 
 \left(\begin{array}{rrr}
  1 & 0 & 0 \\
  0 & -1 & 0 \\
  0 & 0 & \hphantom{-}0
 \end{array}\right),
 \qquad
 \lambda^8 = \sqrt{\frac{1}{3}}\left(\begin{array}{rrr}
  1 & 0 & 0 \\
  0 & \hphantom{-}1 & 0 \\
  0 & 0 & -2
 \end{array}\right),
\end{equation}
we can rewrite the axial part of the weak neutral current~\eqref{eq:fhneutralscalar} as a linear combination of the $J_{\mu}^0$, $J_{\mu}^3$ and $J_{\mu}^8$
\begin{equation}
 \bar{q}\gamma_{\mu}\gamma^5 \lambda^Z q =
 \frac{1}{\sqrt{2}} \left(\frac{J_{\mu}^0}{\sqrt{6}} - \frac{J_{\mu}^8}{\sqrt{3}} - J_{\mu}^3\right)
\end{equation}
and $f_h$ is given by
\begin{equation}
 f_h = f_h^3 + \frac{f_h^8}{\sqrt{3}} -
  \frac{f_h^0}{\sqrt{6}}.
  \label{eq:fh0ChPT}
\end{equation}
For example, $\pi^0$ meson corresponds to $\phi^3$ state in $\chi$PT, so $f_{\pi^0}^0 = f_{\pi^0}^8 = 0$ and Eq.~\eqref{eq:fh0ChPT} gives $f_{\pi^0} = f_{\pi^0}^3 = f_{\pi^+}$ because of isospin symmetry, in full agreement with Eq.~\eqref{eq:relationpia1}. 

For $\eta$ and $\eta'$ application of Eq.~\eqref{eq:fh0ChPT} is not so straightforward.
These mesons are neutral unflavoured mesons with zero isospin and they can oscillate between each other.
So $\eta$ and $\eta'$ do not coincide with any single  $\phi^a$ state.
Rather they are mixtures of $\phi^0$ and $\phi^8$ states.
In real world isospin is not a conserved quantum number, so $\phi^3$ state also should be taken into account, but its contribution is negligible~\cite{Feldmann:1999uf}, so we use $f^3_{\eta}=f^3_{\eta'}=0$.
Another complication is $U(1)$ QCD anomaly for $J^0_{\mu}$ current that not only shifts masses of corresponding mesons but also contributes to $f^0_h$ meson constant.
To phenomenologically take into account the effect of anomaly it was proposed to use two mixing angles scheme~\cite{Leutwyler:1997yr},
\begin{equation}
 \left(\begin{array}{cc}
  f_\eta^8 & f_\eta^0 \\
  f_{\eta'}^8 & f_{\eta'}^0
 \end{array}\right)
 =
 \left(\begin{array}{lr}
  f_8 \cos \theta_8 & \,-f_0 \sin\theta_0 \\
  f_8 \sin \theta_8 & f_0 \cos\theta_0
 \end{array}\right).
\end{equation}
Taking parameter values from the recent phenomenological analysis~\cite{Escribano:2015yup},
\begin{equation}
 f_8 = 1.27(2) f_{\pi},\quad 
 f_0 = 1.14(5) f_{\pi},\quad
 \theta_8 = -21.2(1.9)^{\circ},\quad
 \theta_0 = -6.9(2.4)^{\circ},
\end{equation}
we find
\begin{align}
 f_{\eta} &= 0.63(2) f_{\pi} \approx 81.7(3.1)\text{ MeV},
 \\
 f_{\eta'} &= -0.73(3) f_{\pi} \approx -94.7(4.0)\text{ MeV}.
\end{align}
These numbers should be confronted with the values quoted
in~\cite{Gorbunov:2007ak} and \cite{Atre:2009rg}.

\subsubsection[Decay constant of the $\eta_c$ meson]{Decay constant of $\eta_c$ meson}
\label{sec:fhetac}

The decay constant of $\eta_c$ meson is defined as~\cite{Deshpande:1994mk}
\begin{equation}
  \bra{0}\bar{c}\gamma^\mu\gamma^5 c\ket{\eta_c} \equiv i f_{\eta_c}^{\text{exp}} p^\mu,
\end{equation}
where $f_{\eta_c}^{\text{exp}} = \unit[335]{MeV}$, as measured by CLEO collaboration~\cite{Edwards:2000bb}. Our definition~\eqref{eq:fhneutralscalar} differs by factor $\sqrt{2}$, so $f_{\eta_c} = f_{\eta_c}^{\text{exp}}/\sqrt{2} \approx \unit[237]{MeV}$.

\subsubsection{Decay constant of $\rho$ meson}
\label{sec:rho-decay-constant}
There are 2 parametrizations of the $\rho$ charged current matrix element using $g_\rho$, defined by~(\ref{eq:vectorgh}), or $f_\rho$, which is related to $g_\rho$ are $f_\rho = g_{\rho}/m_{\rho}$. The value of the decay constant can be obtained by 2 methods: from $\rho \to e^+ e^-$ using approximate symmetry~(\ref{eq:relation4even}) or from the $\tau$-lepton decay. Results, obtained in Ref.~\cite{Ebert:2006hj} by these two method differ by about~$5\%$, $f_{\rho,ee}=220(2)$~MeV and $f_{\rho,\tau}=209(4)$~MeV.
We calculate
\begin{align}
	\Gamma(\tau\to \nu\rho) &= 
	\frac{G_F^2 g_{\rho}^2 m_{\tau}^3}{16\pi m_{\rho}^2} |V_{ud}|^2
	\left( 1 + 2 \frac{m_{\rho}^2}{m_\tau^2} \right)
	\left( 1 - \frac{m_{\rho}^2}{m_\tau^2} \right)^2,
	\\
	\Gamma(\rho\to e^+ e^-) &= \frac{e^4 g_\rho^2}{24 \pi m_{\rho}^3},
\end{align}
and get $g_{\rho,\tau} = \unit[0.162]{GeV}^{2}$ and $g_{\rho,ee} = \unit[0.171]{GeV}^{2}$, which corresponds to 
$f_{\rho,\tau} = \unit[209]{MeV}$ and $f_{\rho,ee} = \unit[221]{MeV}$ in full
agreement with~\cite{Ebert:2006hj}. The difference between these results can
be explained by the approximaty of the relation~(\ref{eq:relation4even}). So
we use $g_{\rho,\tau}$ value as more directly measured one. The results of our
analysis agrees with $f_{\rho}$ value in~\cite{Atre:2009rg} (within about
$10\%$), but differ from the value adopted in~\cite{Gorbunov:2007ak} by $\sim 25\%$.

\subsection{Meson form factors of decay into pseudoscalar meson}
\label{sec:meson-form factors}

To describe the semileptonic decays of the pseudoscalar meson into
another pseudoscalar meson one should know the form factors
$f_{+}(q^2)$, $f_{0}(q^2)$, $f_{-}(q^2)$ defined by Eq.~\eqref{eq:22},
only two of which are independent. We  use $f_{+}(q^2)$, $f_{0}(q^2)$
pair for the decay parametrization.

In turn, there are many different parametrizations of meson form factors. One popular parametrization is the Bourrely-Caprini-Lellouch (BCL) parametrization~\cite{Bourrely:2008za} that takes into account the analytic properties of form factors (see e.g.~\cite{Na:2015kha,Aoki:2016frl}),
\begin{equation}
  \label{eq:27}
  f(q^2) = \frac{1}{1-q^2/{M_{\rm pole}^2}} \sum_{n=0}^{N-1} a_n\biggl[\bigl(z(q^2)\bigr)^n - (-1)^{n-N} \frac nN \bigl(z(q^2)\bigr)^N\biggr]
\end{equation}
where 
the function $z(q^2)$ is defined via
\begin{equation}
  \label{eq:28}
  z(q^2) \equiv\frac{\sqrt{t_+ - q^2} - \sqrt{t_+ - t_0}}{\sqrt{t_+ - q^2} + \sqrt{t_+ - t_0}}
\end{equation}
with
\begin{equation}
  \label{eq:16}
  t_+  = \bigl(m_h + m_{h'}\bigr)^2. 
\end{equation}
The choice of $t_0$ and of the pole mass $M_{\text{pole}}$ varies from
group to group that performs the analysis. In this work we follow FLAG collaboration~\cite{Aoki:2016frl} and take
\begin{equation}
 t_0 = \bigl(m_h + m_{h'}\bigr) \bigl(\sqrt {m_h} - \sqrt{m_{h'}}\bigr)^2.
 \label{eq:t0FLAG}
\end{equation}
The coefficients $a_n^+$ and $a_n^0$ are then fitted to the experimental data or lattice results. 

\subsubsection{K meson form factors}

Form factors of $K\to \pi$ transition are well described by the linear approximation~\cite{Yushchenko:2003xz,Lai:2007dx}
\begin{equation}
 f^{K\pi}_{+,0}(q^2) = f^{K\pi}_{+,0}(0)
 \left(
 1 + \lambda_{+,0}\frac{q^2}{m_{\pi^+}^2}
 \right).
 \label{eq:fKpi}
\end{equation}
The best fit parameters are given in Table~\ref{tab:Kscalarformfactors}.

\begin{table}[!th]
\centering
\begin{tabular}{|c|c|c|c|}
\hline
$h,h'$ & $f_{+,0}(0)$ & $\lambda_{+}$ & $\lambda_{0}$  
\\ \hline
$K^0,\pi^+$ & $0.970$ & $0.0267$ & $0.0117 $ \\ \hline
$K^+,\pi^0$ & $0.970$ & $0.0277$ & $0.0183$ \\ 
\hline
\end{tabular}
\caption{Best fit parameters for the form factors~\eqref{eq:fKpi} of $D\to \pi$ and $D\to K$ transitions~\cite{Yushchenko:2003xz,Lai:2007dx,Aoki:2016frl}.}
\label{tab:Kscalarformfactors}
\end{table}

\subsubsection{D meson form factors}

In the recent paper~\cite{Lubicz:2017syv} the form factors for $D\to K$ and $D\to \pi$ transitions are given in the form
\begin{equation}
 f(q^2) = \frac{f(0) - c(z(q^2)-z_0)\left(1+\frac{z(q^2)+z_0}{2}\right)}{1 - P q^2},
 \label{eq:fLubicz}
\end{equation}
where $z_0=z(0)$. The best fit parameter values are given in Table~\ref{tab:Dscalarformfactors}.

\begin{table}[!th]
\centering
\begin{tabular}{|c|c|c|c|}
\hline
$f$ & $f(0)$ & $c$ & $P~(\text{GeV}^{-2})$  \\ \hline
$f_+^{DK}$ & $0.7647$ & $0.066$ & $0.224 $ \\ \hline
$f_0^{DK}$ & $0.7647$ & $2.084$ & $0$ \\ 
\hline
$f_+^{D\pi}$ & $0.6117$ & $1.985$ & $0.1314$ \\ \hline
$f_0^{D\pi}$ & $0.6117$ & $1.188$ & $0.0342$ \\ \hline
\end{tabular}
\caption{Best fit parameters for the form factors~\eqref{eq:fLubicz} of $D\to \pi$ and $D\to K$ transitions~\cite{Lubicz:2017syv}.}
\label{tab:Dscalarformfactors}
\end{table}

Form factors of $D_s\to \eta$ transition read~\cite{Duplancic:2015zna}
\begin{align}
 f_+^{D_s\eta}(q^2) &= 
 \frac{f_+^{D_s\eta}(0)}{\left(1 - q^2/m_{D_s^*}^2\right)\left(1 - \alpha_+^{D_s\eta} q^2/m_{D_s^*}^2\right)},
 \\
 f_0^{D_s\eta}(q^2) &= 
 \frac{f_0^{D_s\eta}(0)}{1 - \alpha_0^{D_s\eta} q^2/m_{D_s^*}^2},
\end{align}
where $f_+^{D_s\eta}(0) = 0.495$, $\alpha_+^{D_s\eta} = 0.198$~\cite{Duplancic:2015zna}, $m_{D_s^*} = \unit[2.112]{GeV}$~\cite{Olive:2016xmw}. Scalar form factor $f_0^{D_s\eta}(q^2)$ is not well constrained by experimental data, so we take $f_0^{D_s\eta}(q^2) = f_+^{D_s\eta}(q^2)$ by Eq.~\eqref{eq:f0def} and $\alpha_0^{D_s\eta} = 0$.

\subsubsection{B meson form factors}

Most of $B$ meson form factors are available in literature in the
form~\eqref{eq:27}, their best fit parameter values are given in Table~\ref{tab:Bscalarformfactors}. 
The form factors for $B_s\to D_s$ are almost the same as for $B\to D$ transition~\cite{Monahan:2017uby}, so we use the same expressions for both cases. 

\begin{table}[!th]
\centering
\begin{tabular}{|c|c|c|c|c|}
\hline
$f$ & $M_{\text{pole}}$~(GeV) & $a_0$ & $a_1$ & $a_2$ \\ \hline
$f_+^{B_{(s)}D_{(s)}}$ & $\infty$ & $0.909$ & $-7.11$ & $66$ \\ \hline
$f_0^{B_{(s)}D_{(s)}}$ & $\infty$ & $0.794$ & $-2.45$ & $33$ \\ \hline
$f_+^{B_s K}$ & $m_{B^*} = 5.325$ & $0.360$ & $-0.828$ & $1.1$ \\ \hline
$f_0^{B_s K}$ & $m_{B^*(0^+)} = 5.65$ & $0.233$ & $0.197$ & $0.18$ \\ \hline
$f_+^{B\pi}$ & $m_{B^*} = 5.325$ & $0.404$ & $-0.68$ & $-0.86$ \\ \hline
$f_0^{B\pi}$ & $m_{B^*(0^+)} = 5.65$ & $0.490$ & $-1.61$ & $0.93$ \\ \hline
\end{tabular}
\caption{Best fit parameters for the form factors~\eqref{eq:27} of $B\to \pi$, $B_{(s)}\to D_{(s)}$ and $B_s\to K$ transitions~\cite{Aoki:2016frl}.}
\label{tab:Bscalarformfactors}
\end{table}

\subsection{Meson form factors for decay into vector meson}

One of the relevant HNL production channel is pseudoscalar meson decay
$h_P\to h'_V\ell_\alpha N$. To compute decay width of this decay one
needs to know the form factors $g(q^2)$, $f(q^2)$, $a_{\pm}(q^2)$, defined by Eqs.~(\ref{eq:24}, \ref{eq:23}). The dimensionless linear combinations are introduced as
\begin{align}
 V^{h h'}(q^2) &= \left(m_h + m_{h'}\right) g^{h h'}(q^2), 
 \\
 A_0^{h h'}(q^2) &= \frac{1}{2 m_{h'}} 
 \left(
 f^{h h'}(q^2) + q^2 a_-^{h h'}(q^2) + \left(m_h^2 - m_{h'}^2 \right) a_+^{h h'}(q^2)
 \right),
 \\
 A_1^{h h'}(q^2) &= \frac{f^{h h'}(q^2)}{m_h + m_{h'}},
 \\
 A_2^{h h'}(q^2) &= - \left(m_h + m_{h'}\right) a_+^{h h'}(q^2).
\end{align}
For these linear combinations the following ansatz is used
\begin{align}
 V^{h h'}(q^{2}) &= \frac{f^{h h'}_{V}}{\left(1-q^{2}/(M^{h}_{V})^{2}\right)\left[ 1-\sigma_{V}^{h h'} q^{2} / (M^{h}_{V})^{2} - \xi_{V}^{h h'} q^{4} / (M^{h}_{V})^{4}\right]},
 \label{eq:Vhh}
\\
 A_{0}^{h h'}(q^{2}) &= \frac{f^{h h'}_{A_{0}}}{\left(1-q^{2}/(M^{h}_{P})^{2}\right)\left[ 1-\sigma_{A_{0}}^{h h'} q^{2} / (M^{h}_{V})^{2} - \xi_{A_{0}}^{h h'} q^{4} / (M^{h}_{V})^{4} \right]},
 \label{eq:A0hh}
\\
 A_{1/2}^{h h'}(q^{2}) &= \frac{f^{h h'}_{A_{1/2}}}{1-\sigma_{A_{1/2}}^{h h'} q^{2} / (M^{h}_{V})^2 - \xi_{A_{1/2}}^{h h'} q^{4} / (M^{h}_{V})^{4}}.
 \label{eq:A12hh}
\end{align}
Best fit values of parameters are adopted from papers~\cite{Melikhov:2000yu,Ebert:2006nz,Faustov:2014dxa}. $f$, $\sigma$ parameters are given in Table~\ref{tab:vector-form factors-constants}, while $\xi$ and the pole masses $M_V$ and $M_P$ are given in Table~\ref{tab:pole-masses}.

\begin{table}[!t]
	\centering
	\begin{tabular}{|c|c|c|c|c|c|c|c|c|}
		\hline  
	    $h,h'$ & $f_{V}^{hh'}$& $f_{A_{0}}^{hh'}$& $f_{A_{1}}^{hh'}$& $f_{A_{2}}^{hh'}$ & $\sigma_{V}^{hh'}$ & $\sigma_{A_{0}}^{hh'}$ & $\sigma_{A_{1}}^{hh'}$ & $\sigma_{A_{2}}^{hh'}$
	    \\
		\hline
		$D, K^{*}$ & 1.03 & 0.76 & 0.66 & 0.49 & 0.27 & 0.17 & 0.30 & 0.67 
		\\
		\hline
		$B, D^{*}$ & 0.76 & 0.69 & 0.66 & 0.62 & 0.57 & 0.59 & 0.78 & 1.40 
		\\
		\hline
		$B, \rho$ & 0.295 & 0.231 & 0.269 & 0.282 & 0.875 & 0.796 & 0.54 & 1.34 
		\\
		\hline
		$B_s, D_s^*$ & 0.95 & 0.67 & 0.70 & 0.75 & 0.372 & 0.350 & 0.463 & 1.04
		\\
		\hline
		$B_s, K^*$ & 0.291 & 0.289 & 0.287 & 0.286 & $-0.516$ & $-0.383$ & 0 & 1.05
		\\
		\hline
	\end{tabular}
	\caption{First part of the table with parameters of meson form
          factors~(\ref{eq:Vhh}-\ref{eq:A12hh}) of decay into vector meson~\cite{Melikhov:2000yu,Ebert:2006nz,Faustov:2014dxa}.}
	\label{tab:vector-form factors-constants}
\end{table}

\begin{table}[!t]
	\centering
	\begin{tabular}{|c|c|c|c|c|c|c|}
		\hline  
		$h,h'$ & $\xi_{V}^{hh'}$ &  $\xi_{A_{0}}^{hh'}$ & $\xi_{A_{1}}^{hh'}$ & $\xi_{A_{2}}^{hh'}$ & $M_{P}^{h}$ (GeV) & $M_{V}^{h}$ (GeV) \\
		\hline
		$D, K^{*}$ & 0 & 0 & 0.20 & 0.16 & $m_{D_{s}} = 1.969$ &$m_{D_s^{*}} = 2.112$ \\
		\hline
		$B, D^{*}$ & 0 & 0 & 0 & 0.41 & $m_{B_{c}} = 6.275$& $m_{B_{c}^{*}}= 6.331$\\
		\hline
		$B, \rho$ & 0 & 0.055 & 0 & $-0.21$ & $m_{B} = 5.279$& $m_{B^{*}}= 5.325$\\
		\hline
		$B_s, D_s^*$ & 0.561 & 0.600 & 0.510 & 0.070 & $m_{B_{c}} = 6.275$& $m_{B_{c}^{*}}= 6.331$\\
		\hline
		$B_s, K^*$ & 2.10 & 1.58 & 1.06 & $-0.074$ & $m_{B_{s}} = 5.367$ & $m_{B_{s}^{*}}= 5.415$\\
		\hline
	\end{tabular}
	\caption{Second part of the table with parameters of meson
          form factors~(\ref{eq:Vhh}-\ref{eq:A12hh}) of decay into vector meson~\cite{Melikhov:2000yu,Ebert:2006nz,Faustov:2014dxa}. Masses of $B_c$,  $D_s$ and $D_s^*$ are taken from~\cite{Olive:2016xmw}, while for $B_c^*$ theoretical prediction~\cite{Mathur:2016hsm} is used.}
	\label{tab:pole-masses}
\end{table}

\section{Production from \texorpdfstring{$\Jpsi$}{J/psi} and \texorpdfstring{$\Upsilon$}{Y} mesons}
\label{sec:production-from-Jpsi}

\subsection{Production from \texorpdfstring{$\Jpsi$}{J/psi}}
\label{sec:production-from-jpsi}

The process $\Jpsi \to N \bar\nu$ allows to creates HNLs with mass up to $M_{\Jpsi}\simeq 3.1\GeV$ and therefore contribute to the production \textit{above} the $D$-meson threshold.

To estimate $\BR(\Jpsi \to N \bar\nu)$ let us first compare the processes $\Jpsi\to e^+ e^-$ and $\Jpsi \to \nu_e \bar{\nu}_e$.
The ratio of their width is given by~\cite{Chang:1997tq}
\begin{equation}
  \label{eq:1}
  \frac{\BR(\Jpsi \to \nu_e \bar{\nu}_e)}{\text{BR}(\Jpsi\to e^+ e^-)} =
  \frac{27 G_F^2 M_{\Jpsi}^4}{256 \pi^2 \alpha^2}\left(1-\frac83\sin^2\theta_W\right)^2
  \sim 4.5\times 10^{-7}
\end{equation}
with the precision of the order of few per cent~\cite{Chang:1997tq}.
Using the measured branching ratio $\BR(\Jpsi \to e^+ e^-) \simeq 0.06$~\cite{Olive:2016xmw}, one can estimate decay into \emph{one flavour of neutrinos}, $\BR(\Jpsi\to \nu_e \bar{\nu}_e)\simeq 2.7\times 10^{-8}$.
The corresponding branching of \Jpsi\ to HNL is additionally suppressed by $U^2$ and
by the phase-space factor $f_{PS}$:
\begin{equation}
  \label{eq:2}
  \sum_\alpha\BR(\Jpsi \to N\bar\nu_\alpha) = U^2 
  f_{PS}(M_N/M_{\Jpsi})
  \BR(\Jpsi \to \nu_e \bar{\nu}_e)
\end{equation}

We estimate this fraction at $M_N = M_D$ (just above the $D$-meson threshold) taking for simplicity $f_{PS}=1$.
Clearly, at masses below $M_D$ the production from $D$-mesons dominates (as the \Jpsi\ production fraction $f(\Jpsi) \simeq 0.01$, see \cite[Appendix A]{Alekhin:2015byh}, reproduced for completeness in Appendix~\ref{sec:heavy-flavour}). Above $D$-meson mass but below $M_\Jpsi$ we should compare with the production from $B$ mesons. We compare the probability to produce HNL from $B$-meson and from $\Jpsi$:
\begin{multline}
  \label{eq:Jpsi}
  \frac{\text{HNLs from }\Jpsi}{\text{HNLs from }B}
  =\frac{X_{c\bar c} \times f(\Jpsi)\times \BR_{\Jpsi\to  N\bar \nu}}{X_{b\bar b} \times f(B) \times \BR_{B\to N X}}=\\
 =3\times 10^{-4} \parfrac{X_{c\bar c}}{10^{-3}} \parfrac{10^{-7}}{X_{b\bar b}}
\end{multline}
where we have adopted $f(B) \times BR(B \to N+X) \sim 10^{-2}$ (c.f.\
Fig.~\ref{fig:Dbranching}, right panel) and used $f(\Jpsi) \sim
10^{-2}$.
The numbers in~\eqref{eq:Jpsi2} are normalized to \ship. We see therefore that \Jpsi\ decays contribute sub-dominantly while $X_{b\bar{b}}/X_{c\bar{c}} \gtrsim 10^{-8}$.

\subsection{Production from \texorpdfstring{$\Upsilon$}{Y}}
\label{sec:upsilon}

The heavy mass of $\Upsilon$ opens up a possibility to produce HNLs up to
$M_N \simeq 10\GeV$.  Similarly to Eq.~\eqref{eq:1} we can find the branching ratio
$\BR(\Upsilon\to \nu\bar{\nu}) = 4\times 10^{-4}\BR(\Upsilon \to e^+
e^-)$~\cite{Chang:1997tq}. Therefore
\begin{equation}
  \label{eq:6}
  \BR(\Upsilon\to N\bar{\nu}_\alpha) = U_\alpha^2 f_{PS}(M_N/M_\Upsilon) \frac{27 G_F^2 M_{\Upsilon}^4}{64 \pi^2 \alpha^2}\left(-1+\frac43\sin^2\theta_W\right)^2 \BR(\Upsilon \to e^+
e^-)
\end{equation}
Using the latest measurement $\BR(\Upsilon \to e^+ e^-)\simeq 2.4\times 10^{-2}$~\cite{Olive:2016xmw} one finds that $\BR(\Upsilon\to \nu\bar{\nu}) \simeq 10^{-5}$.
We do not know the fraction $f(\Upsilon)$ out of all $b\bar b$ pairs, but  one can roughly estimate it equal to the fraction $f(\Jpsi)\sim 1\%$ (see Appendix~\ref{sec:heavy-flavour} in~\cite{Alekhin:2015byh}), so
\begin{equation}
  N_{\Upsilon\to N\bar \nu} \simeq 10^{-10} N_{\Upsilon} \times \parfrac{U^2}{10^{-5}}
\end{equation}
where we have normalized $U^2$ to the current experimental limit for $M_N >
5$~GeV (c.f. Fig.~\ref{fig:HNLbounds}).

\section{Production of heavy flavour at SHiP}
\label{sec:heavy-flavour}

For a particular application of the obtained results we revise the HNL
production at the SHiP experiment. 
The number of mesons produced by $E_p=400$\,GeV proton beam at the SHiP target can be estimated as
\begin{equation}
  N_{h} = 2 \times f(h) \times X_{q\overline{q}} \times N_{PoT}
\end{equation}
where $X_{q\overline{q}}$ represents the $q\bar{q}$ production rate,
$f(h)$ is the meson $h$ production fraction\footnote{$f(h)$ is equal
  to the number of $h$ mesons divided by the number of corresponding
  quarks.} and expected number of protons on target is $N_{PoT}= 2\cdot 10^{20}$. The
following cross sections have been used for the estimates:
\begin{itemize}
\item the proton-nucleon cross section is $\sigma(pN)\simeq 10.7$~mbarn.
\item $X_{ss} \approx 1/7$~\cite{Gorbunov:2007ak}.
\item $\sigma(cc)= 18$~$\mu$barn~\cite{Abt:2007zg} and the fraction
  $X_{cc}=1.7\times 10^{-3}$
\item $\sigma(bb)= 1.7$~nbarn~\cite{Lourenco:2006vw} and the fraction
  $X_{bb}=1.6\times 10^{-7}$
\end{itemize}

\begin{figure}[!htb]
    \centering
      \includegraphics[width=0.49\textwidth]{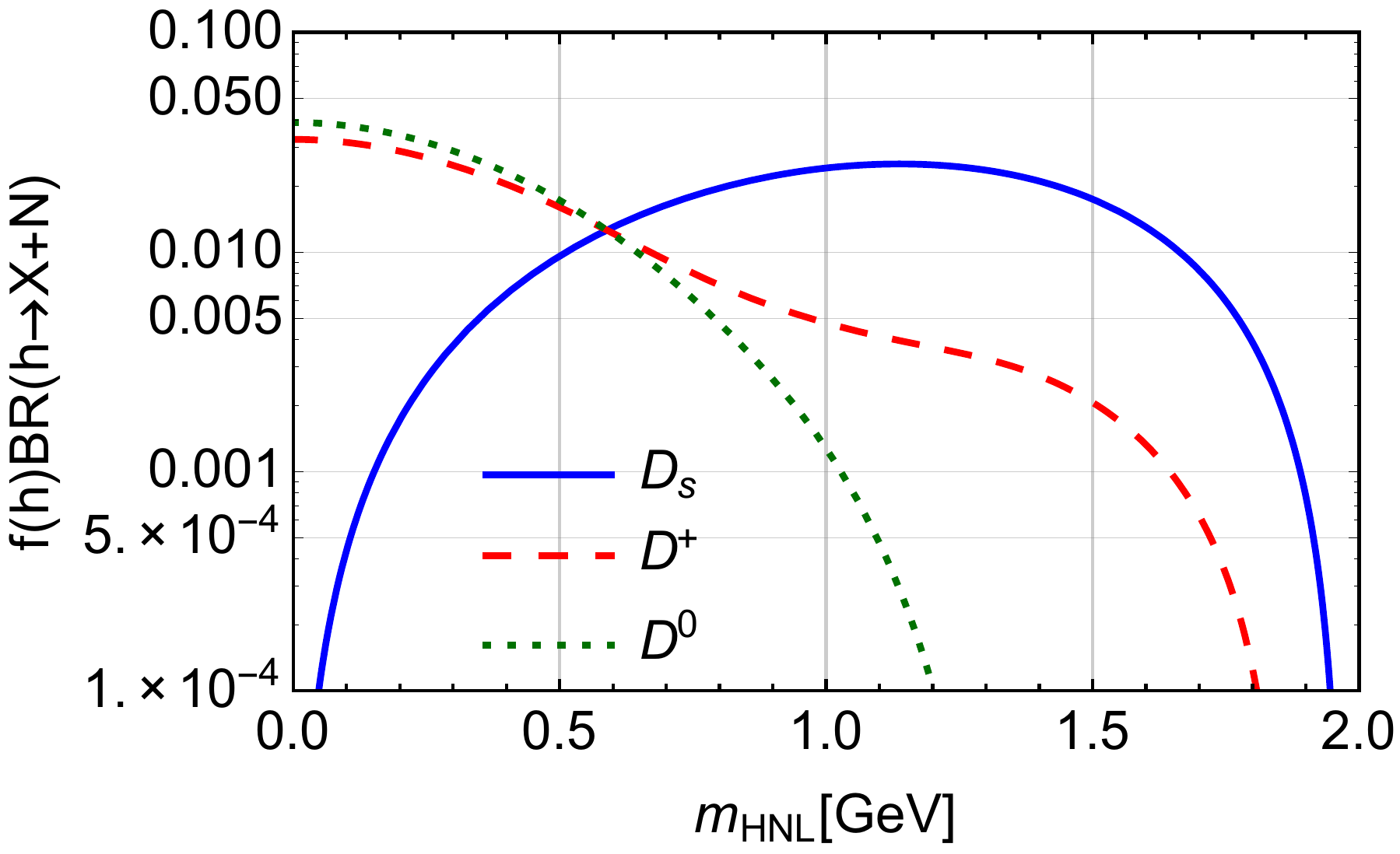}~\includegraphics[width=0.46\textwidth]{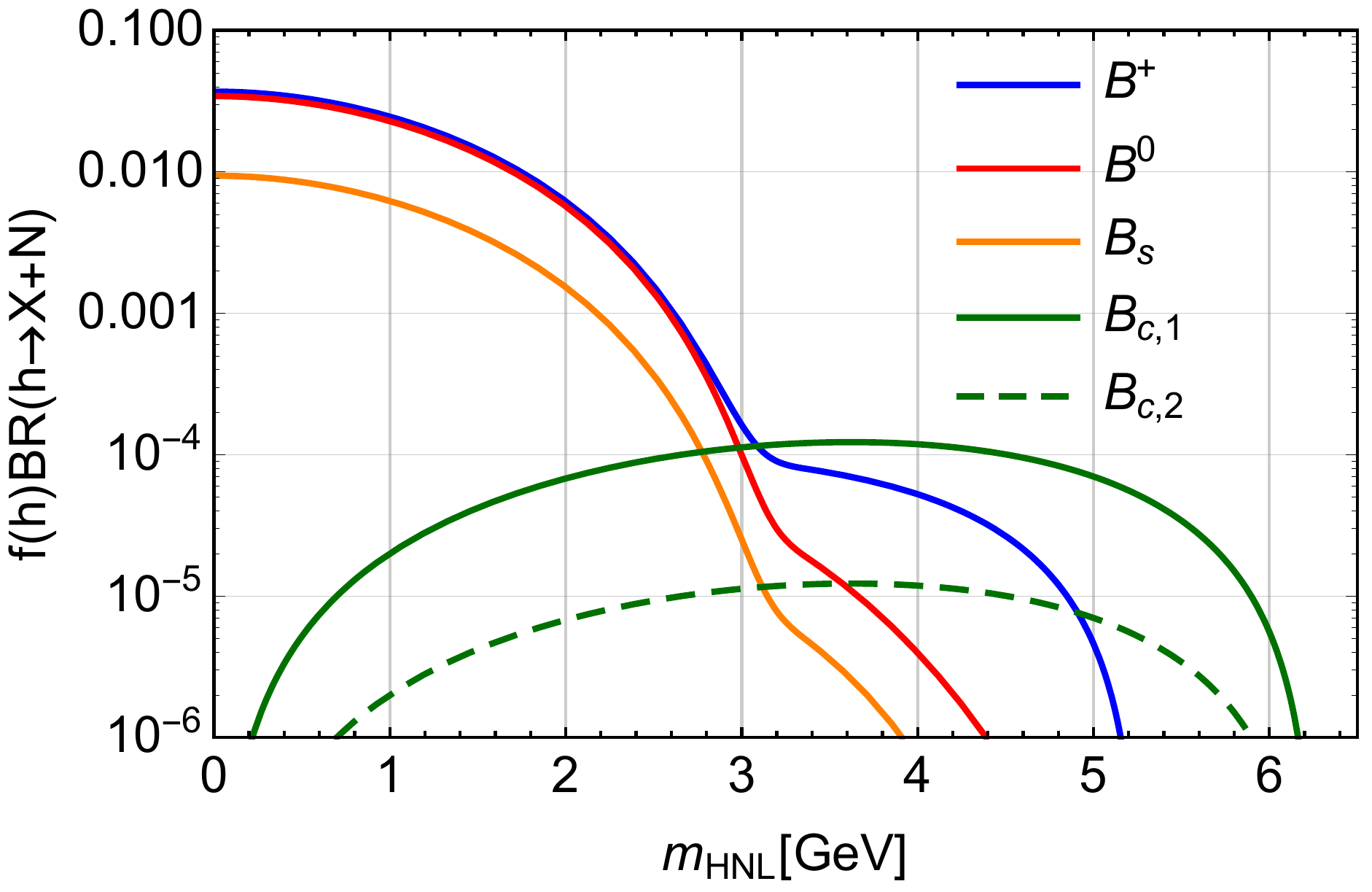}
    \caption{Branching ratio $\times$ Production fraction for charm to HNL (left panel) and for beauty to HNL (right panel) for SHiP experiment and mixing angles $U_e=1$, $U_\mu=U_\tau=0$. For $B_c$ meson $f(B_c)$ is unknown, so the result is shown for two values, $f(B_c) = 2\cdot 10^{-3}$ ($B_{c,1}$ line) and $f(B_c) = 2\cdot 10^{-4}$ ($B_{c,2}$ line).
    }
        \label{fig:ship_production}
  \end{figure}

To calculate the meson production fractions the dedicated simulation
is needed. It should take into account the properties of the target
(materials, geometry) and the cascade processes (birth of the excited
meson states like $D^*$ and its decay into $D$). The values of $f(h)$
for the case of SHiP were calculated in the paper~\cite{Elena}. These
values with the number of different mesons are presented in
Table~\ref{tab:meson_n}. For kaons we do not divide them for
species. Taking into account production fractions of different mesons
the main production channels from charm and beauty quarks for SHiP are
shown in Fig.~\ref{fig:ship_production}.

The expected number of $\tau$-leptons for $N_{PoT}=2\times 10^{20}$ is
$N_{\tau} = 3 \times 10^{15}$.

\begin{table}[t]
\centering
    \begin{tabular}{|c|c|c|}
        \hline
        Meson & $f(h)$ & $N_{h}$ \\
        \hline
        $K$ & $-$ & $5.7\cdot 10^{19}$ \\
        \hline
        $D^{\pm}$ & $0.207$ & $1.4\cdot 10^{17}$ \\
        \hline
        $D^{0}$ & $0.632$ & $4.3\cdot 10^{17}$ \\
        \hline
        $D_{s}$ & $0.088$ & $6.0\cdot 10^{16}$ \\
        \hline
        $\Jpsi$ & $0.01$ & $6.8\cdot 10^{15}$ \\
        \hline
        $B^{\pm}$ & $0.417$ & $2.7\cdot 10^{13}$ \\
        \hline
        $B^{0}$ & $0.418$ & $2.7\cdot 10^{13}$ \\
        \hline
        $B_{s}$ & $0.113$ & $7.2\cdot 10^{12}$ \\
        \hline
    \end{tabular}
    \caption{Production fraction and expected number of different mesons in SHiP.}
    \label{tab:meson_n}
\end{table}

\section{Vector-dominance model}
\label{app:VDM}

Here we provide $F_{\pi}(s)$ formula, given by vector-dominance model~\cite{Lees:2012cj}
\begin{equation}
  F_{\pi}(s) = \frac{
  \text{BW}^{\text{GS}}_{\rho}(s)
  \frac{1 + c_{\omega} \text{BW}^{\text{KS}}_{\omega}(s)}{1 + c_{\omega}} + c_{\rho'}\text{BW}^{\text{GS}}_{\rho'}(s) + c_{\rho''}\text{BW}^{\text{GS}}_{\rho''}(s) + c_{\rho'''}\text{BW}^{\text{GS}}_{\rho'''}(s)
  }{1 + c_{\rho'} + c_{\rho''} +c_{\rho'''}},
\end{equation}
where $c_i = |c_i| e^{i\phi_i}$ are complex amplitudes of the Breit--Wigner (BW) functions. They are different for $\omega$ and $\rho$ mesons. For $\omega$ it is the usual BW function
\begin{equation}
  \text{BW}^{\text{KS}}_{\omega}(s) = \frac{m_{\omega}^2}
  {m_{\omega}^2 - s - i m_{\omega} \Gamma_{\omega}},
\end{equation}
while for $\rho$ mesons Gounaris-Sakurai (GS) model~\cite{Gounaris:1968mw} is taken,
\begin{equation}
  \text{BW}^{\text{GS}}_{\rho_i}(s) = \frac{m_{\rho_i}^2 (1 + d(m_{\rho_i})\Gamma_{\rho_i}/m_{\rho_i})}
  {m_{\rho_i}^2 - s + f(s,m_{\rho_i},\Gamma_{\rho_i})
  - i m_{\rho_i} \Gamma(s, m_{\rho_i},\Gamma_{\rho_i})},
\end{equation}
where
\begin{align}
  \Gamma(s, m,\Gamma) &= \Gamma \frac{s}{m^2} 
  \left(\frac{\beta_{\pi}(s)}{\beta_{\pi}(m^2)}\right)^3, 
  \\
  f(s, m, \Gamma) &= \frac{\Gamma m^2}{k^3(m^2)}
  \left[
  k^2(s)\left( h(s) - h(m^2) \right) +
  (m^2 - s)k^2(m^2)h'(m^2)
  \right]
  \\
  \beta_{\pi}(s) &= \sqrt{1 - \frac{4m_{\pi}^2}{s}}, 
  \\
  d(m) &= \frac{3}{\pi} \frac{m_{\pi}^2}{k^2(m^2)}
  \ln \left( \frac{m^2 + 2 k(m^2)}{2 m_{\pi}} \right) +
  \frac{m}{2\pi k(m^2)} - \frac{m_{\pi}^2 m}{\pi k^3(m^2)}, 
  \\
  k(s) &= \frac{1}{2} \sqrt{s} \beta_{\pi}(s), \\
  h(s) &= \frac{2}{\pi} \frac{k(s)}{\sqrt{s}}
  \ln \left( \frac{\sqrt{s} + 2 k(s)}{2 m_{\pi}} \right)
\end{align}
and $h'(s)$ is a derivative of $h(s)$.


\bibliographystyle{JHEP} %
\bibliography{ship} %

\end{document}